\documentclass[11pt]{article}
\pdfoutput=1

\usepackage{soul}
\usepackage{xkeyval}
\usepackage{graphicx,picture}
\usepackage{amssymb}
\usepackage{amsmath}
\usepackage{bbold}
\usepackage{color}
\usepackage{colordvi}
\usepackage{fancybox, ulem}
\usepackage[footnotesize]{caption2}
\usepackage{graphicx}
\usepackage[center,footnotesize,hang]{subfigure}
\usepackage{bbm}
\usepackage{multirow}
\usepackage{array}
\usepackage{cite}
\usepackage{arydshln}
\usepackage{enumitem}  
\usepackage{slashbox}
\newcommand{\PreserveBackslash}[1]{\let\temp=\\#1\let\\=\temp}
\newcolumntype{C}[1]{>{\PreserveBackslash\centering}p{#1}}
\newcolumntype{R}[1]{>{\PreserveBackslash\raggedleft}p{#1}}
\newcolumntype{L}[1]{>{\PreserveBackslash\raggedright}p{#1}}

\def\thefootnote{\fnsymbol{footnote}}

\newcommand{\LL}{(\overline{L}L)}
\newcommand{\LE}{(\overline{L}E_\RH)}
\newcommand{\QU}{(\overline{Q}U_\RH)}

\newcommand{\FF}{(\overline{F}F)}
\newcommand{\anu}{\overline{\nu}}
\newcommand{\ms}{\mathbf{1}}
\newcommand{\mt}{\mathbf{3}}
\newcommand{\bT}{\mathbb{T}}
\newcommand{\bI}{\mathbb{1}}

\newcommand{\LH}{\text{L}}
\newcommand{\RH}{\text{R}}
\newcommand{\Symm}{\text{S}}
\newcommand{\AntiS}{\text{A}}
\newcommand{\cS}{\mathcal{S}}
\newcommand{\cT}{\mathcal{T}}

\begin{document}

\begin{flushright} 
IPPP/18/4  \\
\end{flushright} 

\begin{center}
{\Large\bf Neutrino non-standard interactions as a portal to test flavour symmetries}
\end{center}

\vspace{0.2cm}

\begin{center}
{\bf TseChun Wang} \footnote{E-mail: tse-chun.wang@durham.ac.uk} \quad {\bf
Ye-Ling Zhou} \footnote{E-mail: ye-ling.zhou@durham.ac.uk}
\\
{Institute for Particle Physics Phenomenology, Department of Physics, \\ Durham University, Durham DH1 3LE, United Kingdom} 
\end{center}

\vspace{1.5cm}

\begin{abstract}

Imposing non-Abelian discrete flavour symmetries to neutrino non-standard interactions (NSIs) is discussed for the first time. 
For definiteness, we choose $A_4$ as the flavour symmetry, which is subsequently broken to the residual symmetry $Z_2$ in the neutrino sector. We provide a general discussion on flavour structures of NSIs from higher-dimensional operators ($d\leqslant8$) without inducing unnecessary tree-level 4-charged-fermion interactions. Both $A_4$- and $Z_2$-motivated NSI textures are obtained. UV completions of higher-dimensional operators lead to extra experimental constraints on NSI textures. 
We study the implementation of matter-effect NSIs in DUNE from phenomenological point of view, and discover that DUNE can test $A_4$ with a high level of statistics. 
We also present exclusion limits of sum rules suggested by UV-complete models. Our result shows that the NSI effects, though predicted to be small for DUNE, could provide useful information that might extend our understanding of the flavour symmetry.

\end{abstract}

\begin{flushleft}
\hspace{0.8cm} PACS number(s): 11.30.Hv, 12.15.Ff, 13.15.+g, 14.60.Pq \\
\hspace{0.8cm} Keywords: non-standard interaction, neutrino oscillation, flavour symmetries
\end{flushleft}

\renewcommand{\thefootnote}{\roman{footnote}}
\setcounter{footnote}{0}


\newpage

\section{Introduction}

Neutrino oscillation experiments have achieved great success in the last two decades \cite{atmospheric,solar,accelerator,reactor}. Two neutrino mass-squared differences ($\Delta m_{21}^2$, $|\Delta m_{31}^2|$) and three mixing angles ($\theta_{12}$, $\theta_{23}$, $\theta_{13}$) have been measured in the standard three-neutrino framework. 
Several next-generation oscillation experiments are proposed, such as the long-baseline accelerator experiments DUNE \cite{Acciarri:2015uup}, T2HK \cite{Hyper-Kamiokande:2016dsw}, the intermediate-baseline reactor experiment JUNO \cite{An:2015jdp,Djurcic:2015vqa}, SBN programme \cite{Antonello:2015lea}, and muon-decay experiments NuSTORM \cite{Adey:2013pio}, MOMENT \cite{Cao:2014bea}, Neutrino Factory \cite{Choubey:2011zzq}, etc. They are aimed at answering the remaining questions in neutrino oscillations: if CP is violated in neutrino oscillations, what the value of the Dirac-type CP-violating phase $\delta$ is, and  which mass ordering ($\Delta m^2_{31}>0$ or $\Delta m^2_{31}<0$) is true. In addition, the already known oscillation parameters can be measured to the percent level and the octant of $\theta_{23}$ ($\theta_{23}<45^\circ$ or $\theta_{23}>45^\circ$) will be determined \cite{future_oscillation_goals,future_oscillation_goals2}. 

These experiments will also test the standard three-neutrino mixing scenario and  might unveil new neutrino couplings beyond the Standard Model (SM). Neutrino nonstandard interactions (NSIs) provide a model-independent framework of studying new physics in neutrino oscillation experiments (for some reviews, see \cite{NSI_review}). 
They are usually considered as effective descriptions of contributions from higher-dimensional operators mediated by heavy mediators \cite{Antusch:2008tz, Gavela:2008ra, Wise:2014oea, Forero:2016ghr},  although they may also be induced by light mediators with very weak couplings  (seeing e.g., \cite{Heeck:2011wj, Farzan:2015doa}). 
In neutrino oscillation experiments, NSIs may appear at neutrino sources, detectors or during neutrino propagation. 
There are no experimental hints for NSIs at the source and the detector \cite{NSI_review, Akimov:2017ade}. Current global-fit results for NSIs during neutrino propagation, i.e., matter-effect NSIs, reach the precision from a few to tens of percentages of the strength of the standard matter effect induced by the weak interaction \cite{Gonzalez-Garcia:2013usa}. 
Due to precision upgrades and because of nonnegligible matter effects, the testability of NSIs in DUNE and T2HK (as well as its alternative T2HKK), and the influences on measurements of mass ordering and CP violation have received a lot of attentions (see, e.g., in \cite{DUNE_NSIs,DUNE_NSIs2,DUNE_NSIs3,DUNE_NSIs4,Blennow:2016jkn}). For NSIs studied in other future experiments, see, e.g., Refs. \cite{NF_NSIs,Tang:2017qen,Rahman:2015vqa,Dasgupta:2012dn,Coloma:2011rq}. 

One important theoretical development promoted by neutrino oscillations is the application of flavour symmetries for understanding lepton flavour mixing. It is directly triggered by the measured values of mixing angles, $\sin^2\theta_{12} \sim 1/3$ and $\sin^2\theta_{23} \sim 1/2$. In the framework of flavour symmetries, it is assumed that an underlying discrete flavour symmetry $G_\text{f}$ exists at some high energy scale. It unifies the three flavours together. After the flavour symmetry is broken at a lower energy scale, special flavour structures arise. The most famous group used as a flavour symmetry is the tetrahedral group $A_4$ \cite{Ma:2001dn}. Most $A_4$ models naturally predict $\sin^2\theta_{12}=1/3$, $\sin^2\theta_{23}=1/2$ but $\sin^2\theta_{13}=0$ \cite{Altarelli:2005yp, Altarelli:2005yx, Lam:2008sh}, i.e., the so-called tri-bimaximal (TBM) mixing \cite{TBM}. 
One important feature of these models are the correspondence between the mixing and  the existence of the residual symmetries $Z_3$ and $Z_2$ after $A_4$ breaking (for some reviews, see e.g. \cite{review}). $Z_3$ and $Z_2$ are subgroups of $A_4$. They are approximately preserved in the charged lepton and neutrino sectors, respectively, acting on charged leptons and neutrinos separately as
\begin{eqnarray}
Z_3: &&~\, e \to e\,,\hspace{3.4cm} \mu \to e^{-i2\pi/3} \mu\,,\hspace{2.2cm}  \tau \to e^{i2\pi/3} \tau \,; \nonumber\\
Z_2: &&\nu_e \to \frac{1}{3} (-\nu_e + 2 \nu_\mu + 2 \nu_\tau)\,,~ \nu_\mu \to \frac{1}{3} (-\nu_\mu + 2 \nu_\tau + 2 \nu_e)\,,~ \nu_\tau \to \frac{1}{3} (-\nu_\tau + 2 \nu_e + 2 \nu_\mu) \,.
\end{eqnarray}
Slight breakings of the residual symmetries provide small corrections to the mixing, specifically generating a non-zero $\theta_{13}$ and making all mixing parameters compatible with oscillation data. The preferred parameters of these models will be tested by the future neutrino oscillation experiments. 

Imposing flavour symmetries may not only influence the flavour mixing measured by neutrino oscillation experiments, but also contribute to other flavour-dependent phenomenological signatures, such as the charged lepton flavour violation (CLFV). The influence of flavour symmetries on CLFV processes has been discussed in \cite{triality,3Higgs_lepton,3Higgs_lepton2,3Higgs_lepton3,Kobayashi:2015gwa, Pascoli:2016wlt, Muramatsu:2017xmn}. In particular, the essential contribution of $A_4$ and $Z_3$ on the CLFV decays of charged leptons have been carefully analysed in \cite{Pascoli:2016wlt}. The branching ratio sum rules of these processes have been obtained therein, which can be regarded as specific features of flavour symmetries. In the neutrino sector, as the couplings are too weak, the phenomenological signatures of flavour symmetries beyond the standard neutrino oscillation measurements have been rarely discussed. 

Previous discussions of NSIs in flavour symmetries are limited only in the Abelian case \cite{Heeck:2011wj, Farzan:2015doa, Farzan:2015hkd, Farzan:2016wym, Babu:2017olk}. In these papers, by assuming a gauged $U(1)$  flavour symmetry, relatively sizeable NSIs are generated via flavour-dependent gauge interaction mediated by a gauge boson with the mass around or below the GeV scale. Note that $U(1)$ symmetries proposed in these works are not supposed to explain the lepton flavour mixing. Thus we do not expect any connection between NSIs and lepton flavour mixing. 

In the non-Abelian case, as $e$, $\mu$ and $\tau$ lepton doublets are arranged as a triplet in the flavour space, which both complicates the NSI construction and strengthens experimental constraints. However, if the non-Abelian discrete symmetry is a true symmetry behind, a combined study of the flavour symmetry and NSIs will be required in the future neutrino experiments. Regarding the $A_4$ case, the measurement of NSIs in neutrino oscillations provides an excellent oppotunity to study the connection with $A_4$ and the residual symmetry $Z_2$ in the neutrino sector, as we will see later.

This work is aimed at discussing how to hint flavour symmetries and residual symmetries in the NSI measurements in neutrino oscillation experiments. We fix the flavour symmetry $A_4$ and residual symmetry $Z_2$ for definiteness. It is a complementarity to studies of $A_4$ and $Z_3$ in CLFV processes and in the standard neutrino oscillation measurements. 
Imposing the flavour symmetry in the fermion sectors, interesting NSI textures or sum rules of NSI parameters are obtained. Both NSIs directly from higher-dimensional operators in the EFT approach with respecting to the electroweak symmetry and those mediated by specified BSM particles will be discussed.  
The rest of this paper is organised as follows. We briefly review the TBM mixing realised in $A_4$ models in Section \ref{sec:FS}. Section \ref{sec:EFT} is devoted to a systematic analysis of how to impose $A_4$ or $Z_2$ to higher-dimensional operators (with the dimension $d\leqslant 8$) which result in NSIs. A class of NSI textures based on $A_4$ and $Z_2$ are obtained, respectively. We only require that the three lepton doublets form a triplet of $A_4$, no requirement for representations of other fermions in the flavour space. In Section \ref{sec:UV}, we consider the UV completion of these operators. New particles in the UV sector impose additional experimental constraints to NSI parameters and thus, some textures are less constrained than the others. We suggest that these textures have a priority to be discussed in the NSI measurement. In Section \ref{sec:DUNE}, based on the DUNE experiment set up, we analyse the potential for the discovery of these textures. We summarise our paper in Section \ref{sec:conclusion}. In the main text of this paper, we focus on NSIs in matter. Connections of flavour symmetries with NSIs at the source and detector are strongly dependent upon representations of the other fermions. 

\section{Flavour symmetries and residual symmetries in lepton mixing \label{sec:FS}}

We briefly review the realisation of the TBM mixing in $A_4$ models and residual symmetries after $A_4$ is broken. 
$A_4$ is generated by two generators $\cS$ and $\cT$ with the requirements $\cS^2=\cT^3=(\cS \cT)^3=1$ and contains 12 elements. It has four irreducible representations: three singlet representations $\mathbf{1}$, $\mathbf{1'}$, $\mathbf{1}''$ and one triplet representation $\mathbf{3}$. Kronecker products of two irreducible representations are reduced in the following way: 
\begin{eqnarray}
&\ms\times\ms^{(\prime,\prime\prime)}=\ms^{(\prime,\prime\prime)}\,,~
\ms'\times\ms'=\ms''\,,~
\ms''\times\ms''=\ms'\,,~ 
\ms'\times\ms''=\ms\,,\nonumber\\
&\mt\times\ms^{(\prime,\prime\prime)}=\mt\,,~
\mt\times\mt=\ms+\ms'+\ms''+\mt_\Symm+\mt_\AntiS \,, 
\end{eqnarray} 
where the subscripts $_\Symm$ and $_\AntiS$ stand for the symmetric and anti-symmetric components, respectively. 

We work in the Altarelli-Feruglio (AF) basis \cite{Altarelli:2005yx}, where $\cT$ and $\cS$ are respectively given by  
\begin{eqnarray}
\cT=\left(
\begin{array}{ccc}
 1 & 0 & 0 \\
 0 & \omega ^2 & 0 \\
 0 & 0 & \omega  \\
\end{array}
\right)\,, \qquad
\cS=\frac{1}{3} \left(
\begin{array}{ccc}
 -1 & 2 & 2 \\
 2 & -1 & 2 \\
 2 & 2 & -1 \\
\end{array}
\right) \,.
\label{eq:generators}
\end{eqnarray}
This basis is widely used in the literature since the charged lepton mass matrix invariant under $\cT$ is diagonal in this basis. The products of each two triplet representations $a=(a_1,a_2,a_3)^T$ and $b=(b_1,b_2,b_3)^T$ can be expressed as
\begin{eqnarray}
\hspace{-5mm}
\begin{array}{c}
(ab)_\mathbf{1}\;\, = a_1b_1 + a_2b_3 + a_3b_2 \,,\\
(ab)_\mathbf{1'}\, = a_3b_3 + a_1b_2 + a_2b_1 \,,\\
(ab)_\mathbf{1''} = a_2b_2 + a_1b_3 + a_3b_1 \,,\\
\end{array} \;\;
(ab)_{\mathbf{3}_\Symm} &=& \frac{1}{2} \left(\begin{array}{c} 2a_1b_1-a_2b_3-a_3b_2\\ 2a_3b_3-a_1b_2-a_2b_1\\ 2a_2b_2-a_3b_1-a_1b_3\end{array} \right) , \;\;
(ab)_{\mathbf{3}_\AntiS} = \frac{1}{2} \left(\begin{array}{c} a_2b_3-a_3b_2\\ a_1b_2-a_2b_1\\ a_3b_1-a_1b_3 \end{array} \right) .
\label{eq:CG2}
\end{eqnarray}

\begin{figure}[h!]
\begin{center}
\includegraphics[width=0.3\textwidth]{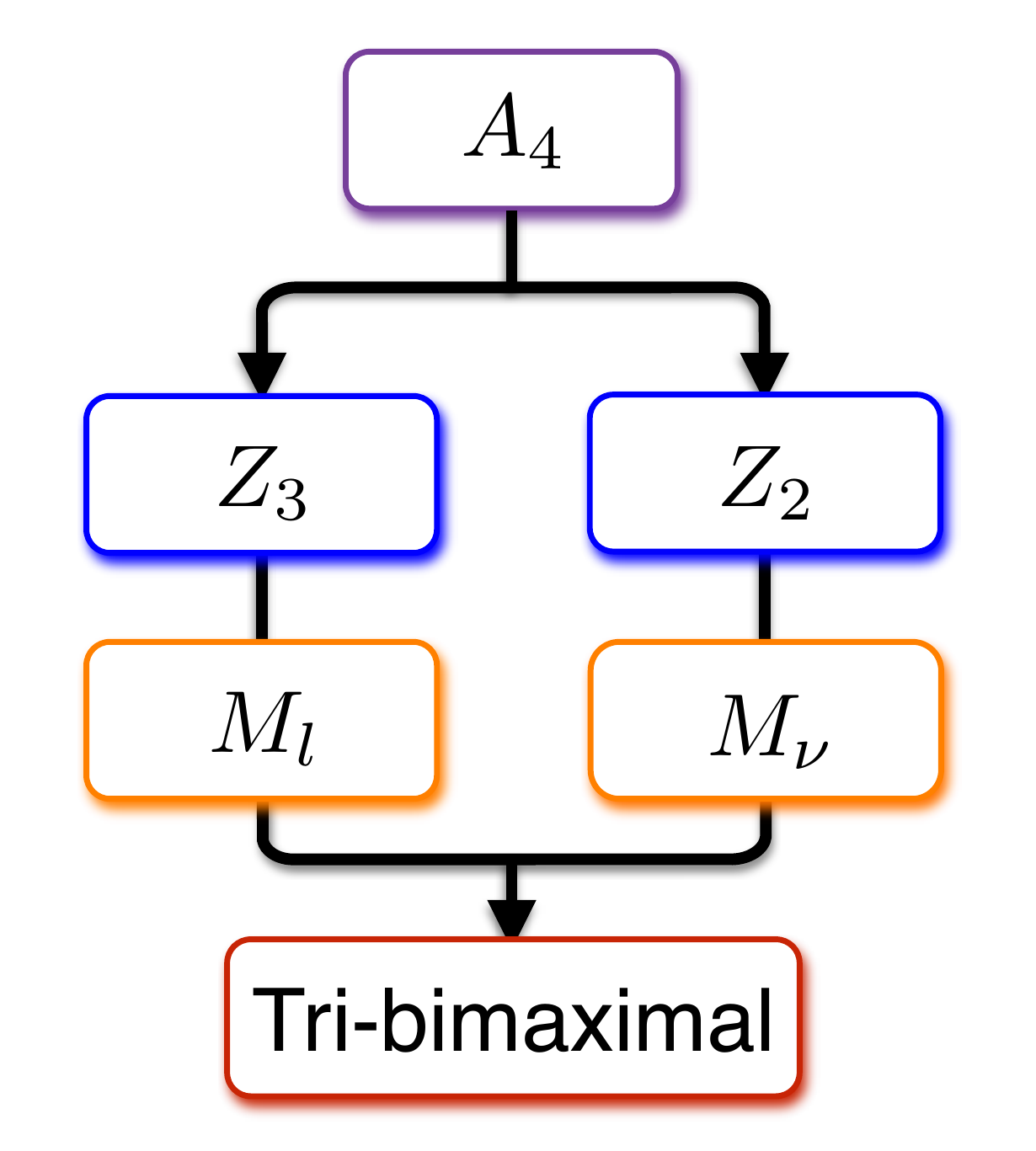}
\caption{A sketch showing how the TBM mixing is generated in $A_4$ models. After $A_4$ is broken, residual symmetries ($Z_3$ in the charged lepton sector and  $Z_2$ in the neutrino sector) are preserved. These symmetries constrains charged lepton and neutrino mass matrices, respectively and finally result in the TBM mixing. The residual symmetries are just approximative symmetries in the model. Besides, there may be additional accidental symmetries in the model, which are not shown here. }
\label{fig:sketch}
\end{center} 
\end{figure}

The $A_4$ symmetry is broken at a certain lower scale. After the $A_4$ breaking, residual symmetries $Z_3$ and $Z_2$, which are generated by $\cT$ and $\cS$, respectively, are approximately preserved in the charged lepton and neutrino sectors, separately. Residual symmetries constrain the lepton mass matrices and lead to the TBM mixing \cite{TBM}. A stekch for how to realise TBM from $A_4$ is shown in Figure~\ref{fig:sketch}.

The Lagrangian terms for generating charged lepton and neutrino masses are effectively realised by some higher-dimensional operators. In the flavour space, the lepton doublets $L_1 = (\nu_{e\LH}, e_\LH)$, $L_2 = (\nu_{\mu\LH}, \mu_\LH)$, $L_3 = (\nu_{\tau\LH}, \tau_\LH)$ are often arranged as a triplet $L \equiv (L_1, L_2, L_3)^T$. This arrangement holds for most flavour models with non-Abelian discrete symmetries, not just for $A_4$ models, in which the flavour symmetry contains a triplet irreducible representation  \cite{review}. In $A_4$ models, the right-handed charged leptons $e_R$, $\mu_R$ and $\tau_R$ are often assigned as singlets $\mathbf{1}$, $\mathbf{1'}$ and $\mathbf{1''}$, respectively \cite{Altarelli:2005yp, Altarelli:2005yx}. The relevant Lagrangian terms are effectively written as  
\begin{eqnarray}
-\mathcal{L}_l &=& \frac{y_e}{\Lambda} (\overline{L} \varphi)_\mathbf{1} e_R H + \frac{y_\mu}{\Lambda} (\overline{L} \varphi)_{\mathbf{1}''} \mu_R H + \frac{y_\tau}{\Lambda} (\overline{L} \varphi)_{\mathbf{1}'} \tau_R H + \text{h.c.} \,, \nonumber\\
-\mathcal{L}_\nu &=& \frac{ y_1}{2\Lambda\Lambda_\text{W}} \big( (\overline{L} \tilde{H}\tilde{H}^T L^c)_{\mathbf{3}_S} \chi  \big)_\mathbf{1}  + \frac{y_2}{2\Lambda_\text{W}}  \big( (\overline{L} \tilde{H}\tilde{H}^T L^c)_\mathbf{1} + \text{h.c.} \,,
\label{eq:Yukawa_coupling} 
\end{eqnarray}
where the Higgs $H \sim \mathbf{1}$ of $A_4$ and $\tilde{H} = i\sigma_2 H^*$. We apply the dimension-5 Weinberg operator $(\overline{L} \tilde{H}\tilde{H}^T L^c)$ to generate neutrino masses and $\Lambda_\text{W}$ is the corresponding UV-complete scale. Operators in Eq.~\eqref{eq:Yukawa_coupling} involve flavons, denoted by $\varphi$ and $\chi$ and a new scale $\Lambda$ corresponding to the decoupling of some heavy $A_4$ multiplets. 

Flavons play the key role in the flavour mixing. They gain VEVs,  leading to the breaking of the flavour symmetry and leaving residual symmetries in the charged lepton and neutrino sectors, respectively. 
The flavon VEVs $\varphi$ and $\chi$ preserving $Z_3$ and $Z_2$, respectively\footnote{In the following, we do not specify notations of flavons with flavon VEVs. }, i.e., 
\begin{eqnarray}
\cT \varphi = \varphi\,,\qquad
\cS \chi = \chi\, \label{eq:Tphi_Schi}
\end{eqnarray}
take the following forms, 
\begin{eqnarray}
\varphi=(1,0,0)^T v_\varphi\,,\qquad
\chi=(1,1,1)^T v_\chi\,. 
\label{eq:VEV}
\end{eqnarray}
The resulting lepton mass matrices are represented as 
\begin{eqnarray}
M_l=\left(
\begin{array}{ccc}
 y_e & 0 & 0 \\
 0 & y_\mu & 0 \\
 0 & 0 & y_\tau \\
\end{array}
\right)\frac{v v_\varphi}{\sqrt{2}\Lambda} \,,\qquad
M_\nu=\left(
\begin{array}{ccc}
 2a+b & -a & -a \\
 -a & 2a & -a+b \\
 -a & -a+b & 2a \\
\end{array}
\right) \,,
\label{eq:lepton_mass}
\end{eqnarray}
where $v=246$ GeV is the Higgs VEV, $a\equiv y_1v_\chi v^2/(4\Lambda\Lambda_\text{W})$ and $b \equiv y_2 v^2/(2\Lambda_\text{W})$. 
It is straightforward to check that the lepton mass matrices $M_l$ and $M_\nu$ satisfy the $Z_3$ and $Z_2$, respectively,
\begin{eqnarray}
\cT M_l M_l^\dag \cT^\dag = M_l M_l^\dag \,, \ \  \cS M_\nu S^T = M_\nu \,.
\end{eqnarray}
They are consistent with the residual symmetries satisfied by the flavon VEVs in Eq.~\eqref{eq:Tphi_Schi}. 
The charged lepton mass matrix $M_l$ is diagonal and the neutrino mass matrix $M_\nu$ is diagonalised by the unitary matrix
\begin{eqnarray}
U_\text{TBM} = \left( \begin{array}{ccc}
 \frac{2}{\sqrt{6}} & \frac{1}{\sqrt{3}} & 0 \\
 -\frac{1}{\sqrt{6}} & \frac{1}{\sqrt{3}} & \frac{1}{\sqrt{2}} \\
 -\frac{1}{\sqrt{6}} & \frac{1}{\sqrt{3}} & -\frac{1}{\sqrt{2}} \\
\end{array} \right)
\end{eqnarray} 
and has eigenvalues $m_1=|3a+b|$, $m_2=|b|$ and $m_3=|3a-b|$. The mixing matrix is identical to $U_\text{TBM}$. This is the so-called TBM mixing pattern,  from which we obtain $\sin\theta_{13}=0$, $\sin\theta_{12}=1/\sqrt{3}$ and $\sin\theta_{23}=1/\sqrt{2}$. We conclude how to realise TBM from $A_4$ in the stekch shown in Figure~\ref{fig:sketch}. 
 
The TBM mixing should be only considered as leading order result since it is not consistent with neutrino oscillation data. Deviations from TBM have to be included in flavour model construction. The deviations are usually obtained from certain subleading interactions which break the $Z_3$ or $Z_2$ residual symmetries. It is crucial to obtain suitable deviations  which are all compatible with current data (For very recent $A_4$ models consistent with current oscillation data, see, e.g., \cite{Pascoli:2016eld, A4_recent} and references therein). These deviations may contribute to NSIs as subleading effects. However, there are various of successful flavour models, and the deviations are usually model-dependent. In addition, these subleading effects are negligible in current NSI measurements. Therefore, we will not consider small corrections to NSIs resulted from small deviations from the TBM mixing.

\section{NSI textures predicted by flavour symmetries in EFT \label{sec:EFT}}

In neutrino oscillation experiments, NSIs may appear in processes of neutrino production at the source, propagation in matter and detection at the detector. 
The matter-effect NSIs are customarily described by a $3\times3$ Hermitian matrix $\epsilon$ added to an effective Hamiltonian $H$ in the flavour basis, 
\begin{eqnarray}
H= \frac{1}{2E} \left\{ 
U \begin{pmatrix} 0 & 0 & 0 \\ 0 & \Delta m^2_{21} & 0 \\ 0 & 0 & \Delta m^2_{31} \end{pmatrix} U^\dag +
A \begin{pmatrix} 1 & 0 & 0 \\ 0 & 0 & 0 \\ 0 & 0 & 0 \end{pmatrix} 
+
A \begin{pmatrix} \epsilon_{ee} & \epsilon_{e\mu} & \epsilon_{e\tau} \\ \epsilon_{\mu e} & \epsilon_{\mu\mu} & \epsilon_{\mu\tau} \\ \epsilon_{\tau e} & \epsilon_{\tau \mu} & \epsilon_{\tau\tau} \end{pmatrix} 
\right\}\,,
\label{eq:Hamiltonian}
\end{eqnarray}
where $\epsilon_{\alpha\beta}=\epsilon^*_{\beta\alpha}$ holds, and $A=2\sqrt{2} G_F N_e E$ is the usual matter effect with $N_e$ the electron number density in the Earth and $E$ the neutrino beam energy. The effective Hamiltonian for antineutrino oscillation is obtained after the replacements $U \to U^*$, $A\to -A$ and $\epsilon_{\alpha\beta} \to \epsilon_{\alpha\beta}^*$. In this section, by assuming NSIs obtained from higher-dimensional operators, we embed $A_4$ or its residual symmetry $Z_2$ to these operators and systematically analyse how to obtain NSI textures from the symmetry. 

\subsection{NSIs from higher-dimensional operators}

We assume that NSIs arise from effective higher-dimensional operators and these operators satisfy the following conditions. 
\begin{itemize}
\item Lorentz invariance and the SM gauge symmetry $SU(2)_\LH\times U(1)_\text{Y}$ around or above the electroweak scale are required. 

\item Since neutrino oscillation experiments cannot test lepton-number-violating (LNV) or baryon-number-violating (BNV) processes, we select lepton- and baryon-number-conserving operators\footnote{This does not mean that the lepton number or baryon number cannot be broken in the UV-complete scale, as will be discussed in the next section.}. 

\item We will only focus on operators in which the number of fermions is 4. The simplest operators have the dimension $d=6$, and the operators with $d>6$ are formed by 4 fermions and $d-6$ Higgs\footnote{Operators modifying neutrino kinetic terms may also contribute to the NSIs through the non-diagonal Z mediation. These effects are small, $\lesssim 10^{-3}$, from the constraints of the PMNS non-unitarity \cite{Fernandez-Martinez:2016lgt, Blennow:2016jkn}, and will not be our case here. }.  In the following, we briefly denote the rest SM fermion contents as 
\begin{eqnarray}
E_\RH = (e_{\RH}, \mu_{\RH}, \tau_{\RH})^T \,,~
U_{\RH} = (u_{\RH}, c_{\RH}, t_{\RH})^T \,,~
D_{\RH} = (d_{\RH}, s_{\RH}, b_{\RH})^T\,,~ 
Q=(Q_1, Q_2, Q_3)^T \,,
\end{eqnarray}
where $Q_1 = (u_{\LH}, d_{\LH})$, $Q_2 = (c_{\LH}, s_{\LH})$, $Q_3 = (t_{\LH}, b_{\LH})$. 

\item For neutrinos propagating in matter, at least
two $L$'s must be involved in the relevant operators. As a comparison, operators for neutrino production and detection involves at least one $L$. 

\item Furthermore, we will impose one more requirement: we only consider NSIs which avoid the strong constraints from 4-charged-fermion interactions, e.g., rare lepton-flavour-violating decays of leptons and hadrons. Since left-handed charged leptons and neutrinos belong to the same electroweak doublet in the SM, any NSI effects from 
higher-dimensional operators are related to an interaction involving at least one charged lepton. Once all 
final and initial states of the latter interaction are electrically charged fermions, 
i.e., charged leptons and quarks, the operator and the relevant NSI parameters should have 
been strongly constrained by these ``visible'' processes. For example, the non-standard 
$\nu_\mu + (e,u,d) \to \nu_e + (e,u,d) $ propagation in matter may be 
constrained by $\mu + (e,u,d) \to e + (e,u,d)$ in the CLFV measurement. 
\end{itemize}

The following classes of operators and their conjugates are allowed by the first four requirements, 
\begin{eqnarray}
\overline{L} E_\RH \overline{D_\RH} Q,~
\overline{L} E_\RH \overline{Q} U_\RH,~
\overline{L} L \overline{F} F \text{ with } F=L,E_\RH,Q,U_\RH,D_\RH
\end{eqnarray}
for $d=6$ and 
\begin{eqnarray}
&&\overline{L} L \overline{D_\RH} U_\RH H^* H^*,~
\overline{L} E_\RH \overline{U_\RH} Q H H,~
\overline{L} E_\RH \overline{Q} D_\RH H H,~
\overline{L} E_\RH \overline{L} E_\RH H H, \nonumber\\
&&\overline{L} E_\RH \overline{D_\RH} Q H^* H,~
\overline{L} E_\RH \overline{Q} U_\RH H^* H,~
\overline{L} L \overline{F} F H^* H \text{ with } F=L,E_\RH,Q,U_\RH,D_\RH
\end{eqnarray} 
for $d=8$. Here we have not written out the necessary $\Gamma$ matrices, gauge indices and flavour indices. The lepton and baryon number conservations forbid any dimension-7 operators involving 4 fermions. 
After the Higgs gets the VEV $\langle H \rangle =(0,1)^T (2\sqrt{2}G_F)^{-1/2}$, these operators classified into two types, those preserving electroweak symmetry and those not. 
Taking the last requirement into account, we extract the following operators:
\begin{itemize}
\item
The first class are explicitly given by
\begin{eqnarray}
&& \varepsilon_{ac}\varepsilon_{bd} (\overline{L_{a\alpha}}\gamma^\mu L_{b\beta} ) (\overline{L_{c\gamma}}\gamma_\mu L_{d\delta} )\,,~
\varepsilon_{ac}\varepsilon_{bd} (\overline{L_{a\alpha}}\gamma^\mu L_{b\beta} ) (\overline{L_{c\gamma}}\gamma_\mu L_{d\delta} ) H^\dag H\,,
\label{eq:d6}
\end{eqnarray} 
where $\alpha,\beta,\gamma,\delta = 1,2,3$ are flavour indices, $a,b,c,d =1,2$ are $SU(2)_\LH$ doublet indices, and non-vanishing entries of $\varepsilon_{ab}$ are given by $\varepsilon_{12}=-\varepsilon_{21}=1$.  Specifically, we denote the flavour indices in the lepton sector as $(1,2,3)=(e,\mu,\tau)$. 
Using the relation $\varepsilon_{ac}\varepsilon_{cd} = \delta_{ab} \delta_{cd} - \delta_{ad} \delta_{bc}$ and the Fierz identity, we expand the first term of the above equation and obtain $(\overline{L_{a\alpha}}\gamma^\mu L_{a\beta} ) (\overline{L_{c\gamma}}\gamma_\mu L_{c\delta} ) - (\overline{L_{a\alpha}}\gamma^\mu L_{a\delta} ) (\overline{L_{c\gamma}}\gamma_\mu L_{c\beta} )$, i.e.,
\begin{eqnarray}
(\overline{\nu_{\alpha \LH}}\gamma^\mu \nu_{\beta \LH} \!) (\overline{E_{\gamma \LH}}\gamma_\mu E_{\delta \LH} \!) \!+\! (\overline{\nu_{\gamma \LH}}\gamma^\mu \nu_{\delta \LH} \!) (\overline{E_{\alpha \LH}}\gamma_\mu E_{\beta \LH} \!) \!-\! (\overline{\nu_{\alpha \LH}}\gamma^\mu \nu_{\delta \LH} \!) (\overline{E_{\gamma \LH}}\gamma_\mu E_{\beta \LH} \!) \!-\! (\overline{\nu_{\gamma \LH}}\gamma^\mu \nu_{\beta \LH} \!) (\overline{E_{\alpha \LH}}\gamma_\mu E_{\delta \LH} \!), \nonumber\\ 
\label{eq:QI}
\end{eqnarray} 
which we denote as $\mathcal{O}^1_{\alpha\beta\gamma\delta}$. Note that $\mathcal{O}^1_{\alpha\beta\gamma\delta} = -\mathcal{O}^1_{\gamma\beta\alpha\delta} = - \mathcal{O}^1_{\alpha\delta\gamma\beta} = \mathcal{O}^1_{\gamma\delta\alpha\beta}$ is satisfied. 
This term can lead to NSIs of neutrino interacting with the electron $\nu_\alpha  e \to \nu_\beta  e$ during the neutrino propagation, but have no influence on 4-charged-lepton interactions such as the scattering $\mu  e \to e  e$ or the rare decay $\mu\to eee$, and thus are not directly constrained by the latter. 
The second term in Eq.~\eqref{eq:d6} gives no more information than $\mathcal{O}^1_{\alpha\beta\gamma\delta}$, which is not necessary to be considered separately. 

\item
The second class of operators are: 
\begin{eqnarray}
&(\overline{L_{\alpha}} \tilde{H} \gamma^\mu \tilde{H}^\dag L_{\beta} ) (\overline{U_{\gamma \RH}} \gamma_\mu U_{\delta \RH}),~
(\overline{L_{\alpha}} \tilde{H} \gamma^\mu \tilde{H}^\dag L_{\beta} ) (\overline{D_{\gamma \RH}} \gamma_\mu D_{\delta \RH}),~
(\overline{L_{\alpha}} \tilde{H} \gamma^\mu \tilde{H}^\dag L_{\beta} ) (\overline{E_{\gamma \RH}} \gamma_\mu E_{\delta \RH}),~\nonumber\\
&(\overline{L_{\alpha}} \tilde{H} \gamma^\mu \tilde{H}^\dag L_{\beta} ) (\overline{Q_{\gamma}} \gamma_\mu Q_{\delta}),~
(\overline{L_{\alpha}} \tilde{H} \gamma^\mu \tilde{H}^\dag L_{\beta} ) (\overline{L_{\gamma}} \gamma_\mu L_{\delta}), \nonumber\\
&(\overline{L_{\alpha}} \tilde{H} \gamma^\mu L_{b\beta} ) (\overline{Q_{b\gamma}} \gamma_\mu \tilde{H}^\dag Q_{\delta}),~
\varepsilon_{bc} (\overline{L_{\alpha}} \tilde{H} \gamma^\mu L_{b\beta} ) (\overline{Q_{\gamma}} H \gamma_\mu Q_{c\delta}), \nonumber\\
&(\overline{L_{\alpha}} \tilde{H} \gamma^\mu H^\dag L_{\beta} ) (\overline{D_{\gamma \RH}} \gamma_\mu U_{\delta \RH}),~
(\overline{L_{\alpha}} \tilde{H} \sigma^{\mu\nu} E_{\beta \RH} ) (\overline{Q_{\gamma}} H \sigma_{\mu\nu} U_{\delta \RH}), 
\nonumber\\
& (\overline{L_{\alpha}} \tilde{H} E_{\beta \RH} ) (\overline{D_{\gamma \RH}} \tilde{H}^\dag Q_{\delta}),~
(\overline{L_{\alpha}} \tilde{H} E_{\beta \RH} ) (\overline{Q_{\gamma}} H U_{\delta \RH}). 
\label{eq:d8}
\end{eqnarray} 
After the Higgs gets the VEV, the above operators are effectively reduced to 11 four-fermion interactions,
\begin{eqnarray}
&(\overline{\nu_{\alpha \LH}} \gamma^\mu \nu_{\beta \LH} ) (\overline{U_{\gamma \RH}} \gamma_\mu U_{\delta \RH}),~ 
(\overline{\nu_{\alpha \LH}} \gamma^\mu \nu_{\beta \LH} ) (\overline{D_{\gamma \RH}} \gamma_\mu D_{\delta \RH}),~
(\overline{\nu_{\alpha \LH}} \gamma^\mu \nu_{\beta \LH} ) (\overline{E_{\gamma \RH}} \gamma_\mu E_{\delta \RH}),\nonumber\\
&
(\overline{\nu_{\alpha \LH}} \gamma^\mu \nu_{\beta \LH} ) (\overline{U_{\gamma \LH}} \gamma_\mu U_{\delta \LH} + \overline{D_{\gamma \LH}} \gamma_\mu D_{\delta \LH}),~
(\overline{\nu_{\alpha \LH}} \gamma^\mu \nu_{\beta \LH} ) (\overline{\nu_{\gamma \LH}} \gamma_\mu \nu_{\delta \LH} + \overline{E_{\gamma \LH}} \gamma_\mu E_{\delta \LH}), \nonumber\\
&\hspace{-3mm} (\overline{\nu_{\alpha \LH}} \gamma^\mu \nu_{\beta \LH} ) (\overline{U_{\gamma \LH}} \gamma_\mu U_{\delta \LH}) \!+\! 
(\overline{\nu_{\alpha \LH}} \gamma^\mu E_{\beta \LH} ) (\overline{D_{\gamma \LH}} \gamma_\mu U_{\delta \LH}),
(\overline{\nu_{\alpha \LH}} \gamma^\mu \nu_{\beta \LH} ) (\overline{D_{\gamma \LH}} \gamma_\mu D_{\delta \LH}) \!-\! (\overline{\nu_{\alpha \LH}} \gamma^\mu E_{\beta \LH} ) (\overline{D_{\gamma \LH}} \gamma_\mu U_{\delta \LH}), \nonumber\\
&(\overline{\nu_{\alpha \LH}} \gamma^\mu E_{\beta \LH} ) (\overline{D_{\gamma \RH}} \gamma_\mu U_{\delta \RH}),~
(\overline{\nu_{\alpha \LH}} \sigma^{\mu\nu} E_{\beta \RH} ) (\overline{D_{\gamma \LH}} \sigma_{\mu\nu} U_{\delta \RH}) , \nonumber\\
& (\overline{\nu_{\alpha \LH}} E_{\beta \RH} ) (\overline{D_{\gamma \RH}} U_{\delta \LH}),~
(\overline{\nu_{\alpha \LH}} E_{\beta \RH} ) (\overline{D_{\gamma \LH}} U_{\delta \RH}). 
\label{eq:QII}
\end{eqnarray} 
In the above operators, the first 5 terms, denoted by $\mathcal{O}_{\alpha\beta\gamma\delta}^{2,3,4,5,6}$, respectively, contribute to NSIs in matter during neutrino propagation. The next 2 terms, denoted by $\mathcal{O}_{\alpha\beta\gamma\delta}^{7,8}$, contribute to both NSIs at the neutrino source and detector, and NSIs for neutrino mediation in matter, and correlate them together. And the final 4 terms, denoted by $\mathcal{O}_{\alpha\beta\gamma\delta}^{9,10,11,12}$, respectively,  contribute to NSIs in the neutrino production and detection processes. For more discussions on textures of NSIs in these processes, please see appendix~\ref{app:NSI}. 
\end{itemize}


\begin{table}[h!]
\newcommand{\tabincell}[2]{\begin{tabular}{@{}#1@{}}#2\end{tabular}}
  \centering
  \begin{tabular}{|c|c|c|c|}\hline
\!\!\!Label\!\!\! & Before EW breaking & After EW breaking & \!\!\!observation\!\!\!\\\hline
$\mathcal{O}^1$ &

\tabincell{c}{$\varepsilon_{ac}\varepsilon_{bd} (\overline{L_{a\alpha}}\gamma^\mu L_{b\beta} ) 
(\overline{L_{c\gamma}}\gamma_\mu L_{d\delta} )$,\\
$\!\!\!\varepsilon_{ac}\varepsilon_{bd} (\overline{L_{a\alpha}}\gamma^\mu L_{b\beta} ) 
(\overline{L_{c\gamma}}\gamma_\mu L_{d\delta} ) H^\dag H\!\!\!$} & 

\tabincell{c}{$\!\!\!(\overline{\nu_{\alpha \LH}}\gamma^\mu \nu_{\beta \LH} ) (\overline{E_{\gamma \LH}}\gamma_\mu 
E_{\delta \LH} ) \!+\! (\overline{\nu_{\gamma \LH}}\gamma^\mu \nu_{\delta \LH} ) (\overline{E_{\alpha \LH}}
\gamma_\mu E_{\beta \LH} )\!\!\!$ \\ $\!\!\!\!- (\overline{\nu_{\alpha \LH}}\gamma^\mu \nu_{\delta \LH} ) 
(\overline{E_{\gamma \LH}}\gamma_\mu E_{\beta \LH} ) \!-\! (\overline{\nu_{\gamma \LH}}\gamma^\mu 
\nu_{\beta \LH} ) (\overline{E_{\alpha \LH}}\gamma_\mu E_{\delta \LH} )\!\!\!\!$}  
& M\\\hline
$\mathcal{O}^2$ & $(\overline{L_{\alpha}} \tilde{H} \gamma^\mu \tilde{H}^\dag L_{\beta} ) (\overline{U_{\gamma \RH}} 
\gamma_\mu U_{\delta \RH})$ & $(\overline{\nu_{\alpha \LH}} \gamma^\mu \nu_{\beta \LH} ) (\overline{U_{\gamma \RH}} \gamma_\mu 
U_{\delta \RH})$ & M \\\hline
$\mathcal{O}^3$ & $(\overline{L_{\alpha}} \tilde{H} \gamma^\mu \tilde{H}^\dag L_{\beta} ) (\overline{D_{\gamma \RH}} 
\gamma_\mu D_{\delta \RH})$ & $(\overline{\nu_{\alpha \LH}} \gamma^\mu \nu_{\beta \LH} ) (\overline{D_{\gamma \RH}} \gamma_\mu 
D_{\delta \RH})$ & M \\\hline
$\mathcal{O}^4$ & $(\overline{L_{\alpha}} \tilde{H} \gamma^\mu \tilde{H}^\dag L_{\beta} ) (\overline{E_{\gamma \RH}} 
\gamma_\mu E_{\delta \RH})$ & $(\overline{\nu_{\alpha \LH}} \gamma^\mu \nu_{\beta \LH} ) (\overline{E_{\gamma \RH}} \gamma_\mu 
E_{\delta \RH})$ & M\\\hline
$\mathcal{O}^5$ & $(\overline{L_{\alpha}} \tilde{H} \gamma^\mu \tilde{H}^\dag L_{\beta} ) (\overline{Q_{\gamma}} 
\gamma_\mu Q_{\delta})$ & $(\overline{\nu_{\alpha \LH}} \gamma^\mu \nu_{\beta \LH} ) (\overline{U_{\gamma \LH}} \gamma_\mu 
U_{\delta \LH} + \overline{D_{\gamma \LH}} \gamma_\mu D_{\delta \LH})$ & M\\\hline
$\mathcal{O}^6$ & $(\overline{L_{\alpha}} \tilde{H} \gamma^\mu \tilde{H}^\dag L_{\beta} ) (\overline{L_{\gamma}} 
\gamma_\mu L_{\delta})$ & $(\overline{\nu_{\alpha \LH}} \gamma^\mu \nu_{\beta \LH} ) (\overline{\nu_{\gamma \LH}} \gamma_\mu
 \nu_{\delta \LH} + \overline{E_{\gamma \LH}} \gamma_\mu E_{\delta \LH})$ & M\\\hline
$\mathcal{O}^7$ & $(\overline{L_{\alpha}} \tilde{H} \gamma^\mu L_{b\beta} ) (\overline{Q_{b\gamma}} \gamma_\mu 
\tilde{H}^\dag Q_{\delta})$ & $\!\!\!(\overline{\nu_{\alpha \LH}} \gamma^\mu \nu_{\beta \LH} ) (\overline{U_{\gamma \LH}} \gamma_\mu 
U_{\delta \LH}) + 
(\overline{\nu_{\alpha \LH}} \gamma^\mu E_{\beta \LH} ) (\overline{D_{\gamma \LH}} \gamma_\mu 
U_{\delta \LH})\!\!\!$ & S,M,D\\\hline
$\mathcal{O}^8$ & $\varepsilon_{bc} (\overline{L_{\alpha}} \tilde{H} \gamma^\mu L_{b\beta} ) (\overline{Q_{\gamma}} 
H \gamma_\mu Q_{c\delta})$ & $\!\!\!(\overline{\nu_{\alpha \LH}} \gamma^\mu \nu_{\beta \LH} ) (\overline{D_{\gamma \LH}} \gamma_\mu 
D_{\delta \LH}) - (\overline{\nu_{\alpha \LH}} \gamma^\mu E_{\beta \LH} ) (\overline{D_{\gamma \LH}} 
\gamma_\mu U_{\delta \LH})\!\!\!$& S,M,D \\\hline
$\mathcal{O}^9$ & $\varepsilon_{bc} (\overline{L_{\alpha}} \tilde{H} \gamma^\mu L_{b\beta} ) (\overline{Q_{\gamma}} 
H \gamma_\mu Q_{c\delta})$ & $(\overline{\nu_{\alpha \LH}} \gamma^\mu E_{\beta \LH} ) (\overline{D_{\gamma \RH}} \gamma_\mu 
U_{\delta \RH})$ & S,D\\\hline
$\mathcal{O}^{10}$ & $(\overline{L_{\alpha}} \tilde{H} \sigma^{\mu\nu} E_{\beta \RH} ) (\overline{Q_{\gamma}} H 
\sigma_{\mu\nu} U_{\delta \RH})$ & $(\overline{\nu_{\alpha \LH}} \sigma^{\mu\nu} E_{\beta \RH} ) (\overline{D_{\gamma \LH}} 
\sigma_{\mu\nu} U_{\delta \RH})$ & S,D\\\hline
$\mathcal{O}^{11}$ & $(\overline{L_{\alpha}} \tilde{H} E_{\beta \RH} ) (\overline{D_{\gamma \RH}} \tilde{H}^\dag 
Q_{\delta})$ & $(\overline{\nu_{\alpha \LH}} E_{\beta \RH} ) (\overline{D_{\gamma \RH}} U_{\delta \LH})$& S,D \\\hline
$\mathcal{O}^{12}$ & $(\overline{L_{\alpha}} \tilde{H} E_{\beta \RH} ) (\overline{Q_{\gamma}} H U_{\delta \RH})$ &
$(\overline{\nu_{\alpha \LH}} E_{\beta \RH} ) (\overline{D_{\gamma \LH}} U_{\delta \RH})$& S,D \\\hline
\end{tabular}
  \caption{\label{tab:operator} Higher-dimensional operators ($d \leqslant 8$) which may  contribute to NSIs in neutrino oscillation experiments. S, M, and D represent NSIs at a source, in matter and at a detector, respectively. }
\end{table}


The effective operators describing neutrino NSIs for neutrino propagation can be expressed as 
\begin{eqnarray}
\mathcal{L}_{\text{NSI}} = 2 \sqrt{2} G_F \sum_{p=1}^{8} c^{p}_{\alpha\beta\gamma\delta} \mathcal{O}_{\alpha\beta\gamma\delta}^{p} + \text{h.c.} \,, 
\label{eq:sum_O}
\end{eqnarray}
where two same flavour indices should be summed. 
Operators in Eqs. \eqref{eq:QI} and \eqref{eq:QII} form a full list of NSI operators with $d\leqslant 8$ before electroweak symmetry breaking. We have checked that all the other NSIs with $d\leqslant 8$ operators can be represented as a linear combination of these $\mathcal{O}_{\alpha\beta\gamma\delta}^{p}$. Matching with the effective NSI matrix $\epsilon$ in Eq.~\eqref{eq:Hamiltonian}, we obtain 
\begin{eqnarray}
\epsilon_{\alpha\beta} = \epsilon^{e}_{\alpha\beta} + \Big(2+\frac{N_n}{N_e}\Big) \epsilon^{u}_{\alpha\beta} + \Big(1+2\frac{N_n}{N_e}\Big) \epsilon^{d}_{\alpha\beta}
\end{eqnarray}
with $N_n$ the neutron number density and
\begin{eqnarray}
\epsilon^e_{\alpha\beta} &=& c^{1}_{\alpha\beta11} + c^{4}_{\alpha\beta11} + c^{6}_{\alpha\beta11} \,, \nonumber\\
\epsilon^u_{\alpha\beta} &=& c^{2}_{\alpha\beta11} + c^{5}_{\alpha\beta11} + c^{7}_{\alpha\beta11} \,, \nonumber\\
\epsilon^d_{\alpha\beta} &=& c^{3}_{\alpha\beta11} + c^{5}_{\alpha\beta11} + c^{8}_{\alpha\beta11} \,.
\end{eqnarray}
For $\mathcal{O}_{\alpha\beta\gamma\delta}^{1}$, it is easy to confirm $c^1_{\alpha\beta\gamma\delta}=-c^1_{\gamma\beta\alpha\delta} = c^1_{\alpha\delta\gamma\beta}$, and thus $c^1_{e\beta11}$ and $c^1_{\alpha e11}$ always vanish. Therefore, $\mathcal{O}_{\alpha\beta\gamma\delta}^{1}$ will not contribute to the first column or first row of $\epsilon$.

\subsection{NSI textures predicted by $A_4$}

We consider how neutrino NSIs from the higher-dimensional operators are constrained by $A_4$. 
We require that the higher-dimensional operators are invariant under the symmetry $A_4$ and consider which kinds of NSI textures we could gain from the symmetry. As we only care about matter-effect NSI textures, we limit our discussion in the operators $\mathcal{O}^{1-8}$. In appendix \ref{app:NSI}, we list the NSI textures at  the source and detector from the operators $\mathcal{O}^{7-12}$.

We follow Section \ref{sec:FS} in which the lepton doublets $L=(L_1,L_2,L_3)^T$ are often arranged as a triplet $\mt$ of $A_4$\footnote{In the AF basis, the conjugate of $L$ should be arranged as $\overline{L} = (\overline{L_1}, \overline{L_3}, \overline{L_2})^T$. }. Besides, we do not specify the representations for the other fermions in the flavour space. In other words, the right-handed charged leptons, left-handed quarks and right-handed quarks could be any irreducible representations of $A_4$, $\ms,\ms',\ms''$ or $\mt$. It is worth noting that we do not specify if $A_4$ can be responsible for the quark mixing in this work. If all quarks are arranged as the singlet representation $\ms$, quark flavour mixing is totally independent of  $A_4$. We scan for all these possibilities, and find the following NSI textures: 
\begin{eqnarray}
&&\bT_{11} \equiv \bI = 
\left(\begin{array}{ccc}
 1 & 0 & 0 \\
 0 & 1 & 0 \\
 0 & 0 & 1 \\
\end{array}\right) \,,  \qquad
\bT_{12} = 
\left(\begin{array}{ccc}
 2 & 0 & 0 \\
 0 & -1 & 0 \\
 0 & 0 & -1 \\
\end{array}
\right) \,, \qquad
\bT_{13} = 
\left(\begin{array}{ccc}
 0 & 0 & 0 \\
 0 & 1 & 0 \\
 0 & 0 & -1 \\
\end{array}
\label{eq:sumruleI}
\right) \,. 
\end{eqnarray}
In the following, we explain how to get these textures. 

The first operator $c^1_{\alpha\beta\gamma\delta}\mathcal{O}^1_{\alpha\beta\gamma\delta}$, i.e., the dimension-6 $\varepsilon_{ac}\varepsilon_{bd} c^1_{\alpha\beta\gamma\delta} (\overline{L_{a\alpha}}\gamma^\mu L_{b\beta} ) (\overline{L_{c\gamma}}\gamma_\mu L_{d\delta} )$, satisfy the anti-permutation property of two $L$'s and two $\overline{L}$'s, as shown in Eq.~\eqref{eq:QI}, which results in $c^1_{e\beta11}=c^1_{\alpha e11}=0$. 
There are 5 independent $A_4$-invariant operators:  
\begin{eqnarray}
\LL_\ms \LL_\ms\,,~ \LL_{\ms'} \LL_{\ms''}\,,~ \LL_{\mt_\Symm} \LL_{\mt_\Symm}\,,~ \LL_{\mt_\AntiS} \LL_{\mt_\AntiS}\,,~ \LL_{\mt_\Symm} \LL_{\mt_\AntiS}\,. 
\end{eqnarray}
Here, we have ignored the unnecessary flavour-independent notations, including the $SU(2)_\LH$ indices, $\Gamma$ matrices and the Higgs field. The representations in the subscripts are understood as in Eq.~\eqref{eq:CG2}. 
Taking account of the CG coefficients in Eq.~\eqref{eq:CG2}, we obtain 
\begin{eqnarray}
c^1_{\mu\mu11} = c^1_{\tau\tau11}\,,~ c^1_{ee 11} = c^1_{\alpha\beta 11} = 0 ~ \text{for}~\alpha\neq\beta
\end{eqnarray}
for the first 4 operators which lead to the NSI texture
\begin{eqnarray}
\bT_{12}' \equiv \left(\begin{array}{ccc}
 0 & 0 & 0 \\
 0 & 1 & 0 \\
 0 & 0 & 1 \\
\end{array}\right) \propto 2\bT_{11}-\bT_{12}\,.
\end{eqnarray} 
The last operator gives vanishing $c^1_{\alpha\beta 11}$ and thus does not contribute to NSIs. 

For the second one in Table~\ref{tab:operator}, $c^2_{\alpha\beta\gamma\delta}\mathcal{O}^2_{\alpha\beta\gamma\delta}$, i.e., the dimension-8 $(\overline{L_{\alpha}} \tilde{H} \gamma^\mu \tilde{H}^\dag L_{\beta} ) (\overline{U_{\gamma \RH}} \gamma_\mu U_{\delta \RH})$, the $A_4$-invariant operators depend on the flavour representation of $U_\RH$: 
\begin{itemize}
\item If $U_{1\RH}$ is arranged to be a singlet $\ms^{(\prime,\prime\prime)}$ of $A_4$, there is only one $A_4$-invariant operator
\begin{eqnarray}
\LL_{\ms}(\overline{U_{1\RH}} U_{1\RH})_\ms\,.
\end{eqnarray} 
It leads to the relation of the coefficients
\begin{eqnarray}
c^2_{ee11} = c^2_{\mu\mu11} = c^2_{\tau\tau11} \,,~ c^2_{\alpha\beta11} = 0~ \text{for } \alpha\neq\beta\,.
\label{eq:relation2}
\end{eqnarray}
Representations of $U_{2\RH}$ and $U_{3\RH}$ are irrelevant for our discussion since $U_{2\RH}$ and $U_{3\RH}$ do not attend to the low energy NSIs. 

\item If $U_\RH = (U_{1\RH}, U_{2\RH}, U_{3\RH})^T$ is a triplet $\mt$ of $A_4$, there are 7 independent $A_4$-invariant operators
\begin{eqnarray}
&\LL_{\ms}(\overline{U_\RH} U_\RH)_\ms\,,~
\LL_{\ms'}(\overline{U_\RH} U_\RH)_{\ms''}\,,~
\LL_{\ms''}(\overline{U_\RH} U_\RH)_{\ms'}\,,~\nonumber\\
&\LL_{\mt_\Symm}(\overline{U_\RH} U_\RH)_{\mt_\Symm}\,,~
\LL_{\mt_\AntiS}(\overline{U_\RH} U_\RH)_{\mt_\Symm}\,,~
\LL_{\mt_\Symm}(\overline{U_\RH} U_\RH)_{\mt_\AntiS}\,,~
\LL_{\mt_\AntiS}(\overline{U_\RH} U_\RH)_{\mt_\AntiS}\,.
\end{eqnarray} 
The first operator gives the same correlation as in Eq.~\eqref{eq:relation2}, $\LL_{\mt_\Symm}(\overline{U_\RH} U_\RH)_{\mt_\Symm}$ and $\LL_{\mt_\AntiS}(\overline{U_\RH} U_\RH)_{\mt_\Symm}$ give rise to 
\begin{eqnarray}
&& c^2_{ee11} = -2 c^2_{\mu\mu11} = -2 c^2_{\tau\tau11} \,,~ c^2_{\alpha\beta11} = 0~ \text{for } \alpha\neq\beta\,; \nonumber\\
&& c^2_{\mu\mu11} = - c^2_{\tau\tau11} \,,~ c^2_{ee11} = c^2_{\alpha\beta11} = 0~ \text{for } \alpha\neq\beta\,,
\label{eq:relation2p}
\end{eqnarray}
respectively, where all non-vanishing values are real. The rest, $\LL_{\ms'}(\overline{U_\RH} U_\RH)_{\ms''}$, $\LL_{\ms''}(\overline{U_\RH} U_\RH)_{\ms'}$, $\LL_{\mt_\Symm}(\overline{U_\RH} U_\RH)_{\mt_\AntiS}$, and $\LL_{\mt_\AntiS}(\overline{U_\RH} U_\RH)_{\mt_\AntiS}$ have no contribution to $c^2_{\alpha\beta11}$. 

\end{itemize}
The correlations of the coefficients $c^2_{\alpha\beta11}$ directly determine the flavour structure of matter-effect NSIs. In detail, Eq.~\eqref{eq:relation2} directly gives rise to $\bT_{11}$, and Eq.~\eqref{eq:relation2p} leads to $\bT_{12}$ and $\bT_{13}$.  
The discussion of $\mathcal{O}^2_{\alpha\beta\gamma\delta}$ applies to $\mathcal{O}^{3-8}_{\alpha\beta\gamma\delta}$. In other words, the NSI textures $\bT_{11}$, $\bT_{12}$ and $\bT_{13}$ can be derived from 
\begin{eqnarray}
\LL_\ms \FF_\ms\,,~ \LL_{\mt_\Symm} \FF_{\mt_\Symm} \,,~  \LL_{\mt_\AntiS} \FF_{\mt_\Symm}\,,
\end{eqnarray}
respectively, where $F$ represents any fermions in the SM.

\subsection{NSI textures predicted by the residual symmetry of $A_4$}

In order to break $A_4$ and obtain residual symmetries, we include the flavon VEV in the NSI operators. 
We consider that the operators $c^p_{\alpha\beta\gamma\delta}\mathcal{O}^p_{\alpha\beta\gamma\delta}$ are effectively realised via\footnote{Since the conjugates of $\varphi$ and $\chi$ are identical with $\varphi$ and $\chi$, respectively, it is not necessary to write out operators realised by $\varphi^*$ or $\chi^*$ separately. }
\begin{eqnarray} 
c^{\varphi, p}_{\alpha'\alpha\beta\gamma\delta}\frac{\varphi_{\alpha'}}{v_\varphi}\mathcal{O}^p_{\alpha\beta\gamma\delta} ~\text{ or }~
c^{\chi, p}_{\alpha'\alpha\beta\gamma\delta}\frac{\chi_{\alpha'}}{v_\chi}\mathcal{O}^p_{\alpha\beta\gamma\delta} \,.
\end{eqnarray}
These operators are $A_4$-invariant before flavons get VEVs. Taking the VEVs in Eq.~\eqref{eq:VEV}, we obtain $c^p_{\alpha\beta\gamma\delta}\mathcal{O}^p_{\alpha\beta\gamma\delta}$ with
\begin{eqnarray}
c^p_{\alpha\beta\gamma\delta} = c^{\varphi, p}_{1\alpha\beta\gamma\delta} ~\text{ or }~
c^{\chi, p}_{1\alpha\beta\gamma\delta} + c^{\chi, p}_{2\alpha\beta\gamma\delta} + c^{\chi, p}_{3\alpha\beta\gamma\delta} \,.
\end{eqnarray}
They are not $A_4$-invariant any more, but preserves only a $Z_3$ or $Z_2$ symmetry, since $\varphi$ and $\chi$ preserve $Z_3$ and $Z_2$ symmetries, respectively. The $Z_3$-invariant operators $\varphi \mathcal{O}$ will not give nothing new, but Eq.~\eqref{eq:sumruleI}. The reason is that the generator of $Z_3$, $\cT$, is diagonal, and the predicted NSI textures must be also diagonal. In the following, we will not consider the $Z_3$-invariant operator $\varphi \mathcal{O}$ anymore. 

Now we focus on the $A_4$-breaking $Z_2$-invariant operators  $\chi \mathcal{O}$.  We first define the following non-diagonal textures: 
\begin{eqnarray}
&&\bT_{21} = \left(
\begin{array}{ccc}
 ~0~ & ~1~ & ~1~ \\
 ~1~ & ~0~ & ~1~ \\
 ~1~ & ~1~ & ~0~ \\
\end{array}
\right) \,,  \qquad
\bT_{22} = \left(
\begin{array}{ccc}
 0 & -1 & -1 \\
 -1 & 0 & 2 \\
 -1 & 2 & 0 \\
\end{array}
\right)
 \,, \qquad
\bT_{23} = \left(
\begin{array}{ccc}
 0 & -1 & 1 \\
 -1 & 0 & 0 \\
 1 & 0 & 0 \\
\end{array}
\right)
 \,, \nonumber\\
&&
\bT_{31} = 
\left(
\begin{array}{ccc}
 0 & -i & i \\
 i & 0 & -i \\
 -i & i & 0 \\
\end{array}
\right) \,,\; \qquad
\bT_{32} = 
\left(
\begin{array}{ccc}
 0 & i & -i \\
 -i & 0 & -2 i \\
 i & 2 i & 0 \\
\end{array}
\right) \,,\; \qquad
\bT_{33} = 
\left(
\begin{array}{ccc}
 ~0~ & ~i~ & ~i~ \\
 -i & 0 & 0 \\
 -i & 0 & 0 \\
\end{array}
\right)\,. \hspace{1.4cm}
\label{eq:sumruleII}
\end{eqnarray}
$\bT_{2n}$ represent non-diagonal real NSI textures, while $\bT_{3n}$ represent pure imaginary NSI textures. 

For $c^{\chi, 1}_{\alpha'\alpha\beta\gamma\delta}\chi_{\alpha'}\mathcal{O}^1_{\alpha\beta\gamma\delta}$, 
there are 9 $Z_2$-invariant operators that can contribute to NSIs: 
\begin{eqnarray}
&\chi \LL_{\mt_\Symm} \LL_\ms,\, \chi \LL_{\mt_\Symm} \LL_{\ms'},\, \chi \LL_{\mt_\Symm} \LL_{\ms''}, \nonumber\\
&\chi \LL_{\mt_\AntiS} \LL_\ms, \,\chi \LL_{\mt_\AntiS} \LL_{\ms'},\, \chi \LL_{\mt_\AntiS} \LL_{\ms''}, \nonumber\\
&\chi \big( \LL_{\mt_\Symm} \LL_{\mt_\Symm} \big)_{\mt_\Symm}\,,~ 
\chi \big( \LL_{\mt_\AntiS} \LL_{\mt_\AntiS} \big)_{\mt_\Symm}\,, \nonumber\\
&\chi \big( \LL_{\mt_\Symm} \LL_{\mt_\AntiS} \big)_{\mt_\Symm}\,,~
\chi \big( \LL_{\mt_\Symm} \LL_{\mt_\AntiS} \big)_{\mt_\AntiS}\,.
\end{eqnarray}
Due to the antisymmetric property between $\alpha$ and $\gamma$ and that between $\beta$ and $\delta$, $c^1_{e\beta11}=c^1_{\alpha e11}=0$ for all cases. The other coefficients satisfy the following relations, respectively. 
Taking the CG coefficients in Eq.~\eqref{eq:CG2} into account, we obtain 
\begin{eqnarray}
&&2c^1_{\mu\mu11} = 2c^1_{\tau\tau11} = c^1_{\mu\tau11} = c^1_{\tau\mu11} 
\end{eqnarray}
for $\chi \big( \LL_{\mt_\Symm} \LL_{\ms,\ms',\ms''} \big)_{\mt}$, $\chi \big( \LL_{\mt_\Symm} \LL_{\mt_\Symm} \big)_{\mt_\Symm}$, $\chi \big( \LL_{\mt_\AntiS} \LL_{\mt_\AntiS} \big)_{\mt_\Symm}$ and 
\begin{eqnarray}
c^1_{\mu\mu11} = -c^1_{\tau\tau11} \,,~c^1_{\mu\tau11} = c^1_{\tau\mu11}=0
\end{eqnarray}
for $\chi \big(\LL_{\mt_\AntiS} \LL_{\ms,\ms',\ms''} \big)_{\mt}$, $\chi \big( \LL_{\mt_\Symm} \LL_{\mt_\AntiS} \big)_{\mt_\Symm}$, $\chi \big( \LL_{\mt_\Symm} \LL_{\mt_\AntiS} \big)_{\mt_\AntiS}$. 
The first two relations give  
\begin{eqnarray}
\frac{1}{3}(2\bT_{11} - \bT_{12} + 2\bT_{21} + 2\bT_{23} )
=\left(
\begin{array}{ccc}
 0 & 0 & 0 \\
 0 & 1 & 2 \\
 0 & 2 & 1 \\
\end{array}
\right) 
\end{eqnarray}
and $\bT_{13}$, respectively. 

For $c^{\chi,2}_{\alpha'\alpha\beta\gamma\delta}\chi_{\alpha'}\mathcal{O}^2_{\alpha\beta\gamma\delta}$, i.e., the first dimension-8 operator $(\overline{L_{\alpha}} \tilde{H} \gamma^\mu \tilde{H}^\dag L_{\beta} ) (\overline{U_{\gamma \RH}} \gamma_\mu U_{\delta \RH})$, depending on the representation of $U_\RH$, there are several $Z_2$-invariant operators: 
\begin{itemize}
\item If $U_{1\RH}$ is a trivial singlet $\ms$, $\ms'$, or $\ms''$ of $A_4$, there are two $Z_2$-invariant operators
\begin{eqnarray}
\chi \LL_{\mt_\Symm}(\overline{U_{1\RH}} U_{1\RH})_\ms\,, \chi \LL_{\mt_\AntiS}(\overline{U_{1\RH}} U_{1\RH})_\ms\,.
\end{eqnarray} 
They lead to the correlations of the coefficients
\begin{eqnarray}
&c^2_{ee11} = c^2_{\mu\tau11} = c^2_{\tau\mu11} = -2 c^2_{\mu\mu11} = - 2c^2_{\tau\tau11} = -2c^2_{e\mu11} = -2c^2_{\mu e11} = -2c^2_{e\tau11} = -2c^2_{\tau e11} \,; \nonumber\\
&-c^2_{\mu\mu11} = c^2_{\tau\tau11} = c^2_{e\mu11} = c^2_{\mu e11} = -c^2_{e\tau11} = -c^2_{\tau e11}\,,~ c^2_{ee11} = c^2_{e\tau11} = c^2_{\tau e11} = 0 \,,
\label{eq:relationII2} 
\end{eqnarray} 
respectively. They give rise to two textures $\bT_2\equiv\bT_{12}+\bT_{22}$ and $\bT_3\equiv \bT_{13}+\bT_{23}$, respectively. 

\item If $U_{1\RH}$ is arranged as one component of a triplet $U_\RH=(U_{1\RH}, U_{2\RH}, U_{3\RH})^T \sim \mt$ of $A_4$, there are 6 independent $Z_2$-invariant operators contributing to NSIs, 
\begin{eqnarray}
&\chi\LL_{\mt_\Symm}(\overline{U_\RH} U_\RH)_\ms\,,~
\chi\LL_{\mt_\AntiS}(\overline{U_\RH} U_\RH)_\ms\,,~
\chi \big( \LL_{\mt_\Symm}(\overline{U_\RH} U_\RH)_{\mt_\Symm}\big)_{\mt_\Symm}\,,~\nonumber\\
&\chi \big( \LL_{\mt_\Symm}(\overline{U_\RH} U_\RH)_{\mt_\Symm}\big)_{\mt_\AntiS}\,,~
\chi \big( \LL_{\mt_\AntiS}(\overline{U_\RH} U_\RH)_{\mt_\Symm} \big)_{\mt_\Symm}\,,~
\chi \big( \LL_{\mt_\AntiS}(\overline{U_\RH} U_\RH)_{\mt_\Symm} \big)_{\mt_\AntiS}\,.~
\end{eqnarray} 
The first two give the two correlations as in Eq.~\eqref{eq:relationII2}. The rest four give rise to 
\begin{eqnarray}
& c^2_{ee11} = -2 c^2_{\mu\mu11} = -2 c^2_{\tau\tau11} = -2 c^2_{\mu\tau11} = -2 c^2_{\tau\mu11} = 4c^2_{e\mu11} = 4c^2_{\mu e11} = c^2_{e\tau11} = 4c^2_{\tau e11} \,;~ \nonumber\\
&c^2_{\mu\mu11} = -c^2_{\tau\tau11} = 2 c^2_{e\mu11} = 2 c^2_{\mu e11} = -2 c^2_{e\tau11} = 2 c^2_{\tau e11} \,,~ c^2_{ee11} = c^2_{e\tau11} = c^2_{\tau e11} = 0 \,; \nonumber\\
&ic^2_{\mu\tau11} = -ic^2_{\tau\mu11} = -2i c^2_{e\mu11} = 2i c^2_{\mu e11} = 2i c^2_{e\tau11} = -2i c^2_{\tau e11}\,,~ c^2_{ee11} = c^2_{\mu\mu11} = c^2_{\tau\tau11} = 0 \,; \nonumber\\
&ic^2_{e\mu11} = -i c^2_{\mu e11} = ic^2_{e\tau11} = -i c^2_{\tau e11} \,,~ c^2_{ee11} = c^2_{\mu\mu11} = c^2_{\tau\tau e11} = c^2_{\mu\tau e11} = c^2_{\tau\mu11} = 0 \,,
\label{eq:relationII2p}
\end{eqnarray}
respectively, where all non-vanishing values are real, required by the Hermitean of the Lagrangian. They give rise to 
\begin{eqnarray}
&&
2\bT_{12}-\bT_{22}  = 
\left(
\begin{array}{ccc}
 4 & 1 & 1 \\
 1 & -2 & -2 \\
 1 & -2 & -2 \\
\end{array}
\right) \,, \quad 2\bT_{13}-\bT_{23} = \left(
\begin{array}{ccc}
 0 & 1 & -1 \\
 1 & 2 & 0 \\
 -1 & 0 & -2 \\
\end{array}
\right)  \,,
\label{eq:sumruleIIp}
\end{eqnarray}
and $\bT_{32}$ and $\bT_{33}$, respectively. 

\end{itemize}
The similiar discussion applies to $\mathcal{O}^{3-8}$ and the same textures as predicted by $\mathcal{O}^{2}$ are obtained from these operators. 

Nine textures $\bT_{mn}$ in Eqs.~\eqref{eq:sumruleI} and \eqref{eq:sumruleII} form a complete basis for a Hermitian $3\times3$ matrix. Any two of these textures are orthogonal in the Hilbert-Schmidt inner product, $\text{tr}(\bT^\dag_{mn} \bT_{m'n'}) \propto \delta_{mm'}\delta_{nn'}$. Matter-effect NSIs contribute to the effective Hamiltonian term via the matrix
\begin{eqnarray}
\epsilon & \equiv &
\left(
\begin{array}{ccc}
\epsilon_{ee} & \epsilon_{e\mu} & \epsilon_{e\tau}\\
\epsilon_{\mu e} & \epsilon_{\mu\mu} & \epsilon_{\mu\tau}\\
\epsilon_{\tau e} & \epsilon_{\tau\mu} & \epsilon_{\tau\tau}
\end{array}
\right)
\equiv\left(
\begin{array}{ccc}
\epsilon_{ee} & |\epsilon_{e\mu}|\mathrm{e}^{i\phi_{e\mu}} & 
|\epsilon_{e\tau}| \mathrm{e}^{i\phi_{e\tau}}\\ 
|\epsilon_{\mu e}| \mathrm{e}^{-i\phi_{e\mu}} & \epsilon_{\mu\mu} 
& |\epsilon_{\mu\tau}| \mathrm{e}^{i\phi_{\mu\tau}}\\
|\epsilon_{e\tau}| \mathrm{e}^{-i\phi_{e\tau}} & |\epsilon_{\mu\tau}| 
\mathrm{e}^{-i\phi_{\mu\tau}} & \epsilon_{\tau\tau}
\end{array}
\right) 
= \sum_{m,n=1,2,3} \alpha_{mn} \bT_{mn}/N_{mn},
\label{eq:standard_para}
\end{eqnarray}
where $N_{mn}$ are normalization factor $N_{11}=\sqrt{3}$, $N_{12}=\sqrt{6}$, $N_{13}=\sqrt{2}$, $N_{21}=N_{31}=\sqrt{6}$, $N_{22}=N_{32}=2\sqrt{3}$ and $N_{23}=N_{33}=2$. The relations between $\epsilon_{\alpha\beta}$ and $\alpha_{mn}$ are shown in Table~\ref{tab:prob_coeff_texture}, and the following properties are satisfied
\begin{eqnarray} 
\text{tr}(\epsilon \epsilon^{\dag}) = \sum_{\alpha,\beta=e,\mu,\tau} |\epsilon_{\alpha\beta}|^2 
=\sum_{m,n=1,2,3} \alpha_{mn}^2\,.
\end{eqnarray}
Note that $\bT_{11}\equiv \bI$ is unobservable in neutrino oscillations experiments. 

We list all $A_4$- and $Z_2$-motivated matter-effect NSI textures predicted by $A_4$- and $Z_2$-invariant operators $\mathcal{O}^p$ and $\chi \mathcal{O}^p$ in Table~\ref{tab:NSI_A4}, where $\chi$ is the flavon VEV inducing $A_4$ breaking to $Z_2$. As seen in the table, an NSI texture predicted by an $A_4$-invariant ($Z_2$-invariant) operator usually does not preserve $A_4$ ($Z_2$). This is because the matter-effect NSIs have specified the first-generation charged fermions. These charged fermions, if not arranged as an singlet $\ms$ of $A_4$, is not invariant in $A_4$ ($Z_2$), and thus the NSI texture does not respect $A_4$ ($Z_2$). In a specific $A_4$ model, the NSI matrix $\epsilon$ could be a linear combinations of $\bT_{mn}$. However, it is notable that $\bT_{31}$ cannot be obtained directly from the above analysis. 
The analysis based on higher-dimensional operators cannot determine which texture is more important and dominant in oscillation experiments. However, as what we will discuss in the next section, once we consider UV completion for these textures and include experimental constraints, some of them are suppressed and cannot be measured in neutrino experiments.


\begin{table}[h!]
\renewcommand\arraystretch{1}
\addtolength{\tabcolsep}{-2pt}
\begin{center}
\begin{tabular}{|l|l|L{7.6cm}|C{2.5cm}|}
\hline\hline

& Representations & $A_4$-invariant operators  &NSI textures \\ \hline

\multirow{2}{*}{$\mathcal{O}^1$} & \multirow{2}{*}{$L\sim\mt$} &
$\LL_\ms \LL_\ms$, $\LL_{\ms'} \LL_{\ms''}$, $\LL_{\mt_\Symm} \LL_{\mt_\Symm}$, $\LL_{\mt_\AntiS} \LL_{\mt_\AntiS}$ & 
\multirow{2}{*}{$2\bT_{11}- \bT_{12}$}   \\\hline

\multirow{3}{*}{$\mathcal{O}^{2-8}$} & $L\sim\mt, F\sim \ms,\ms',\ms'',\mt$ &
$\LL_\ms \FF_\ms$ & 
$\bT_{11}$  \\\cline{2-4}

& \multirow{2}{4cm}{$L\sim\mt, F\sim\mt$} &
$\LL_{\mt_\Symm} \FF_{\mt_\Symm}$ &
$\bT_{12}$  \\\cline{3-4}

& & 
$\LL_{\mt_\AntiS} \FF_{\mt_\Symm}$ &
$\bT_{13}$  \\\hline\hline


& Representations & $Z_2$-invariant operators  &NSI textures \\ \hline

\multirow{3}{*}{$\chi\mathcal{O}^1$} & \multirow{3}{*}{$\chi\sim\mt, L\sim\mt$} &
$\chi \big( \LL_{\mt_\Symm} \LL_{\ms,\ms',\ms''} \big)_{\mt}$, $\chi \big( \LL_{\mt_\Symm} \LL_{\mt_\Symm} \big)_{\mt_\Symm}$, $\chi \big( \LL_{\mt_\AntiS} \LL_{\mt_\AntiS} \big)_{\mt_\Symm}$ & $\frac{1}{3}(2\bT_{11} \!-\! \bT_{12}\!$ $+\!2\bT_{21}\!+\!2\bT_{23}) $ \\\cline{3-4} 

& & $\chi \big(\LL_{\mt_\AntiS} \LL_{\ms,\ms',\ms''} \big)_{\mt}$, $\chi \big( \LL_{\mt_\Symm} \LL_{\mt_\AntiS} \big)_{\mt_\Symm}$ & $\bT_{13}$   \\ \hline

\multirow{6}{*}{$\chi\mathcal{O}^{2-8}$} & \multirow{2}{*}{$\chi\sim\mt, L\sim\mt, F\sim \ms,\ms',\ms'',\mt$} &
$\chi\LL_{\mt_\Symm} \FF_\ms$ & 
$\bT_{12}+\bT_{22}$  \\\cline{3-4}

& &
$\chi\LL_{\mt_\AntiS} \FF_\ms$ & 
$\bT_{13}+\bT_{23}$  \\\cline{2-4}

& \multirow{4}{4cm}{$\chi\sim\mt, L\sim\mt, F\sim\mt$} &
$\chi \big( \LL_{\mt_\Symm} \FF_{\mt_\Symm} \big)_{\mt_\Symm}$ &
$2\bT_{12}-\bT_{22}$  \\\cline{3-4}

& &
$\chi \big( \LL_{\mt_\AntiS} \FF_{\mt_\Symm} \big)_{\mt_\Symm}$ &
$2\bT_{13}-\bT_{23}$  \\\cline{3-4}

& &
$\chi \big( \LL_{\mt_\Symm} \FF_{\mt_\Symm} \big)_{\mt_\AntiS}$ &
$\bT_{32}$  \\\cline{3-4}

& & 
$\chi \big( \LL_{\mt_\AntiS} \FF_{\mt_\Symm} \big)_{\mt_\AntiS}$ &
$\bT_{33}$  \\\hline\hline


\end{tabular}
\caption{\label{tab:NSI_A4}NSI Textures in matter predicted by $A_4$ and the residual symmetry $Z_2$, where $F$ represents any SM fermion. The textures $\bT_{1n}$ are defined in Eq.~\eqref{eq:sumruleI}, $\bT_{2n}$ and $\bT_{3n}$ are defined in Eq.~\eqref{eq:sumruleII}, and $\chi$ is defined in Eq.~\eqref{eq:VEV}. }
\end{center}
\end{table}


\section{NSI textures realised in renormalisable flavour models \label{sec:UV}} 

In this section, we consider how to realise higher-dimensional operators in UV-complete models. We follow the widely used technique in \cite{Antusch:2008tz, Gavela:2008ra}, where the dimension-6 operator is mediated by singly-charged gauge-singlet scalars and the dimension-8 operators can be realised with the help of singly-charged gauge-singlet scalars and neutral fermions. Imposing the $A_4$ symmetry differs the analysis in the following ways: 1) It requires to extend the heave particles as  relevant multiplets of $A_4$. 2) Mass matrices of these particles gain special structures constrained by $A_4$ or $Z_2$ (if the $Z_2$-invariant flavon VEV $\chi$ is included), which further contribute the NSI structure. 3) Although experimental constraints to the heavy particles have been studied in \cite{Antusch:2008tz, Gavela:2008ra} and later work, e.g., \cite{Wise:2014oea,Antusch:2014woa}, the non-Abelian flavour symmetry connects channels of different flavours together and may result in stronger constraints. Due to these differences, NSIs with  $A_4$-invariant UV completion deserve a careful consideration. 

\subsection{UV completion of the dimension-6 operator}

We first consider the UV completion of $\mathcal{O}^1$, $\varepsilon_{ac}\varepsilon_{bd} (\overline{L_{a\alpha}}\gamma^\mu L_{b\beta} ) 
(\overline{L_{c\gamma}}\gamma_\mu L_{d\delta} )$. The only way is to introduce a singly charged scalar $S$ which is a $SU(2)_\LH$ singlet with $Y=+1$ and assume that it couples to $L$ in an ``antisymmetric'' form \cite{Antusch:2008tz}. Together with the kinetic and mass term of $S$, we write down the renormalisable Lagrangian terms as 
\begin{eqnarray}
\mathcal{L}_S&=& (D_\mu S)^\dag (D^\mu S) - (M_S^2)_{\alpha\beta} S^*_\alpha S_\beta +  \lambda_{\alpha\beta\gamma} \varepsilon_{ab} \overline{L_{a\alpha}^\text{C}} L_{b\beta} S_\gamma + \text{h.c.}\,,
\label{eq:D6_UV}
\end{eqnarray} 
where $\lambda_{\alpha\beta\gamma}=-\lambda_{\beta\alpha\gamma}$. In the framework of $A_4$, $S$ cannot be arranged as a singlet representation $\ms,\ms'$ or $\ms''$ of $A_4$ since the symmetric CG coefficients of $A_4$ and the anti-symmetric property of $\lambda$ lead to $S(\overline{L^\text{C}} L)_{\ms^{(\prime\prime,\prime)}}\equiv 0$. Similarly by arranging $S\sim \mt$, we obtain $S(\overline{L^\text{C}} L)_{\mt_\Symm}= 0$. The only term that can contribute to the operator in Eq.~\eqref{eq:D6_UV} is $S(\overline{L^\text{C}} L)_{\mt_\AntiS}$ for $S\sim \mt$. All non-vanishing coefficients satisfy
\begin{eqnarray}
\lambda_{123} = \lambda_{231} = \lambda_{312} = - \lambda_{132} = - \lambda_{213} = - \lambda_{321} \equiv \lambda_0\,.
\label{eq:lambda_A4}
\end{eqnarray}
After $S$ decouples and by using the Fierz identity, we obtain $\mathcal{O}^1$ and the resulted NSI parameters are obtained as 
\begin{eqnarray}
\epsilon^e_{\alpha\beta} = \frac{1}{\sqrt{2}G_F} \lambda_{\beta e} (M_S^2)^{-1} \lambda_{\alpha e}^\dag \,,
\end{eqnarray}
where each $\lambda_{\alpha\beta}$ is the $1\times 3$ matrix given by $\lambda_{\alpha\beta}=(\lambda_{\alpha\beta1}, \lambda_{\alpha\beta2}, \lambda_{\alpha\beta3})$. 

The structures of $\epsilon_{\alpha\beta}^e$ are fully determined by the flavour structure of $M_S^2$. We will see how to constrain the $M_S^2$ structure. 
\begin{itemize}
\item
An $A_4$-invariant mass term for the charged scalar can only take the form $\mu_S^2 (S^* S)_\mathbf{1} = \mu_S^2 \sum_\alpha S_\alpha^*S_\alpha$ with $\mu^2_S>0$, leading to the charged scalar mass matrix $M_S^2 = \mu_S^2 \bI$. From this mass matrix, we obtain the texture $\epsilon^e =\alpha_0 \bT_{12}'$ with $\alpha_0=\frac{\mu_S^2}{\sqrt{2}G_F}$. 

\item 
In order to obtain non-vanishing off-diagonal NSI entries, $A_4$ has to be broken. As shown in the last section, the key is to introduce a flavon with the $Z_2$-preserving VEV $\chi$. We add the following renormalisable couplings to the Lagrangian, 
\begin{eqnarray}
\frac{\mu_S^2}{v_\chi} \left[ \frac{2}{3}h_\Symm\Big(\chi(S^* S)_{\mt_\Symm}\Big)_\ms - \frac{2}{\sqrt{3}}h_\AntiS\Big(\chi(S^* S)_{\mt_\AntiS}\Big)_\ms \right] \,,
\label{eq:S_mass}
\end{eqnarray}
where $h_\Symm$ and $h_\AntiS$ are real dimensionless coefficients as required by the Hermiticity of the Lagrangian. Then, the $S$ mass matrix is non-diagonal and the resulted NSI matrix becomes
\begin{eqnarray}
\epsilon^e =\alpha_0 \left[ \bT_{12}' + 
\frac{1}{3} \left(
\begin{array}{ccc}
 0 & 0 & 0 \\
 0 & h_\Symm-h_\Symm^2 & 2 h_\Symm + h_\Symm^2 \\
 0 & 2 h_\Symm + h_\Symm^2 & h_\Symm - h_\Symm^2 \\
\end{array}
\right) 
+
\frac{1}{3} \left(
\begin{array}{ccc}
 0 & 0 & 0 \\
 0 & \sqrt{3}h_\AntiS-h_\AntiS^2 & h_\AntiS^2 \\
 0 & h_\AntiS^2 & -\sqrt{3}h_\AntiS-h_\AntiS^2 \\
\end{array}
\right) \right]\,,
\end{eqnarray}
where $\alpha_0 = |\lambda_0|^2 / [\sqrt{2}G_F \mu_S^2 (1-h_\Symm^2-h_\AntiS^2)]$. 
$\epsilon^e$ contains three real parameters $\epsilon_{\mu\mu}$, $\epsilon_{\tau\tau}$ and $|\epsilon_{\mu\tau}|$. The renormalisable quartic terms $\Big((\chi \chi)_{\mt_\Symm}(S^* S)_{\mt_\Symm}\Big)_\ms$ and $\Big((\chi \chi)_{\mt_\Symm}(S^* S)_{\mt_\AntiS}\Big)_\ms$ are also allowed by the symmetry, such terms do not modify the flavour structures of $M_S^2$ and $\epsilon^e$ except redefinitions of $h_\Symm$ and $h_\AntiS$. 
\end{itemize}

However, sizeable NSI textures are hard to be realised in this approach due to the strong constraint from the radiative charged LFV  measurements. Although the tree-level 4-charged-fermion interactions have been avoided, radiative decays $E_\alpha \to E_\beta \gamma$ involving $S$ and neutrinos in the loop are triggered by the interaction $\overline{L^\text{C}} L S$, and the relative branching ratios are $\propto |G_F^{-1} \lambda_{\alpha \gamma} (M_S^2)^{-1} \lambda_{\beta\gamma}^\dag|^2$, where $\gamma\neq\alpha,\beta$. General upper bounds of $\tau\to e \gamma$ and $\tau \to \mu \gamma$ branching ratios are around $10^{-8}$ \cite{Belle} and \cite{BaBar}, and that of $\mu \to e \gamma$ is $4.2\times 10^{-13}$ \cite{MEG}. Without flavour symmetries, the coefficients $\lambda_{\alpha\beta\gamma}$ and mass terms $(M_S^2)_{\alpha\beta}$ are free parameters, and $\tau\to e \gamma$ and $\mu \to e \gamma$ do not provide direct constraints to NSIs \cite{Antusch:2008tz}. Once the flavour symmetry is included, relations such as Eqs.~\eqref{eq:lambda_A4} and \eqref{eq:S_mass} are satisfied. In the limit $h_\Symm, h_\AntiS \to 0$, all radiative decays are forbidden. However, off-diagonal NSIs are also forbidden in this case, becoming less interesting in oscillation experiments. On the other hand, by assuming $h_\Symm$ or $h_\AntiS \sim \mathcal{O}(1)$, very strong constraint, $|\epsilon_{\alpha\beta}^e| < 7 \times 10^{-5}$, is obtained from the upper limit of $\mu\to e \gamma$.

\subsection{UV completions of dimension-8 operators \label{sec:UV_dim8}}

In the following, we will only consider NSIs from UV completions of dimension-8 operators. Before performing a detailed analysis, we directly point out our main result that, in UV-complete models with the $Z_2$ residual symmetry, only linear combinations of the following NSI textures are worth for phenomenological studies in neutrino oscillation experiments, 
\begin{eqnarray}
\hspace{-3mm}
\bT_1= \frac{1}{3} \left(
\begin{array}{ccc}
 2 & \!-1 & \!-1 \\
 \!-1 & 2 & \!-1 \\
 \!-1 & \!-1 & 2 \\
\end{array}
\right) ,  \,
\bT_{2} = \frac{1}{3} \left(
\begin{array}{ccc}
 2 & \!-1 & \!-1 \\
 \!-1 & \!-1 & 2 \\
 \!-1 & 2 & \!-1 \\
\end{array}
\right)
 , \,
\bT_{3} = \frac{1}{\sqrt{3}} \left(
\begin{array}{ccc}
 0 & \!-1 & 1 \\
 \!-1 & 1 & 0 \\
 1 & 0 & \!-1 \\
\end{array}
\right)
 , \,
\bT_4 = \frac{1}{\sqrt{3}}
\left(
\begin{array}{ccc}
 0 & \!-i & i \\
 i & 0 & \!-i \\
 \!-i & i & 0 \\
\end{array}
\right) .
\label{eq:majorNSI}
\end{eqnarray}
We refer them to major NSI textures. 
They are combinations of some $\bT_{mn}$, $\bT_{1}=\frac{1}{3}(2\bT_{11}-\bT_{21})$, $\bT_{2}=\frac{1}{3}(\bT_{12}+\bT_{22})$, $\bT_{3}=\frac{1}{\sqrt{3}}(\bT_{13}+\bT_{23})$, and $\bT_4= \frac{1}{\sqrt{3}}\bT_{31}$. 
As discussed later in this section, the rest NSI textures $\bT_{12}$, $\bT_{13}$, $\bT_{32}$, $\bT_{33}$ and their combinations are strongly constrained by non-oscillation data. Therefore, we call them  `minor NSI texture'. Here, we classify them into `major' and `minor' due to their testability. In the former case, although they are small, we may still have the opportunity to detect them, while in the later case, we will have no chance to test them in the next-generation neutrino experiments. Throughout this paper, we will put our focus on the `major NSIs texture'.

\subsubsection*{Major NSI textures realised in UV-complete $A_4$ models}\label{sec:UV_dim8_major}

We consider how to realise the major NSI textures in the renormalisable $A_4$ models and consider their experimental constraints. %
Before electroweak symmetry breaking, the operators $\mathcal{O}^{2-6}$ take the form as dimension-8 operator $(\overline{L}\tilde{H}\gamma^\mu \tilde{H}^\dag L) (\overline{F}\gamma_\mu F)$. A popular way to realise large NSIs is introducing a vector boson $Z'$. Then, the 4-charged-fermion interaction $(\overline{F}\gamma^\mu F) (\overline{F}\gamma_\mu F)$ is unavoidable. In order to be consistent with experimental data, the coupling must be very small. 
Here, we will carefully avoid the 4-charged-fermion interactions newly introduced after the decouple of the new particles in the UV sector. Thus, interactions mediated by $Z'$ will not be considered. 

We focus on $\mathcal{O}^4$ by using a singly charged scalar $\phi$ and a neutral fermion $N$ to realise major NSI textures. 
The renormalisable interactions are given by 
\begin{eqnarray}
\mathcal{L}_{\phi, N}&=& (D_\mu \phi)^\dag (D^\mu \phi) - (M_\phi^2)_{\alpha\beta} \phi^*_\alpha \phi_\beta +\overline{N} i \partial\!\!\!/ N - M_{N\alpha\beta} \overline{N_{\alpha \RH}} N_{\beta \LH} \nonumber\\ &&- \kappa_{\alpha\beta\gamma} \overline{E_{\alpha \RH}} N_{\beta \LH} \phi_\gamma^*- y_{\alpha\beta} \overline{L_\alpha} \tilde{H} N_{\beta \RH} + \text{h.c.}\,, 
\label{eq:D8_model}
\end{eqnarray} 
where $D_\mu = \partial_\mu + ie A_\mu$. 
The charged scalar is a $SU(2)_\LH$ singlet with $Y=-1$. In order to distinguish it from $S$ in the last subsection, we denote it as $\phi$. There is no lepton-number-violating (LNV) coupling in the above interactions. For the neutral fermion $N$, we require a vector-like mass term $M_N \overline{N_\RH} N_\LH$ as shown in the above. If there is an additional small LNV mass term $\mu \overline{N_\LH^\text{C}} N_\LH$ and hierarchical masses $y /\sqrt{G_F} \ll M_N$, we recover the inverse seesaw model \cite{inverse}. But here we do not specify if $N$ is related to the origin of active neutrino masses. No matter whether there is a small LNV mass term, we can always arrive at a dimension-8 operator $\sim \frac{\kappa^2 y^2}{M_\phi^2 M_N^2} (\overline{L}\tilde{H}E_\RH) (\overline{E_\RH}\tilde{H}^\dag L) $ after the decouple of the charged scalar and sterile neutrinos, from which we obtain $\mathcal{O}^4$. 
Once the flavour structure is included, the $3\times 3$ NSI parameter matrix $\epsilon^e$ is expressed as  
\begin{eqnarray}
\epsilon^e = \frac{1}{8 G_F^2} (y M_N^{-1} \kappa_e) (M_\phi^2)^{-1} (y M_N^{-1} \kappa_e)^\dag \,,
\label{eq:NSI_matrix}
\end{eqnarray}
where $\kappa_e$ is a $3\times3$ matrix defined via $(\kappa_\alpha)_{\beta\gamma} = \kappa_{\alpha\beta\gamma}$ for $\alpha=e,\mu,\tau$. 


\begin{figure}[t]
\centering
\includegraphics[width=.9\textwidth]{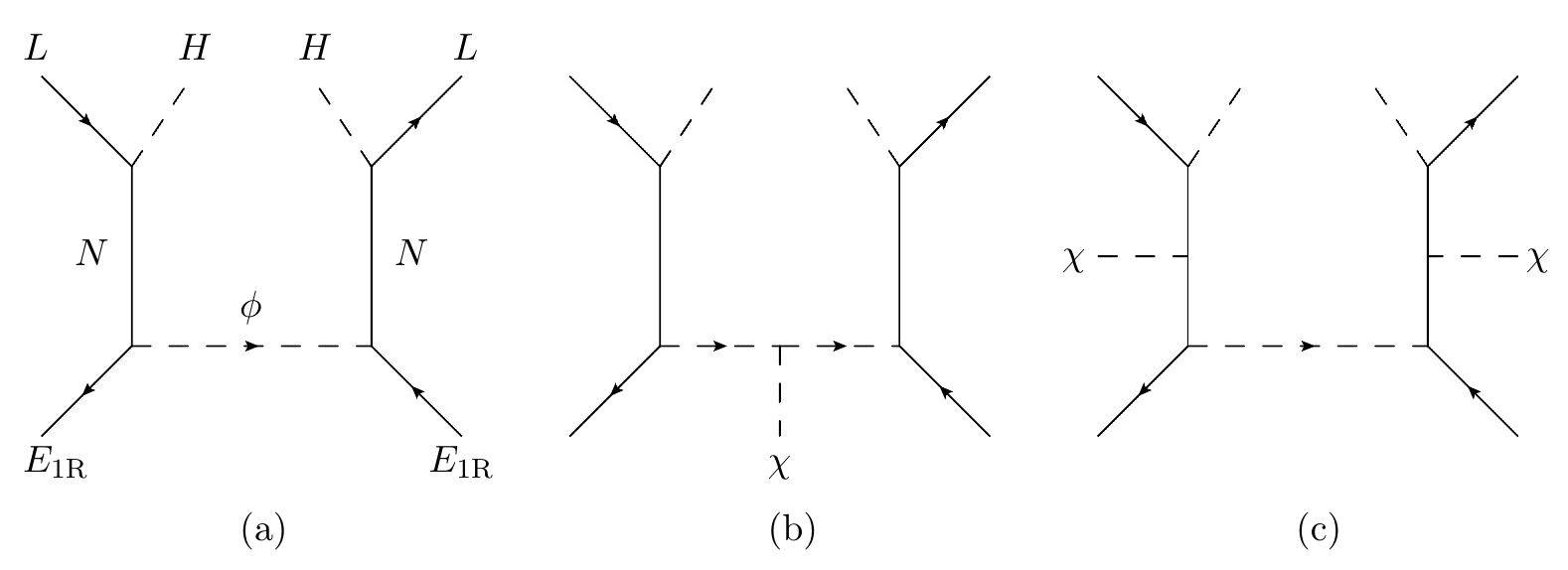}
\caption{\label{fig:dis_epsilon} Diagrams to realise sizeable NSI textures corresponding to dimension-8 operator $\mathcal{O}^4$ in leptonic $A_4$ models. }
\end{figure}


We will discuss how the $A_4$ symmetry can constrain NSIs originating from this renormalisable model. We first consider $A_4$-motiviated NSI textures without the involvement of flavons. 
In the flavour space, since we have arranged $L\sim \mt$, the fields $N_\LH$, $N_\RH$ and $\phi$ must be triplets to ensure the invariance of Lagrangian in $A_4$. We follow the setup of most $A_4$ models that $E_{1\RH}$ is fixed as a singlet $\mathbf{1}$ of $A_4$. 
An $A_4$-invariant mass term for the charged scalar can only take the form $\mu_\phi^2 (\phi^* \phi)_\mathbf{1} = \mu_\phi^2 \sum_i\phi_i^*\phi_i$ with $\mu^2_\phi>0$, i.e., the charged scalar mass matrix $M_{\phi}^2 = \mu_\phi^2 \bI$. Similarly, to be invariant under transformations of $A_4$, the Dirac mass matrix of the sterile neutrinos $M_N$ and the Yukawa coupling between $L$ and $N_\RH$, $y$ is also proportional to an identity matrix, $M_{N} = \mu_N \bI$, $y=y_0 \bI$. The structures of the couplings $y$ and $\kappa$ depend on representations of $E_\RH$. Interactions involving $\phi$ and $N$ are given by
\begin{eqnarray}
\kappa_0 \overline{E_{1\RH}} (N_\LH \phi^*)_\ms + y_0 (\overline{L} \tilde{H} N_\RH)_\ms + \text{h.c.}\,. 
\end{eqnarray}
Thus, both coupling matrices $\kappa$ and $y$ appear to be proportional to the identity matrix, $\kappa = \kappa_0 \bI$, $y=y_0 \bI$. 
After $\phi$ and $N$ are integrated out from the Lagrangian, we obtain that the $\mathcal{O}^4$ takes the $\LL_\ms \FF_\ms$ form as listed in Table~\ref{tab:NSI_A4} for $F=E_\RH$. Finally, we obtain the NSI texture $\epsilon^e = \alpha_0 \bI$, where 
\begin{eqnarray}
\alpha_0 = \frac{ |y_0 \kappa_0|^2}{8G_F^2 \mu_N^{2} \mu_\phi^2} \,.
\end{eqnarray}
Since $\bI$ is an identity matrix, $\epsilon^e$ in this special case has no observable signatures in neutrino oscillation experiments. 

The involvement of $\chi$ breaks $A_4$ to $Z_2$ and modifies the correlation relations of NSI parameters. In order to realise relatively large and measurable NSI effects, we only consider the contribution of renormalisable couplings of $\chi$. There are cases, as shown in Figure~\ref{fig:dis_epsilon} (b) and (c), where $\chi$ couples to $\phi$ and $N$, modifying their mass matrices, respectively.  
\begin{itemize}
\item
The charged scalar $\phi$ mass matrix modified by the coupling between $\chi$ and $\phi$. We add the following renormalisable coupling to the Lagrangian, 
\begin{eqnarray}
\frac{\mu_\phi^2}{v_\chi} \left[ \frac{2}{3}f_\Symm\Big(\chi(\phi^* \phi)_{\mt_\Symm}\Big)_\ms - \frac{2}{\sqrt{3}}f_\AntiS\Big(\chi(\phi^* \phi)_{\mt_\AntiS}\Big)_\ms \right] \,,
\end{eqnarray}
where $f_\Symm$ and $f_\AntiS$ are real dimensionless coefficients as required by the Hermiticity of the Lagrangian. The relevant higher-dimensional operators after $\phi$ and $N$ integrated out take the forms as $\chi\LL_{\mt_\Symm} \FF_\ms$ and $\chi\LL_{\mt_\AntiS} \FF_\ms$, respectively. The modified $\phi$ mass matrix turns out to be
\begin{eqnarray}
M_\phi^2 /\mu_\phi^2 =  \bI + f_\Symm \bT_{2} + f_\AntiS \bT_{3} \,.
\label{eq:phi_mass}
\end{eqnarray}
Terms such as $\big((\chi\chi)_{\mt_\Symm}(\phi^* \phi)_{\mt_\Symm}\big)_\ms$, $\big((\chi\chi)_{\mt_\Symm}(\phi^* \phi)_{\mt_\AntiS}\big)_\ms$ are also renormalisable and should be considered for completeness. These terms will not induce new structures different from Eq.~\eqref{eq:phi_mass}. 

\item
The Dirac mass matrix of $N$ is modified by couplings between $\chi$ and $N$. The related renormalisable Lagrangian term is given by 
\begin{eqnarray}
\frac{\mu_N}{v_\chi} \left[ \frac{2}{3}g_\Symm \big(\chi (\overline{N_\LH} N_\RH)_{\mt_\Symm}\big)_\ms - \frac{2}{\sqrt{3}} g_\AntiS \big(\chi (\overline{N_\LH} N_\RH)_{\mt_\AntiS}\big)_\ms \right] + \text{h.c.}\,,
\end{eqnarray}
where $g_\Symm$ and $g_\AntiS$ are in general complex parameters. Dirac mass matrix $M_N$ is modified to
\begin{eqnarray}
M_N/\mu_N =  \bI + g_\Symm \bT_{2} + g_\AntiS \bT_{3}\,.
\label{eq:N_mass}
\end{eqnarray}

\end{itemize} 
Taking the flavon-modified mass matrices of $\phi$ and $N$ into account, we state that the final detectable (i.e., ignoring the undetectable $\bI$)  NSI matrix $\epsilon^e$ in Eq.~\eqref{eq:NSI_matrix} is always a linear combination of $\bT_i$ for $i=1,2,3,4$. This is guaranteed by the algebra of $\bT_i$ and can be straightforwardly proven by implying Eqs.~\eqref{eq:AB} and \eqref{eq:invA} in Appendix \ref{sec:math}. 
From Table~\ref{fig:dis_epsilon}, one can expect that the textures $\bT_2$ and $\bT_3$ will be predicted. 
The other two textures, $\bT_1$ and $\bT_4$, which are not predicted from higher-dimensional operators, are obtained from the inverse transformations of $M_\phi^2$ and $M_N$, and matrix product $\bT_2 \bT_3 = - i \bT_4$. $\bT_1$ and $\bT_4$ appear at the second order of $f_\Symm,f_\AntiS$ and $g_\Symm, g_\AntiS$. If $f_\Symm,f_\AntiS,g_\Symm, g_\AntiS\ll 1$ is satisfied, the $\bT_1$ and $\bT_4$ parts are negligible compared with the $\bT_2$ and $\bT_3$ parts. However, these coefficients, as coefficients of renormalisable terms, may take $\mathcal{O}(1)$ values, and thus in this case, $\bT_1$ and $\bT_4$ may have comparable NSI effects to $\bT_2$ and $\bT_3$. 

The flavour structures of NSIs can be further discussed in the following scenarios, dependent on the role of the flavon VEV $\chi$: 
\begin{itemize}

\item
With the assumption of additional symmetries, $\chi$ may only couple to $\phi$, not to $N$, i.e., $g_\AntiS,g_\Symm=0$.
The resulted detectable NSI matrix is explicitly expressed as
\begin{eqnarray}
\epsilon^e =\alpha_0 \left[ (f_\Symm^2+f_\AntiS^2)\bT_{1} -f_\Symm \bT_{2} -f_\AntiS \bT_{3} \right]\,.
\end{eqnarray}
Here, only $\bT_{1}$, $\bT_{2}$ and $\bT_{3}$ appear, and $\alpha_0$ has been redefined. 

\item
On the other hand, if $\chi$ only couple to $N$, we obtain the following NSI matrix
\begin{eqnarray}
\epsilon^e &=& \alpha_0 \left\{ \big[ - (2+|g_\Symm|^2+|g_\AntiS|^2) (|g_\Symm|^2+|g_\AntiS|^2) + 4 \text{Re} (g_\Symm^2 + g_\AntiS^2) + 4 [\text{Im}(g_\Symm^* g_\AntiS)]^2 \big] \bT_{1} \right. \nonumber\\ 
&& \hspace{2cm} \left. - 2 \text{Re}(g_\Symm) \bT_{2} - 2 \text{Re}(g_\AntiS) \bT_{3} - 2 \text{Im}(g_\Symm^* g_\AntiS) \bT_4 \right\}\,.
\end{eqnarray}
where $\alpha_0$ has been redefined. It is a linear combination of all four $\bT_i$, but $\bT_4$ is important only if both $|g_\Symm|$ and $|g_\AntiS|$ are sizeable and there is a relative phase between $g_\Symm$ and $g_\AntiS$.  

\item
If the anti-symmetric couplings $f_\AntiS$ and $g_\AntiS$ are forbidden, the NSI matrix can be simplified to a linear combination of $\bT_1$ and $\bT_2$. On the other hand, if the symmetric couplings $f_\Symm$ and $g_\Symm$ are forbidden, the NSI matrix is a linear combination of $\bT_1$ and $\bT_3$. These two cases are valid if the group $A_4$ is replaced by larger groups. For example, in the hexahedron group $S_4$ \cite{S4}, there are two triplet irreducible representations, and the symmetric and anti-symmetric products $\mt_\Symm$ and $\mt_\AntiS$ correspond to two different representations. By arranging $\chi$ to be one of the triplets, the anti-symmetric (or symmetric) products can be forbidden, and thus only the symmetric (or anti-symmetric) couplings are left. 
\end{itemize}

Naively, one may expect that NSIs from the UV completion of the dimension-8 operator is more constrained than that of the dimension-6 operator, but this is not the case in the framework of the flavour symmetry. 
First of all, no tree-level CLFV interactions have been introduced from the Lagrangian in Eq.~\eqref{eq:D8_model} as required. Although radiative CLFV processes are induced by the coupling $\overline{E_\RH} N_\LH \phi$, they essentially rely on the coupling with the second or third generation charged lepton $E_{2\RH}$ or $E_{3\RH}$. By arranging $E_{1\RH}$, $E_{2\RH}$ and $E_{3\RH}$ as different singlets of $A_4$, the relevant coefficients are theoretically independent of those involving in matter NSIs \cite{Biggio:2009kv, Biggio:2009nt}. Constraints on CLFV do not apply to NSIs. 
On the side of collider searches, with a careful treatment of $\phi$ decaying to $e/\mu$ plus missing transverse momentum or $\tau$ plus missing transverse momentum, the existing LEP and LHC data still allow a singlet charged scalar as light as 65 GeV \cite{Cao:2017ffm}. 
The main constraint in this model is the bound of the non-unitarity of the lepton mixing. 
The decouple of sterile neutrinos contributes to the active neutrino kinetic mixing as $
\frac{y^2}{M_N^2} (\overline{L} \tilde{H}) \partial\!\!\!/ (\tilde{H}^\dag L)$. 
After rescaling the kinetic terms of active neutrinos, non-unitarity of the PMNS matrix is
\begin{eqnarray}
\eta \equiv V_{\text{PMNS}}^\dag V_{\text{PMNS}} - \mathbf{1} = \frac{1}{2\sqrt{2}G_F} (yM_N^{-1})(yM_N^{-1})^\dag \,.
\end{eqnarray} 
The non-unitarity bound from a global analysis of LFV decays, probes of the universality of weak interactions, CKM unitarity bounds and electroweak precision data is around $\eta \sim 10^{-3}$ \cite{Fernandez-Martinez:2016lgt}. 
Combining with the above constraints, we see that it is still possible to achieve the major NSI textures with coefficients $ \sim \eta /(G_F M_\phi^2)$ at $10^{-2}$ or $10^{-3}$ level. These values may be potentially measured by the next-generation accelerator neutrino oscillation experiments.

In the above, we have constructed UV-complete models for $\mathcal{O}^4$ and $\chi \mathcal{O}^4$. A similar discussion can be directly extended to the $\mathcal{O}^{2,3,5}$ and $\chi \mathcal{O}^{2,3,5}$ by replacing the singly-charged scalar $\phi$ by $\phi_{U_\RH,D_\RH,Q}$ which are $SU(2)_\LH$ gauge singlet, single and doublet with hypercharge $Y=-2/3,+1/3$ and $-1/6$, respectively, and replacing the singlet $F=E_{1\RH}$ with $F=U_{1\RH}$, $D_{1\RH}$ and $Q_{1}$, respectively. The resulted NSI matrix is also a linear combination of the textures $\bT_1$, $\bT_2$, $\bT_3$ and $\bT_4$. The textures $\bT_1$, $\bT_2$, $\bT_3$ and $\bT_4$ are obtained by assuming the charged fermion as singlets of $A_4$. This treatment can avoid strong constraints from the second- and third-generation charged fermions. These textures are less constrained than the other textures discussed below and thus, we call them major NSI textures.

\subsubsection*{Minor NSI textures realised in UV-complete $A_4$ models}

The minor NSI textures $\bT_{12}$, $\bT_{13}$, $\bT_{32}$, $\bT_{33}$ and their combinations cannot be realised in the above discussions. This is compatible with Table~\ref{tab:NSI_A4}, where the minor textures are obtained by setting $F\sim \mt$. 
To achieve these textures, as shown in Table~\ref{tab:NSI_A4}, $F$ has to be assumed to be a triplet of $A_4$. Then $F$ cannot be chosen as right-handed charged leptons and not realised in the $\mathcal{O}^4$ and $\chi \mathcal{O}^4$ series. We will discuss how to realise them in UV-complete $A_4$ models as a complement. 

To realise the $A_4$-motivated $\bT_{12}$ and $\bT_{13}$, we choose $F=U_\RH \equiv (U_{1\RH}, U_{2\RH}, U_{3\RH})^T \sim \mathbf{3}$ of $A_4$ and consider the UV completion of $\mathcal{O}^{2}$. The latter is obtained by replacing the singly charged scalar $\phi$ with a fractionally charged scalar $\phi_{U_\RH}$, i.e., a scalar leptoquark, with the hypercharge $Y=-2/3$, and coupling to $N_\LH$ and  $U_\RH$. The renormalisable couplings are given by
\begin{eqnarray}
\kappa_\Symm^{U_\RH} ((\overline{U_\RH} N_\LH)_{\mt_\Symm} \phi_{U_\RH}^*)_\ms + \kappa_\AntiS^{U_\RH} ((\overline{U_\RH} N_\LH)_{\mt_\AntiS} \phi_{U_\RH}^*)_\ms + \text{h.c.}\,. 
\end{eqnarray}
Then, coupling matrix $\kappa$ is modified to $\kappa_{U_\RH} = \kappa_\Symm^{U_\RH} \bT_{12} + \kappa_\AntiS^{U_\RH} \bT_{13}$ and the $A_4$-preserved NSI texture 
\begin{eqnarray}
\epsilon^u \equiv \frac{1}{8 G_F^2} (y M_N^{-1} \kappa_{U_\RH}) (M_{\phi_{U_\RH}}^2)^{-1} (y M_N^{-1} \kappa_{U_\RH})^\dag 
\label{eq:NSI_matrix_v2}
\end{eqnarray}
is obtained as a linear combination of $\bT_{12}$ and $\bT_{13}$. 
Finally, we include the $A_4$-breaking effect in the $\phi_{U_\RH}$ and $N$ mass matrices, as in Eqs.~\eqref{eq:phi_mass} and \eqref{eq:N_mass}. Non-zero $\bT_{32}$ and $\bT_{33}$ can be extracted out in principle. 

The minor textures $\bT_{12}$, $\bT_{13}$, $\bT_{32}$ and $\bT_{33}$ are expected to receive stronger constraints. The main reason is that $U_\RH=(U_{1\RH}, U_{2\RH}, U_{3\RH})$ is arranged as a triplet of $A_4$ and constraints from the second- and third-generation charged fermions should be included. The neutrino kinetic mixing leads to coupling $\overline{U_\RH} \nu_\LH \phi^*_{U_\RH}$. It further modifies  processes, e.g., (semi-)leptonic decays $U_\alpha \to U_\beta \nu\overline{\nu}$ at tree level, radiative decays $U_\alpha \to U_\beta \gamma \gamma$ at loop level and FCNC processes $U_\alpha \to U_\beta \overline{U_\gamma} U_\delta$ at loop level,  from their SM predictions. As a consequence, precision measurements of charm mesons and baryons can give strong constraints to $\epsilon^u$. A detailed discussion of these constraints is not our subject in this paper. 
Realisations of sizeable NSI textures $\bT_{12}$, $\bT_{13}$, $\bT_{32}$ and $\bT_{33}$ via UV completions of the other dimension-8 operators are also hard. Those via $\mathcal{O}^{3,5,7,8}$ gain strong constraints from $K$ and $B$ decays, and those via $\mathcal{O}^{6}$ gain constraints from $E_\alpha \to E_\beta \gamma$ decays again. Since it is hard to generate sizeable NSI for textures $\bT_{12}$, $\bT_{13}$, $\bT_{32}$, $\bT_{33}$ or their combinations, we refer them to minor NSI textures.

\section{Testing NSI textures at LBL experiments \label{sec:DUNE}} 

The long baseline experiment with the wide-band beam and sizeable matter effects is expected to measure more than one $\epsilon_{\alpha\beta}$, which implies that the flavour dependence of NSIs $\epsilon_{\alpha\beta}$ can be tested.
As a result, an experiment of this kind is possible to study the flavour symmetry model through the operators $\mathcal{O}^{1-8}$. 
In this section we will study the matter NSI effects for DUNE experiment under the flavour symmetry $A_4$ or $Z_2$.
We summarise the connection of texture parameters $\alpha_{mn}$ to the conventional parameters $\epsilon_{\alpha\beta}$ in Table~\ref{tab:prob_coeff_texture}.
%
%
Some benefits can be seen to consider matter-effect NSIs under flavour symmetries. When we assume that $A_4$ symmetry is not broken, only two types of NSIs could be seen, and they are both flavour-conserving ones. If $A_4$ is broken and the residual $Z_2$ symmetry is preserved, there is no such benefits as all textures are predicted under this symmetry, until we impose an UV complete model.
Therefore, we expect a good performance for DUNE to figure out these scenarios.
We will test the NSI textures from the $A_4$ symmetry without assuming any UV complete model in Section~\ref{sec:test_A4}. In section~\ref{sec:test_Z2}, we will study on the $Z_2$ testing, following with the discussion in Section~\ref{sec:UV_dim8_major}.
The approximation to oscillation probabilities with NSI matter effects is presented in Appendix~\ref{sec:NSIprob}; the true value used for oscillation parameters through out the simulation in this section is given in Table~\ref{tab:global_fit_parameters}.


\begin{table}[h]
\newcommand{\tabincell}[2]{\begin{tabular}{@{}#1@{}}#2\end{tabular}}
  \centering
  \begin{tabular}{|c|c|}\hline
$\tilde{\epsilon}_{ee}(\equiv \epsilon_{ee}-\epsilon_{\mu\mu})$ &  
$3\alpha_{12}/\sqrt{6}-\alpha_{13}/\sqrt{2}$
\\\hline
$\tilde{\epsilon}_{\tau\tau}(\equiv \epsilon_{\tau\tau}-\epsilon_{\mu\mu})$ & 
$-2\alpha_{13}/\sqrt{2}$
\\\hline
$\epsilon_{e\mu}$ & 
$\alpha_{21}/\sqrt{6}-\alpha_{22}/\sqrt{12}-\alpha_{23}/2
+i\left(-\alpha_{31}/\sqrt{6}+\alpha_{32}/\sqrt{12}+\alpha_{33}/2\right)$
 \\\hline
$\epsilon_{e\tau}$ & 
$\alpha_{21}/\sqrt{6}-\alpha_{22}/\sqrt{12}+\alpha_{23}/2
+i\left(\alpha_{31}/\sqrt{6}-\alpha_{32}/\sqrt{12}+\alpha_{33}/2\right)$ \\\hline
$\epsilon_{\mu\tau}$ &  
$\alpha_{21}/\sqrt{6}+2\alpha_{22}/\sqrt{12}+i\left(-\alpha_{31}/\sqrt{6}
-\alpha_{32}/\sqrt{12}\right)$ 
\\\hline
\end{tabular}
  \caption{\label{tab:prob_coeff_texture}Expressions of conventional parameters $\epsilon_{\alpha\beta}$ in terms of texture parameters $\alpha_{mn}$ according to Eqs.~(\ref{eq:sumruleI}), (\ref{eq:sumruleII}), (\ref{eq:standard_para}).}
\end{table}


The current global fit for matter-effect NSIs \cite{Gonzalez-Garcia:2013usa} includes solar, atmospherical, reactor and LBL neutrino data. With the assumption that all NSIs coming entirely from up quark or down quark to avoid NSIs at the source and the detector, 
the current global fit to standard NSI parameters $\epsilon^u_{\alpha \beta}$ and $\epsilon^d_{\alpha \beta}$ has been performed in \cite{Gonzalez-Garcia:2013usa}, respectively. 
We adopt these results to estimate the bounds for $\alpha^{u,d}_{mn}$. We only take the bound for each $\epsilon^{u,d}_{\alpha\beta}$, i.e., the results of 1-D projection. 
Furthermore, we neglect underlying corrections between any two or among more than two parameters, which are $\epsilon_{\alpha\beta}$, or mixing angles, mass-squared differences. 
Assuming Gaussian distributions, taken  $90\%$ C.L. limits from \cite{Gonzalez-Garcia:2013usa}, bounds on $\epsilon^{u,d}_{\alpha\beta}$ at $1\sigma$ are shown in Table \ref{tab:NSIbound_standard}. Since in their analysis the imaginary part is assumed to be $0$ or $\pi$, we directly translate their bounds to $\alpha_{1n}^{u,d}$ and $\alpha_{2n}^{u,d}$ by setting the imaginary $\alpha^{u,d}_{3n}=0$, and the results are shown in Table\ \ref{tab:NSIbound_texture}. NSIs with down quarks $\epsilon^{u,d}_{\alpha\beta}$ have very similar constraints as those with $\epsilon^{u,d}_{\alpha\beta}$. 
As we neglect some correlations among parameters, our results can be viewed as optimal. In Table \ref{tab:NSIbound_texture}, we see that most parameters are constrained around or below the percent level of weak interactions, except for $\alpha^{u,d}_{12}$, for which $1\sigma$ bounds are around $15\%$. 


\begin{table}[h!]
\newcommand{\tabincell}[2]{\begin{tabular}{@{}#1@{}}#2\end{tabular}}
  \centering
\begin{tabular}{|l|l|l|l|}
\hline
\multicolumn{4}{|l|}{$1\sigma$ bounds of global fit results}                                              \\ \hline
$\tilde{\epsilon}^u_{ee}$       & $[0.188, 0.376]$   & $\tilde{\epsilon}^d_{ee}$       & $[0.203, 0.384]$  \\ \hline
$\tilde{\epsilon}^u_{\tau\tau}$ & $[-0.003, 0.012]$ & $\tilde{\epsilon}^d_{\tau\tau}$ & $[-0.003, 0.012]$ \\ \hline
$\epsilon^u_{e\mu}$             & $[-0.046, 0.002]$ & $\epsilon^d_{e\mu}$             & $[-0.048,0 ]$     \\ \hline
$\epsilon^u_{e\tau}$            & $[-0.038, 0.065]$ & $\epsilon^d_{e\tau}$            & $[-0.036, 0.066]$ \\ \hline
$\epsilon^u_{\mu\tau}$          & $[-0.004, 0.003]$ & $\epsilon^d_{\mu\tau}$          & $[-0.004, 0.003]$ \\ \hline
\end{tabular}  \caption{\label{tab:NSIbound_standard}Taken from the current global fit results \cite{Gonzalez-Garcia:2013usa} for $\epsilon^u_{\alpha\beta}$ and $\epsilon^d_{\alpha\beta}$. In these results, the authors assume that off-diagonal elements $\epsilon_{\alpha\ne\beta}$ are real, consider that NSIs is only contributed by $u$ ($d$) quarks for $\epsilon^u_{\alpha\beta}$ ($\epsilon^d_{\alpha\beta}$), but do not include NSIs at the source and the detector.}
\end{table}
\begin{table}[h!]
\centering
\begin{tabular}{|l|l|l|l|}
\hline
\multicolumn{4}{|l|}{$1\sigma$ bounds by global fit results}               \\ \hline
$\alpha^u_{12}$ & $[0.089, 0.247]$  & $\alpha^d_{12}$ & $[0.099, 0.26]$    \\ \hline
$\alpha^u_{13}$ & $[-0.003, 0.007]$ & $\alpha^d_{13}$ & $[-0.003, 0.007]$  \\ \hline
$\alpha^u_{21}$ & $[-0.045, 0.049]$ & $\alpha^d_{21}$ & $[-0.045, 0.047]$  \\ \hline
$\alpha^u_{22}$ & $[-0.037, 0.03]$  & $\alpha^d_{22}$ & $[-0.035, 0.0302]$ \\ \hline
$\alpha^u_{23}$ & $[-0.019, 0.096]$ & $\alpha^d_{23}$ & $[-0.0154, 0.096]$ \\ \hline\end{tabular}
  \caption{\label{tab:NSIbound_texture} The $1\sigma$ bounds for $\alpha^u_{12}$ ($\alpha^d_{12}$), $\alpha^u_{13}$ ($\alpha^d_{13}$) and  $\alpha^u_{2i}$ ($\alpha^d_{2i}$), with fixed $\alpha^u_{3i}=0$ ($\alpha^d_{3i}=0$), by global fit results \cite{Gonzalez-Garcia:2013usa} shown in Table\ \ref {tab:NSIbound_standard}. More details can be seen in the text.}
\end{table}


The smallness of matter effect NSIs is predicted as we see in Table \ref{tab:NSIbound_standard}. Fortunately, DUNE can improve the sensitivity and is possible to detect these effects. In this section, our goal is to see if these minor features\footnote{Assuming an equal amount of NSI effects with $u$, $d$ quarks and electrons, the $1\sigma$ size of total NSI matter effects in the earth is roughly $3$ times of the $1\sigma$ region shown Table \ref{tab:NSIbound_standard}. This estimate will be applied in the following (Tables \ref{Tab:A4b_dof} and \ref{Tab:Z2_dof}) for the comparison.} appearing in DUNE can provide any extra information for the flavour symmetry. 
We firstly discuss how matter-effect NSIs $\alpha_{mn}$ affect neutrino oscillations in DUNE and then, study the physics capacity for DUNE to test $A_4$ symmetry and $Z_2$ residual symmetry via the NSI measurement. We emphasis that the results in Sections \ref{sec:test_A4} and \ref{sec:test_Z2} are in the general point of view; we consider all possible correlations by using the conventional parametrisation (3 mixing angles, 1 Dirac CP phase and 2 mass-squared differences), instead of implementing any possible flavour model for oscillation parameters. The final note is that for a given model that consistently predicts values for both oscillation and NSI parameters, we should further adopt the Wilks' theorem that the $\Delta \chi^2$ value for nested hypothesis testing asymptotically follows the $\chi^2$-distribution with the degrees of freedom that equals to the difference in the number of free parameters between two models\cite{Wilks:1938dza}. Therefore, we will further study two cases with the maximum and the minimum of the possible degrees of freedom for $\chi^2$-distribution.

\subsection{Oscillation probabilities in DUNE}

As mentioned in the introduction, matter-effect NSIs in DUNE have been widely discussed. Because of the propagation in such long distance ($1300$ km) of neutrino in the earth, nonnegligible matter density, and the GeV-energy-scale neutrino beam, matter effects play a substantial role in the oscillation. Before discussing the physics potential for understanding any flavour symmetries, we firstly study the impact of $\alpha_{mn}$ on the oscillation probability for DUNE.

The DUNE experiment consists of a neutrino source known as Long Baseline
Neutrino Facility (LBNF), a detector based at Fermilab and a LArTPC detector complex located in SURF a distance of 1300 km away. The beam design is considered based on both LBNE (reference design) and LBNF studies (optimised design). The optimisation is according to the physics capability of $\delta$ discovery. Over $1$ MW power generates large amount of $\nu_\mu$ (POT/year $\sim 10^{21}$) to $1300$ km away. On the other side, the detector configuration is planned of four 10-kiloton LArTPC detectors. LArTPC technology has a particularly strong particle identification capability as well as good energy resolution which are both crucial in providing high efficiency searches and low backgrounds. DUNE covers the 1st maximum of appearance channel ($0.5\sim5$ GeV); with the wide-band design and LArTPC technology, it allows to read the behaviour of $P(\nu_\mu\rightarrow\nu_e)$ in energy around the $1$st maximum for the appearance channel with the high precision.

\begin{figure}[h!]
\centering
\includegraphics[width=0.4\textwidth, clip, trim = 0 0 0 0]{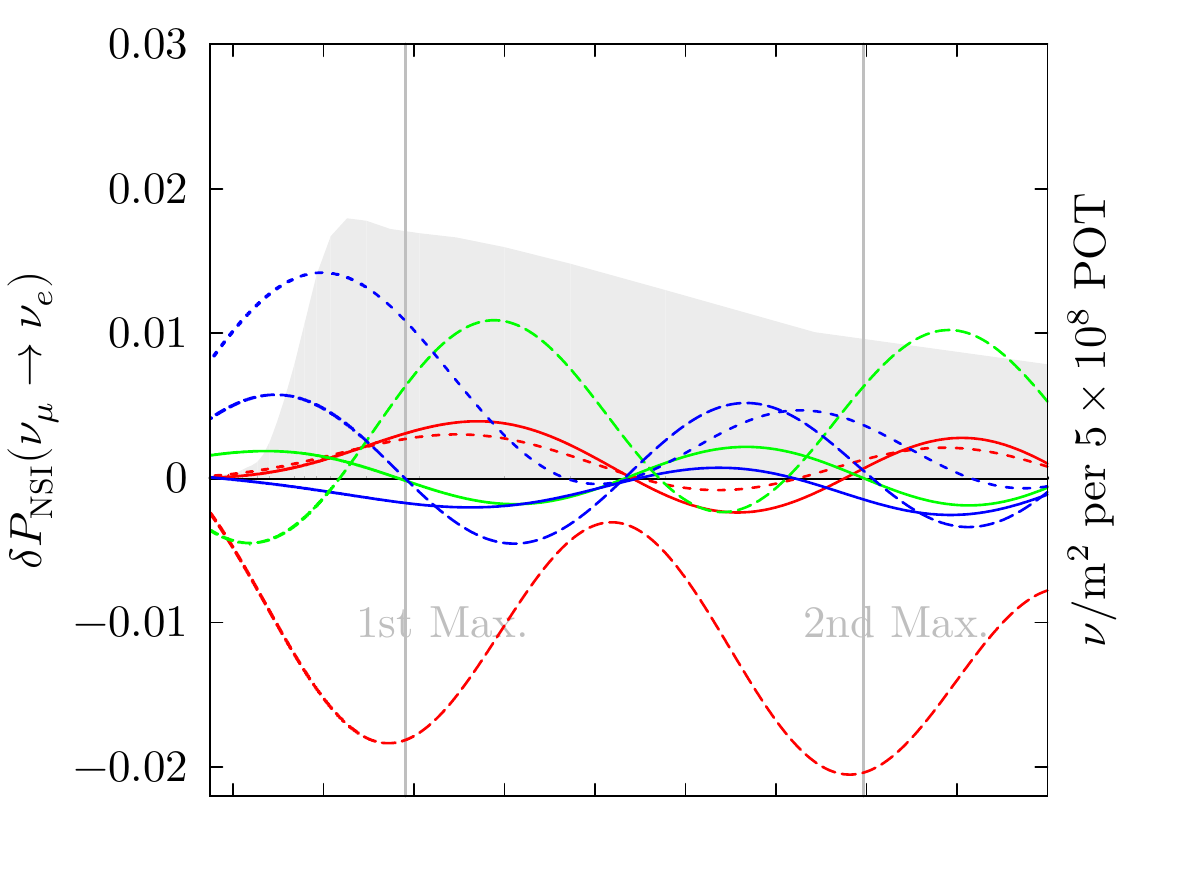}
\includegraphics[width=0.4\textwidth, clip, trim = 0 0 0 0]{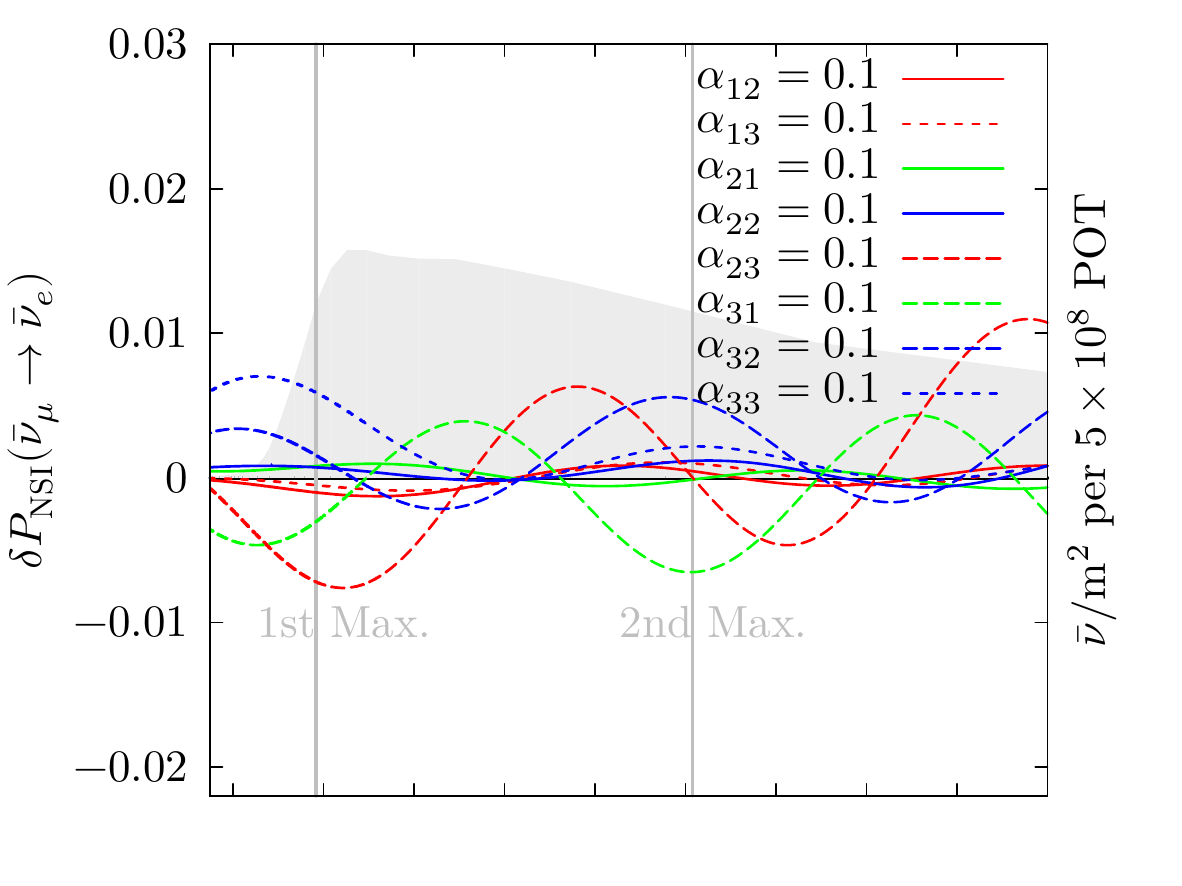}
\includegraphics[width=0.4\textwidth, clip, trim = 0 0 0 0]{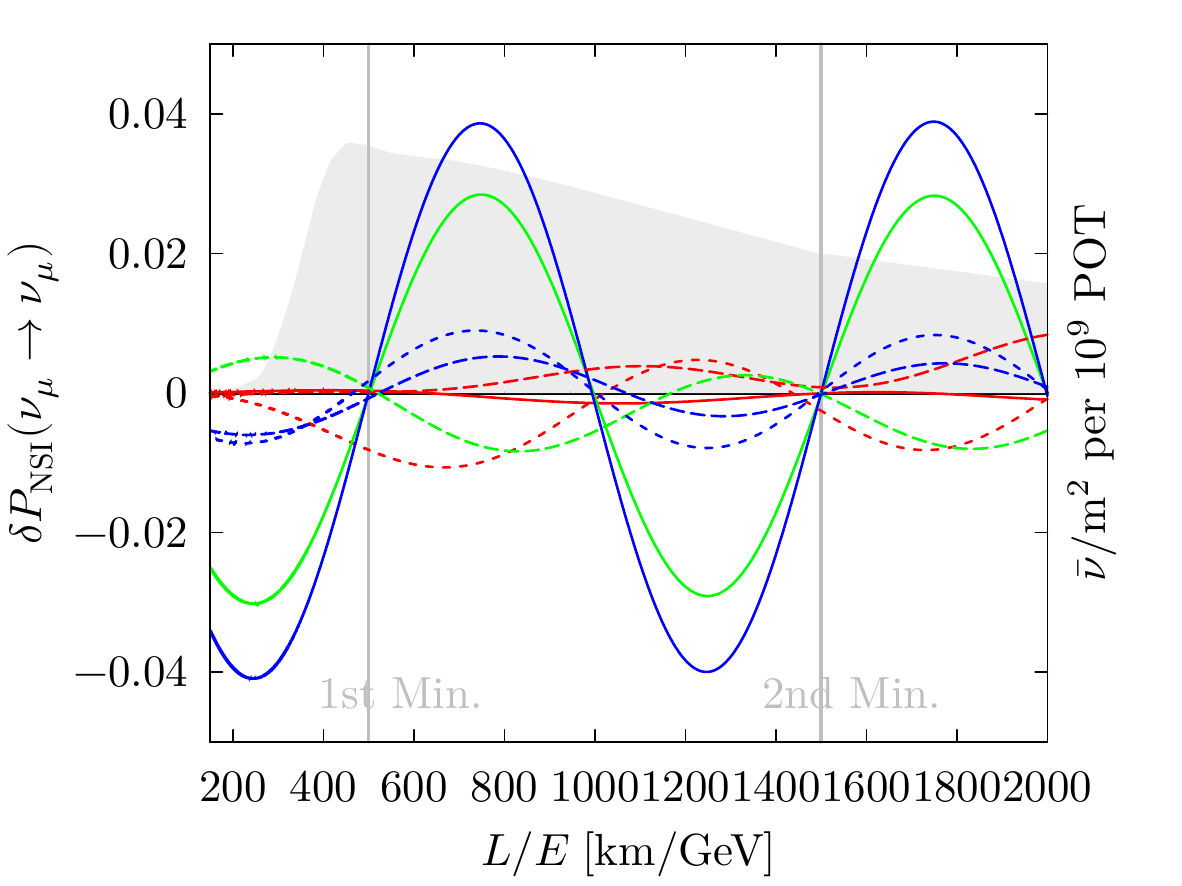}
\includegraphics[width=0.4\textwidth, clip, trim = 0 0 0 0]{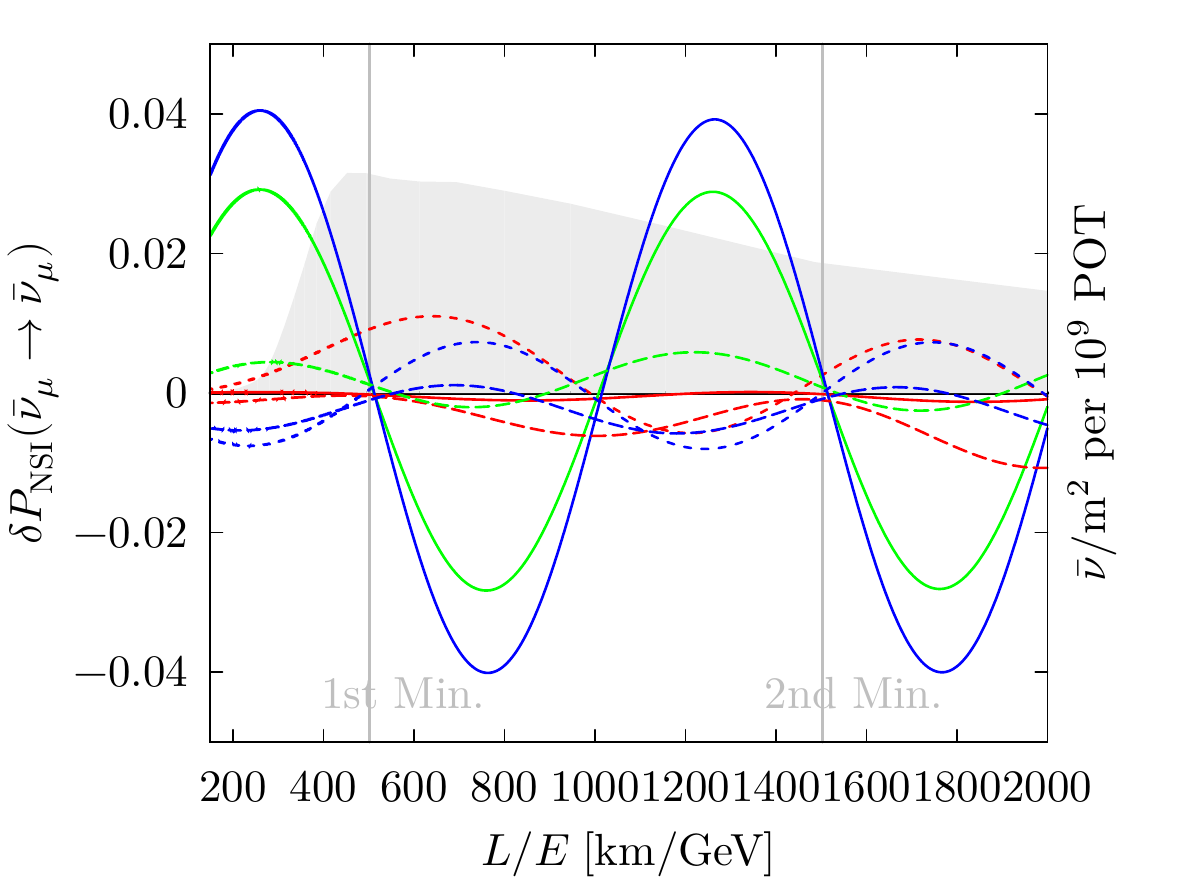}
\caption{\label{fig:dP} Oscillation probabilities  $\delta P_{\text{NSI}}(\nu_\mu\rightarrow\nu_e)$ (upper left), $\delta P_{\text{NSI}}(\bar{\nu}_\mu\rightarrow\bar{\nu}_e)$ (upper right), $\delta P_{\text{NSI}}(\nu_\mu\rightarrow\nu_\mu)$ (lower left) and $\delta P_{\text{NSI}}(\bar{\nu}_\mu\rightarrow\bar{\nu}_\mu)$ (lower right) against $L/E$ [km/GeV] for the case with one $\alpha_{mn}$, fixed at $0.1$. The oscillation parameters are used the current global fit results \cite{Esteban:2016qun} (shown in Table~\ref{tab:global_fit_parameters}) for the normal ordering with $\delta=270^\circ$, and the oscillation baseline is considered $1300$ km. In the left (right) panels, the grey shadow shows $\nu$ ($\anu$) flux of the 2-horn-optimised design for DUNE at the far detector without oscillations.}
\end{figure}

We show the difference of oscillation probabilities with one nonzero $\alpha_{mn}$ and those without NSIs, $\delta P_{\text{NSI}} (\nu_\alpha \to \nu_\beta) \equiv P (\nu_\alpha \to \nu_\beta) - P_0 (\nu_\alpha \to \nu_\beta)$ in Figure \ref{fig:dP}. The coefficient $\alpha_{mn}$ is fixed at $0.1$, but the other NSI parameters are fixed at zeros. The Dirac phase $\delta=270^\circ$ and the normal mass ordering are assumed. 



For appearance channels in the upper 2 panels of Figure \ref{fig:dP}, we see that the NSI parameters non-trivially modify the oscillation probability. NSIs modify the amplitude of oscillation probability and distort the oscillation behaviour against $L/E$. $\alpha_{23}$, $\alpha_{31}$ and $\alpha_{33}$ have larger impacts on $\delta P_{\text{NSI}}$ than the other NSI parameters, and $\delta P_{\text{NSI}}$ around the 1st maximum reaches up to or over $0.01$ for the neutrino mode. 
These impacts are slightly larger in the neutrino mode than the antineutrino mode, and this is due to our assumption of the normal mass ordering. 
DUNE with the wide-band-beam fluxes (the grey shadows) reads the variation of $\delta P_{\text{NSI}}$ around the 1st maximum. As a result, the sophisticated behaviour in the appearance channel around the 1st maximum plays a role of distinguishing different textures.

\begin{figure}[h!]
\centering
\includegraphics[width=0.4\textwidth, clip, trim = 0 0 0 0]{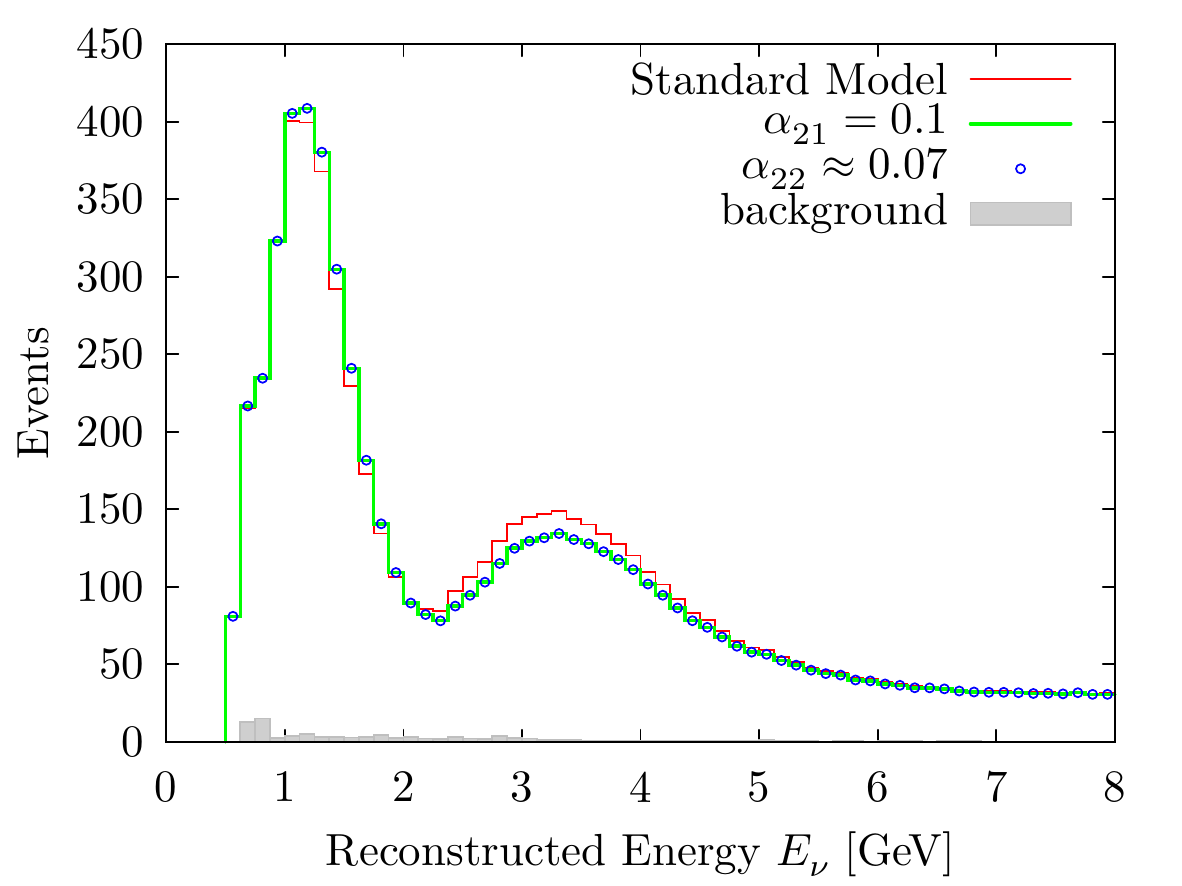}
\includegraphics[width=0.4\textwidth, clip, trim = 0 0 0 0]{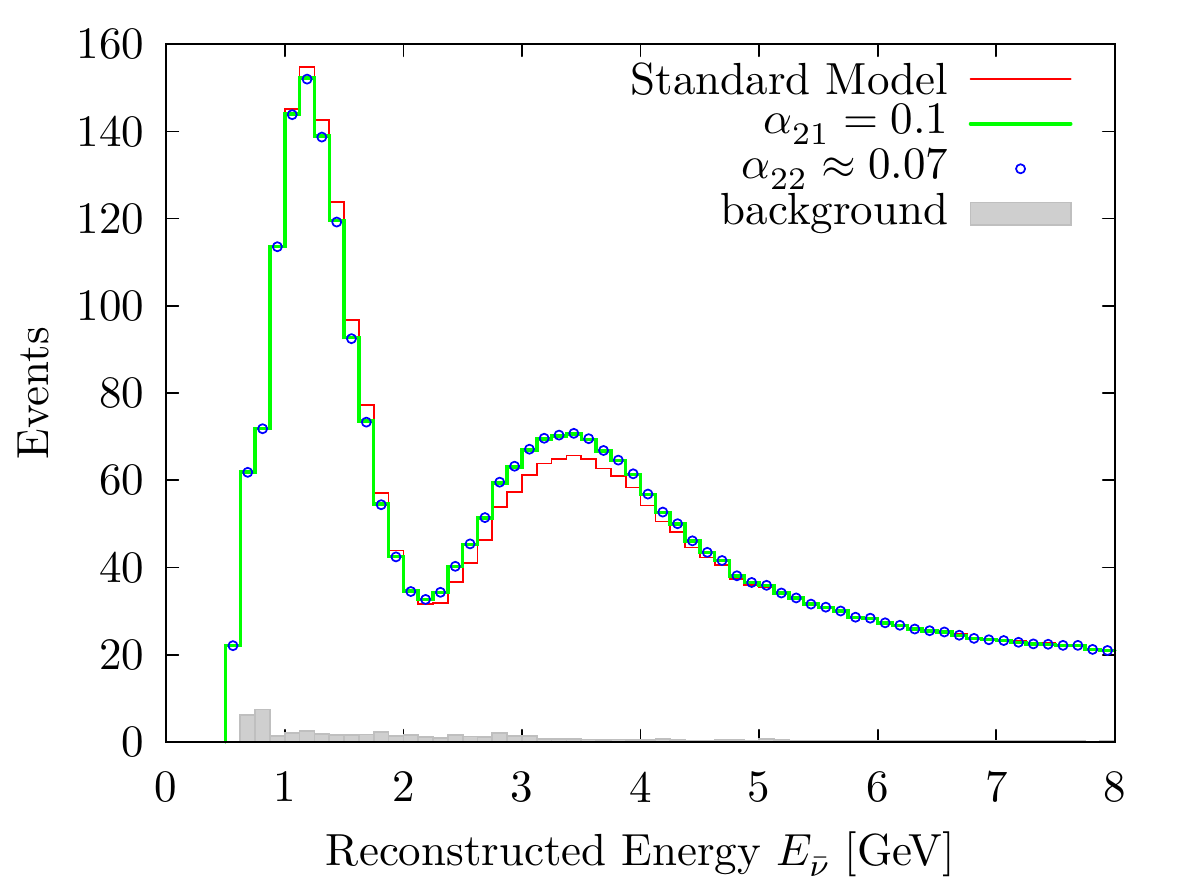}
\caption{\label{fig:spectrum}The event spectral with $\alpha_{21}=0.1$ (green line), $\alpha_{22}\approx0.07$ (blue circles), and the case without NSIs (red line). The overlapping between green and blue curves; this present the correlation between $\alpha_{21}$ and $\alpha_{22}$.}
\end{figure}

In lower 2 panels of Figure~\ref{fig:dP}, we observe the oscillation behaviour of $\delta P_{\text{NSI}}$ in $L/E$ in disappearance channels, and except for $\alpha_{13}$ it goes to $0$ at the 1st and 2nd minimums. As a result, this is clear that we would not see the NSI effects if we focus on the first minimum, where roughly peaks of DUNE fluxes are. As the grey shadows shown in this figure, the wide-band beam feature of DUNE provides more information about how much $\alpha_{mn}$ affects on the disappearance channels around the $1$st minimum.
Further, it is obvious that the disappearance channels can be sensitive to $\alpha_{21}$ and $\alpha_{22}$ as their impacts $\delta P_{\text{NSI}}$ are significantly larger than the others. 
An interesting feature is that for neutrino and antineutrino modes $\delta P_{\text{NSI}}$ behaves oppositely, i.e.\ $\delta P_{\text{NSI}}(\nu_\mu\rightarrow\nu_\mu)\cong-\delta 
P_{\text{NSI}}(\bar{\nu}_\mu\rightarrow\bar{\nu}_\mu)$. This is because $P(\nu_\mu\rightarrow\nu_\mu;\delta,A)\cong P(\bar{\nu}_\mu\rightarrow\bar{\nu}_\mu;-\delta,-A)$, and also due to the fact that the contribution of $\alpha_{mn}$ is proportional to $A$ in the leading approximation for the disappearance channel. We see this correlation in Figure \ref{fig:spectrum}, in which the event rates with $\alpha_{21}=0.1$ (green spectra), $\alpha_{22}\approx0.7$ (blue circle) and that without NSIs (red spectra) are presented in $\nu$ and $\bar{\nu}$ disappearance channels. We see the overlapping between the blue circles and the green spectra demonstrates the difficulty of distinguishing $\alpha_{21}$ and $\alpha_{22}$ in disappearance channels. 




We conclude that the wide-band-beam feature of DUNE is an advantage to detect NSI textures. Different NSI textures result in different distortions of the probabilities in the appearance channel. Therefore, we can distinguish different textures by reading out the variation of $P(\nu_\mu\rightarrow\nu_e)$ along energy. In addition, this feature helps us to measure the size of NSI effects in the disappearance channel.

\subsection{Testing ``$A_4$ symmetry'' in DUNE \label{sec:test_A4}}

Matter NSI effects predicted by $A_4$-invariant operators only allow diagonal entries. After the breaking of $A_4$ by the $Z_2$-preserving flavon VEV $\chi$, textures $\mathbb{T}_{2n}$, $\mathbb{T}_{3n}$, or their linear combinations is involved in the NSI matrix $\epsilon$. Eqs.~\eqref{Prob_disapp_NSI} and \eqref{Prob_app_NSI} indicate us that accelerator LBL experiments can be sensitive to off-diagonal terms in  $\epsilon$, because of the fact that $\epsilon_{\mu\tau}$ is the leading term in the disappearance channel, and $\epsilon_{e\mu}$, $\epsilon_{e\tau}$ are for the appearance channel. As a result, experiments of this kind can test the conservation of $A_4$ symmetry. 

Through out the study in this section, we adopt General Long Baseline Experiment Simulator (GLoBES) library \cite{Huber:2004ka,Huber:2007ji}. To simulate probabilities with matter-effect NSIs, we modify the default probability engine of GLoBES, by simply adding the matrix $A\epsilon$ in the Hamiltonian. For the simulation for DUNE, we implement the simulation package in Ref.\ \cite{Alion:2016uaj}, with run time fixed by $7$ years total (corresponding to $300$ MW$\times$kton$\times$years) and 2-horn optimised beam design with $80$ GeV protons. The other sets for oscillation parameters are described in Appendix~\ref{app:OSC}. 

We study the capacity for DUNE to rule out the ``$A_4$ symmetry'' hypothesis. 
The statistics quantity that we study is
\begin{eqnarray}\label{chi2_A4b}
\Delta\chi^2_{A_4}\equiv\left.\chi^2\right|_{\alpha_{2n}=\alpha_{3n}=0}-\chi^2_{b.f.},
\end{eqnarray}
where $\left.\chi^2\right|_{\alpha_{2n}=\alpha_{3n}=0}$ is the $\chi^2$ value with 
the assumption of $\alpha_{2n}=\alpha_{3n}=0$ ($n=1,2,3$), and $\chi^2_{b.f.}$ is the $\chi^2$ value for the best fit. 
The expression of $\chi^2$ is used
\begin{eqnarray}\label{chi2}
\chi^2=\min_{\Theta, \xi=\{\xi_s,\xi_b\}}\left[2\sum_i
\left(\eta_i(\Theta,\ \xi)-n_i+n_i\ln\frac{n_i}{\eta_i(\Theta,\xi)}\right)+p(\xi,\sigma)+\mathrm{P}(\Theta_{\mathrm{OSC}.})\right].
\end{eqnarray}
The sum in this expression is over the $i$ energy bins of the
experimental configuration, with simulated true event rates of $n_i$ and
simulated event rates $\eta_i(\Theta,\xi)$ for the hypothesis parameters
$\Theta\equiv\{\theta_{ij},\Delta m^2_{ij}, \text{NSI parameters}\}$ and systematic error parameters $\xi$. Based on different conventions or assumptions, we may adopt the different parametrisation for NSI parameters; in this subsection, we use $\alpha_{mn}$. The systematic errors of the
experiments are treated using the method of pulls, parameterized as $\xi_s$ for
the signal error and $\xi_b$ for the background error. These parameters are
given Gaussian priors which form the term
$p(\xi,\sigma)=\xi_s^2/\sigma_s^2+\xi_b^2/\sigma_b^2$, where
$\sigma=\{\sigma_s,\sigma_b\}$ are the sizes of the systematic errors given in Ref.\ \cite{Alion:2016uaj}. 
$\mathrm{P}(\Theta_{\mathrm{OSC}.})$ comprises a sum of Gaussian priors for oscillation parameters $\Theta_{\mathrm{OSC}.}$, except for $\delta$.
The central values and widths are respectively used the best fit and $1\sigma$ width of NuFit results, and are given in Tab.~\ref{tab:global_fit_parameters}.
The value of $\chi^2_{b.f.}$ is always $0$, as the best fit is exactly the true value.
In the following results, we allow $\alpha_{12}$ and $\alpha_{13}$ to be free to vary.
While varying the true value for one of $\{\alpha_{21}, \alpha_{22}, \alpha_{23}, \alpha_{31}, \alpha_{32}, \alpha_{33}\}$, we set true values of $\alpha_{12}$ and $\alpha_{13}$ to be 0. 


\begin{figure}[h!]
\centering
\includegraphics[scale=.9]{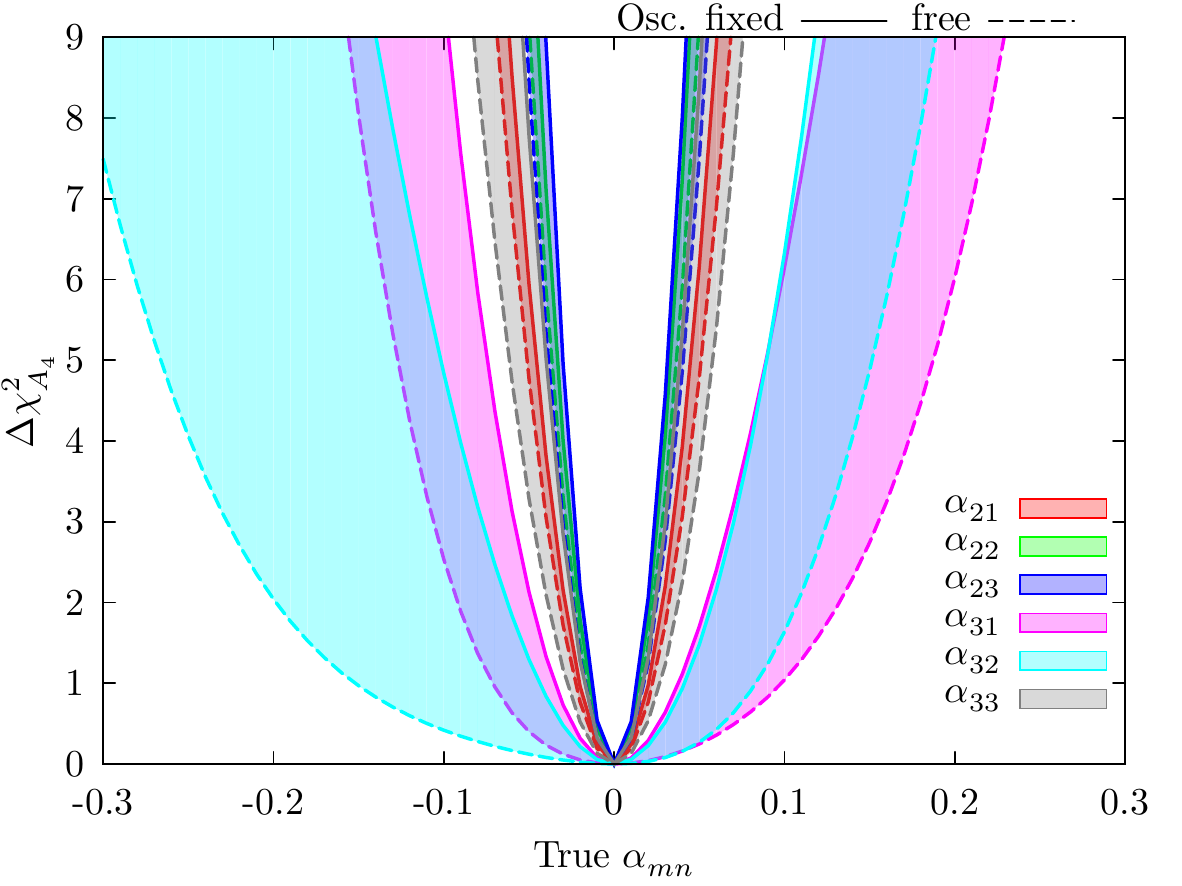}
\caption{\label{fig:A4b} $\Delta\chi^2_{A_4}$ to exclude the ``$A_4$ symmetry'' hypothesis ($\alpha_{2n}=\alpha_{3n}=0$) over the true value from $-0.3$ to $0.3$. $\alpha_{2n}$ or $\alpha_{3n}$ are forbidden under the flavour symmetry $A_4$. Normal mass ordering with $\delta=270^\circ$ is assumed. For generality, the solid (dashed) curves are presented for fixed (free) all oscillation parameters, which can been seen as the case with minimum (maximum) correlations with oscillation parameters. More details of the setting can be seen in Table \ref{A4b}. The oscillation parameters are taken from the current global fit results \cite{Esteban:2016qun} (shown in Table~\ref{tab:global_fit_parameters}).}
\end{figure}


We scan all possible true values for the targeted parameter to test the ``$A_4$ symmetry'' hypothesis, 
i.e. $\alpha_{2n}=\alpha_{3n}=0$ (for $n=1,2,3$) in Figure~\ref{fig:A4b}. 
The solid curves and dashed curves correspond to oscillation parameters fixed at their best-fit values and values varying in $1\sigma$ ranges as given in Appendix \ref{app:OSC}. 
The solid (dashed) curve can been seen as the case with minimum (maximum) correlations with oscillation parameters. 
This is for considering all possible correlation between or among parameters. For any flavour model consistent with oscillation data, the $\Delta\chi^2_{A_4}$ value is located between these two curves. 
We summarise the above setting in Appendix \ref{app:A4b_tab}.
The larger $\Delta\chi^2_{A_4}$ values are seen for $\alpha_{21}$, $\alpha_{22}$, $\alpha_{23}$ and $\alpha_{33}$. 
For the other two parameters $\alpha_{31}$ and $\alpha_{32}$, with a worse performance, a minor asymmetry feature is seen. $\alpha_{31}<0$ has the slightly higher significance than $\alpha_{31}>0$.  At $\alpha_{31}=0.1$, the exclusion level can reach $1 \leqslant \Delta\chi^2_{A_4} \leqslant 6$; however, at $\alpha_{31}=-0.1$, $\Delta\chi^2_{A_4}$ ranges from $2.5$ to $9.5$. The asymmetry is in the opposite way for $\alpha_{32}$, as $1.6 \leqslant \Delta\chi^2_{A_4} \leqslant 6.3$ ($0.4 \leqslant \Delta\chi^2_{A_4} \leqslant 4.8$) at $\alpha_{32}=0.1$ ($-0.1$).


\begin{table}[h!]
\centering
\begin{tabular}{|c|c|c|c|} 
\hline
\backslashbox[20mm]{d.o.f.}{Parameter}  
 & $\alpha_{21}$ & $\alpha_{22}$ & $\alpha_{23}$ \\ \hline
$6$ & $4.8\sigma\sim5.7\sigma$ & $4.8\sigma\sim5.5\sigma$ & $7.8\sigma\sim10.2\sigma$ \\ \hline
$12$ & $3.7\sigma\sim4.6\sigma$ & $3.7\sigma\sim4.4\sigma$ & $6.9\sigma\sim9.4\sigma$ \\ \hline
\end{tabular}
\caption{\label{Tab:A4b_dof} The averaged statistics significance to exclude $A_4$ symmetry at the $1\sigma$ bounds in Table \ref{tab:NSIbound_texture} for two possible degrees of freedom (d.o.f.), with adopting Wilks' theorem. These two cases are considered the maximum and minimum of the possible degres of freedom. The range is for all possible correlations. The maximum (minimum) number of d.o.f. corresponds to the case that 6 free oscillation parameters and 8 free NSI parameters compared to the $A_4$ symmetry preserved case with 0 (6) free oscillation parameters and 2 free NSI parameters; $|(6+8)-(0+2)|=12$ for the maximum, while for the minimum $|(6+8)-(6+2)|=6$.
}
\end{table}


To understand the statistics meaning of the result in Figure~\ref{fig:A4b}, we need to see Table \ref{Tab:A4b_dof}. 
Giving a flavour model that predicts both oscillation and NSI parameters, we should adopt Wilks' theorem. Considering the maximum and minimum of possible degrees of freedom for the $\chi^2$-distribution, in Table \ref{Tab:A4b_dof} we show the average statistical significance $N\sigma$ to exclude the $A_4$ symmetry by simply using Wilks' theorem in the case with the matter effect corresponding to the $1\sigma$ bounds in Table \ref{tab:NSIbound_texture}. 
The exclusion level for $\alpha_{23}$ is from $7\sigma$ to about $10\sigma$, while that for $\alpha_{21}$ and $\alpha_{22}$ is ranged from $\sim4\sigma$ to $\sim6\sigma$. 

We conclude this subsection that DUNE has a high potential to test textures predicted by the ``$A_4$ symmetry'' hypothesis, which predicts only diagonal entries of $\epsilon$. 


\subsection{\label{sec:test_Z2}Testing ``$Z_2$ symmetry'' in DUNE}

From the EFT point of view, combining dimension-8 operators with $Z_2$-preserving flavon VEV can predict plenty of off-diagonal NSI textures. Therefore, testing the ``$Z_2$ symmetry'' by using $Z_2$-motivated NSI textures is more complicated than testing the ``$A_4$ symmetry''. Fortunately, some of them have stronger constraints than the others if UV completions of these operators are accounted, and only $\bT_1$, $\bT_2$, $\bT_3$ and $\bT_4$ may reach the percent level, as shown in Section~\ref{sec:UV_dim8}. To simplify our discussion, we will only focus on these textures. We re-parametrise their linear combination as follows
\begin{eqnarray}\label{eq:YL_structure}
\left(
\begin{array}{c c c}
-x & x+y-z-iw\ & x+y+z+iw\\
x-z+iw & z & y-iw \\
x+z-iw & y+iw & -z 
\end{array}
\right)
\end{eqnarray}
for the phenomenological benefit, where $x\equiv \alpha_2$, 
$y\equiv -\frac{\alpha_1}{3}+\frac{2\alpha_2}{3\sqrt{2}}$, $z\equiv\frac{\alpha_3}{\sqrt{3}}$ and $w\equiv \frac{\alpha_{31}}{\sqrt{6}}$. This parametrisation piles two strong constraints $\Delta\epsilon_{\mu\tau}$ and $\Delta\tilde{\epsilon}_{\tau\tau}$ to $y$ and $z$ respectively.As we will see later, this helps us to focus on a simple but not highly excluded structure for the NSI matrix. 


\begin{table}[h!]
\centering
\begin{tabular}{|l|l|l|l|l|l|}
\hline
\multicolumn{2}{|l|}{Global Fit} & \multicolumn{2}{l|}{Global Fit} & \multicolumn{2}{l|}{DUNE sensitivity} \\ \hline
$w^u$  & \multicolumn{1}{c|}{--} & $w^d$ & \multicolumn{1}{c|}{--} & $w$        & $[-0.013, 0.025]$        \\ \hline
$x^u$  & $[-0.034, 0.013]$       & $x^d$ & $[-0.035, 0.012]$       & $x$        & $[-0.1, 0.1]$            \\ \hline
$y^u$  & $[-0.004, 0.003]$       & $y^d$ & $[-0.004, 0.003]$       & $y$        & $[-0.01, 0.01]$          \\ \hline
$z^u$  & $[-0.002, 0.005]$       & $z^d$ & $[-0.002,0.005]$        & $z$        & $[-0.007, 0.017]$        \\ \hline
\end{tabular}
  \caption{\label{tab:wxyzbound} The $1\sigma$ bounds for $x^{u,d}$, $y^{u,d}$, and $z^{u,d}$, by global fit \cite{Gonzalez-Garcia:2013usa} 
  shown in Table~\ref {tab:NSIbound_standard}, and expected $1\sigma$ bounds $w$, $x$, $y$, and $z$, for DUNE with fixed oscillation parameters, 
  assuming true values $w=x=y=z=0$. The upper-scripts $u$, $d$ denote NSIs only with $u$ and $d$ quarks, respectively.
  For both fitting, we allow the other NSI parameters to vary, except for $w$ in the fit using global fit results.  
  To avoid conflicting to the ``real $\epsilon_{\alpha\ne\beta}$'' assumption of global fit, we set $w=0$ in the second and fourth columns.}

\end{table}


Table~\ref{tab:wxyzbound} shows the $1\sigma$ constraint on $x$, $y$, $z$, $w$ in Eq.~\eqref{eq:YL_structure} translated from Table~\ref{tab:NSIbound_standard}, and predicted sensitivity for DUNE with fixed oscillation parameters, assuming $w=x=y=z=0$. For both cases, we test one parameter and allow the others to vary, except for $w$ in the fitting with global fit results. Keeping in mind that a rough factor `3' should be multiplied to $x^{u,d}$, $y^{u,d}$ and $z^{u,d}$ when comparing with $x$, $y$ and $z$, we find the precision on $x$, $y$, and $z$ for DUNE is competitive to current global fit results. Besides, DUNE is sensitive to the imaginary part $w$, which however is assumed at zero in the global fit.

We find the result in Table~\ref{tab:wxyzbound} imposes very restricting bounds for $y$ and $z$ around zeros through the elements $\tilde{\epsilon}_{\tau\tau}$ and $\epsilon_{\mu\tau}$, and the possibility of nonzero $x$.
This result motivates us the structure
\begin{eqnarray}\label{eq:YL_structure_x}
\epsilon= \left(
\begin{array}{c c c}
0 & x & x\\
x & x & 0 \\
x & 0 & x 
\end{array}
\right).
\end{eqnarray}
Two sum rules can be read out from Eq.~\eqref{eq:YL_structure_x},
\begin{eqnarray}\label{eq:sumrule_YL1}
&\epsilon_{e\mu}=\epsilon_{e\tau}=-\tilde{\epsilon}_{ee} \,, \\
\label{eq:sumrule_YL2}
&\epsilon_{\mu\tau}=\tilde{\epsilon}_{\tau\tau}=0 \,.
\end{eqnarray} 

In the following, we study the exclusion level for DUNE to exclude the matter-effect NSIs in the form of Eq.~\eqref{eq:YL_structure_x}.
The statistics quantity that we study is
\begin{eqnarray}\label{chi2_Z2}
\Delta\chi^2_{Z_2}\equiv\left.\chi^2\right|_{x}-\chi^2_{b.f.}\,,
\end{eqnarray}
where $\left.\chi^2\right|_{x}$ is the $\chi^2$ value defined in Eq.\eqref{chi2}, assuming $\epsilon$ satisfies the structure Eq.~\eqref{eq:YL_structure_x}. Thus for  $\left.\chi^2\right|_{x}$ we use $x$ for the NSI parameters, while for $\chi^2_{b.f.}$, the parametrisation $\epsilon_{\alpha\beta}$ is used. 


\begin{figure}[h!]
\centering
\includegraphics[scale=.9]{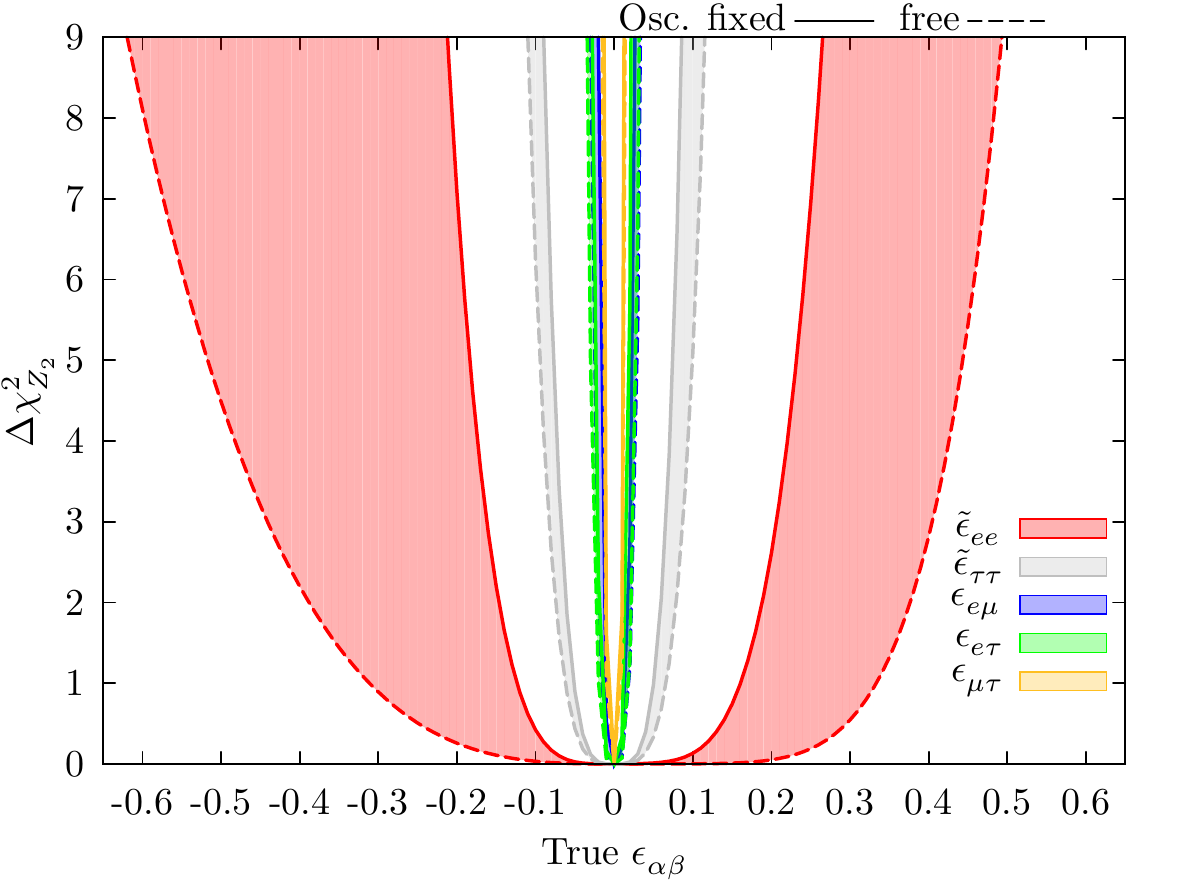}
\caption{\label{fig:SR_osc}$\Delta\chi^2_{Z_2}$ value (defined in Eq.~\eqref{chi2_Z2}) to exclude sum rules in Eqs.~\eqref{eq:sumrule_YL1} and \eqref{eq:sumrule_YL2} over true value of $-0.65<\epsilon_{\alpha\beta}<0.65$, for normal mass ordering with $\delta=270^\circ$. For generality, the solid (dashed) curves are presented for fixed (free) all oscillation parameters, which can been seen as the case with minimum (maximum) correlations with oscillation parameters. Also, we consider all possible numbers of degree of freedom; in the right panel we show how much the average statistics significance $N\sigma$ to exclude this model by simply using Wilks' approximation at the $1\sigma$ bounds in Table \ref{tab:NSIbound_standard}.}
\end{figure}

In Figure~\ref{fig:SR_osc}, we show $\Delta\chi^2_{Z_2}$ for all possible correlations from $\epsilon_{\alpha\beta}$ or $\epsilon_{\alpha\beta}=-0.65$ to $0.65$. We vary the true value of one certain $\epsilon_{\alpha\beta}$, but fix the others to be zero.
We use the same experimental setting and the same oscillation parameters values as those in Section~\ref{sec:test_A4}. For the first sum rule, in Eq.~\eqref{eq:sumrule_YL1}, within $[-0.05, \ +0.05]$, $\epsilon_{e\mu}$ and $\epsilon_{e\tau}$ can reach the significance $\Delta\chi^2_{Z_2}>10$. The performance for the $ee$ component is the worst one.
For the second sum rule, in Eq.~\eqref{eq:sumrule_YL2}, ``$\Delta\chi^2_{Z_2}<1$'' significance covers roughly $-0.05<\tilde{\epsilon}_{\tau\tau}<0.05$ and $-0.03<\epsilon_{\mu\tau}<0.03$.  

\begin{table}[h!]
\centering
\begin{tabular}{|c|c|c|c|c|c|}
\hline
\backslashbox{d.o.f.}{Parameter}  
 & $\tilde{\epsilon}_{ee}$ & $\tilde{\epsilon}_{\tau\tau}$ & $\epsilon_{e\mu}$ & $\epsilon_{e\tau}$ & $\epsilon_{\mu\tau}$ \\ \hline
$7$ & $2.2\sigma\sim4.7\sigma$ & $\sim0$ & $3.1\sigma\sim6.1\sigma$ & $5.7\sigma\sim9.4\sigma$ & $\sim0$ \\ \hline
$13$ & $1.1\sigma\sim3.7\sigma$ & $\sim0$ & $2\sigma\sim5.1\sigma$ & $4.7\sigma\sim8.6\sigma$ & $\sim0$ \\ \hline
\end{tabular}
\caption{\label{Tab:Z2_dof} The averaged statistics significance to exclude the model Eq.~\eqref{eq:YL_structure_x} for the value of $\tilde{\epsilon}_{\alpha\alpha}$ or $\epsilon_{\alpha\beta}$ corresponding to $1\sigma$ bounds in Table \ref{tab:NSIbound_standard} for two possible degrees of freedom, approximated by adopting Wilks' theorem. These two cases are considered the maximum and minimum of the possible degrees of freedom. The range is for all possible correlations. For the number of d.o.f., the maximum (minimum) is the case that 6 free oscillation parameters and 8 free NSI parameters compared to the hypothesis holding pattern Eq.\eqref{eq:YL_structure_x} for NSIs with 0 (6) free oscillation parameters and 1 free NSI parameters; $|(6+8)-(0+1)|=13$ for the maximum, while for the minimum $|(6+8)-(6+1)|=7$.}
\end{table}

As discussed in Section~\ref{sec:test_A4}, we show the statistics significance for every element of NSI matrix with two possible degrees of freedom, at the value of $\tilde{\epsilon}_{\alpha\alpha}$ and $\epsilon_{\alpha\beta}$ corresponding to $1\sigma$ bounds in Table \ref{tab:NSIbound_standard}. These two cases again are for the maximum and minimum of possible degrees of freedom. We find that for $\tau\tau$ and $\mu\tau$ elements, there is no chance to exclude this model. This is because of tight constraint for these two elements in global fit results. We see the high exclusion level for $\epsilon_{e\tau}$; it ranges from $4.7\sigma$ to $9.4\sigma$. In the following for $\epsilon_{e\mu}$, the significance is expected from $2\sigma$ to $6.1\sigma$. For the $ee$ element, we also see a high significance from $1.1\sigma$ to $4.7\sigma$.

\section{Conclusion \label{sec:conclusion}}

Non-Abelian discrete flavour symmetries, as originally proposed to explain lepton flavour mixing, may contribute to other phenomenological signatures beyond the standard case of 3-generation neutrino oscillations. 
The test of flavour symmetries have been discussed for a while in the charged lepton sector, but has not been mentioned in the neutrino sector by far. In this paper, under the assumption of an $A_4$ flavour symmetry, we investigate the constraints on matter-effect non-standard interactions (NSIs) imposed by $A_4$ and, after its breaking, those imposed by the residual symmetry $Z_2$. 
We establish connections between NSIs and flavour symmetries at two levels: the effective field theory level  and the UV completion level. 

At the effective field theory level, we impose $A_4$ symmetry to higher-dimensional operators ($d \leqslant 8$) which result in NSIs in neutrino oscillations. We only consider operators involving 4 SM fermions. We have carefully removed those operators introducing tree-level 4-charged-fermion interactions to avoid the strong constraints from the relevant flavour-violating processes. There are only one dimension-6 operator $\mathcal{O}^1=\varepsilon_{ac}\varepsilon_{bd} (\overline{L_{a\alpha}}\gamma^\mu L_{b\beta} ) (\overline{L_{c\gamma}}\gamma^\mu L_{d\delta} )$ and seven dimension-8 operators $\mathcal{O}^{2,3,4,5,6}=(\overline{\nu_{\alpha \LH}} \gamma^\mu \nu_{\beta \LH} ) (\overline{F_{\gamma}} \gamma_\mu F_{\delta})$ (for $F=U_\RH, D_\RH, E_\RH, Q, L$), $\mathcal{O}^7=(\overline{L_{\alpha}} \tilde{H} \gamma^\mu L_{b\beta} ) (\overline{Q_{b\gamma}} \gamma_\mu 
\tilde{H}^\dag Q_{\delta})$ and $\mathcal{O}^8=\varepsilon_{bc} (\overline{L_{\alpha}} \tilde{H} \gamma^\mu L_{b\beta} ) (\overline{Q_{\gamma}} 
H \gamma_\mu Q_{c\delta})$ contributing to matter-effect NSIs, shown in Table \ref{tab:operator}. Following the general approach used in flavour models, the three lepton doublets $L_1$, $L_2$ and $L_3$ are arranged as a triplet of $A_4$. For any other SM fermions, we  perform a scan of all possible representations in the flavour space. Including a flavon with a $Z_2$-preserving VEV, $A_4$ is broken to $Z_2$, and we can obtain $Z_2$-motivated NSI textures. Both $A_4$-motivated textures and $Z_2$-motivated textures have been systematically searched in this work, with the main result listed in Table \ref{tab:NSI_A4} 

Then, we consider how to realise these operators by introducing new particles in renormalisable models of $A_4$. 
The dimension-6 operator is realised by introducing electroweak singly-charged scalars as mediators. However, this case is strongly suppressed since couplings for $L_1$, $L_2$ and $L_3$ in $A_4$ are correlated with each other, and thus strong constraints from CLFV measurements cannot be avoided. Dimension-8 operators are realised by including heavy sterile neutrinos and charged scalars. The operators $\mathcal{O}^{2,3,4,5}$ involve extra fermions $F=U_\RH, D_\RH, E_\RH, Q$. By arranging $F$ as singlets of $A_4$, couplings for different generation fermions, i.e., $F_i$ and $F_j$ (for $i \neq j$), are not correlated with each other, and the constraints from CLFV or quark flavour violating processes do not apply to NSIs. Imposing $A_4$ does not give interesting observable NSI textures. After $A_4$ is broken to $Z_2$, four interesting textures $\bT_1$, $\bT_2$, $\bT_3$ and $\bT_4$, are obtained, as shown in Eq.~\eqref{eq:majorNSI}. We regard them as major textures. The main constraints to these textures are from the measurement of the non-unitary effect of the lepton mixing. Including the experimental constraints, coefficients of these textures may maximally reach $10^{-2}$ or $10^{-3}$ level. Arranging $F$ as triplets of $A_4$ gives additional NSI textures, all strongly constrained by experiments, and we refer them to minor textures.


To understand what we can do with NSI texture in the near future, we apply the $A_4$- and $Z_2$-motivated NSI textures to analyse how to test the flavour symmetry by measuring NSIs in DUNE. We consider all possible correlations and the maximum and minimum numbers of free parameters, which affect the corresponding statistics significance.
Two applications are studied. One is testing ``$A_4$ symmetry''. The off-diagonal entries of the NSI matrix are forbidden by $A_4$ symmetry, i.e., $\alpha_{21}=\alpha_{22}=\alpha_{23}=\alpha_{31}=\alpha_{32}=\alpha_{33}=0$. Excluding this hypothesis can be used to exclude the ``$A_4$ symmetry''.  We foresee that DUNE has the outstanding performance on it. 
For the case with the maximum and minimum of all possible degree of freedom for the $\chi^2$-distribution, in Table \ref{Tab:A4b_dof} we show the average statistical significance $N\sigma$ to exclude the $A_4$ symmetry by simply using Wilks' theorem in the case with the matter effect corresponding to the $1\sigma$ bounds in Table \ref{tab:NSIbound_texture}. The exclusion level for $\alpha_{23}$ is from $7\sigma$ to about $10\sigma$, while that for $\alpha_{21}$ and $\alpha_{22}$ is ranged from $\sim4\sigma$ to $\sim6\sigma$. High exclusion levels for $\alpha_{3n}$ ($n=1,2,3$) are also expected.
DUNE can constrain NSI parameters competitively with current global data. In particular, it can measure the imaginary part, labelled as $w$, with the percentage precision. We also suggest to test two sum rules of NSI parameters as shown in Eqs.~\eqref{eq:sumrule_YL1} and \eqref{eq:sumrule_YL2}. 
We show the statistics significance for excluding the model Eq.~\eqref{eq:YL_structure_x} for every elements of NSI matrix at the value corresponding to $1\sigma$ bounds in Table \ref{tab:NSIbound_standard}, in the cases with the maximum and minimum of possible degrees of freedom. We find that though for $\tau\tau$ and $\mu\tau$ elements, there is no way to exclude this model, the high exclusion level for $\epsilon_{e\tau}$; it ranges from $4.7\sigma$ to $9.4\sigma$. For $\epsilon_{e\mu}$ and $\tilde{\epsilon}_{ee}$, the significance is expected from $2\sigma$ to $6.1\sigma$ ($1.1\sigma$ to $4.7\sigma$). We now see a good performance on both applications for DUNE. 

To summarise, NSIs in neutrino oscillations have been studied in the framework of non-Abelian discrete flavour symmetries for the first time. Textures of NSIs are predicted by flavour symmetries. Measuring these textures can in principle provide a new way to test flavour symmetries and residual symmetries. It is a complimentary to the studies of flavour symmetries in standard neutrino oscillation measurements and CLFV processes. Our simulation result shows that even though matter NSI effects are predicted to be small for DUNE in general, these could provide extra informations that might extend our understanding of the flavour symmetry. And, we show how useful they are. What we raise up in this article is not only the theoretical feature of a flavour symmetry, but also the idea that we cannot wast these small but useful effects. We especially point out that if $A_4$ is conserved at the NSI level, it could be hard to see matter-effect NSIs in DUNE. This is because DUNE is less sensitive to those flavour-conserving ones. Therefore, the null result of matter-effect NSIs in DUNE could mean that `$A_4$ symmetry' is conserved at the NSI level. And this could still extend our knowledge on the symmetry of flavour at the higher energy.



\section*{Acknowledgements}
We thank S.~Pascoli for very useful discussions and improvements on the manuscript. We are also grateful to A.~Olivares-Del Campo for double checking the probability approximation, N.~W.~Prouse for sharing T2HK simulations and J.~Zhang for useful discussions. 
This work has been supported by the European Research Council under ERC Grant NuMass (FP7-IDEAS-ERC ERC-CG 617143), H2020 funded ELUSIVES ITN (H2020-MSCA-ITN-2015, GA-2015-674896-ELUSIVES), InvisiblePlus (H2020-MSCA-RISE-2015, GA-2015-690575-InvisiblesPlus) and the Science and Technology Facilities Council (STFC).
\begin{appendix}

\section{Neutrino oscillation parameters\label{app:OSC}}

In the standard case, neutrino oscillations are described by mass-squared differences $\Delta m^2_{21}$, $\Delta m^2_{31}$ and $\Delta m^2_{32}$ with $\Delta m^2_{ji} = m^2_j - m^2_i$ and the mixing matrix $U$ which is parametrised by three mixing angles $\theta_{ij}$ and a CP-violating phase $\delta$ as
\begin{eqnarray}
U\equiv 
\left(
\begin{array}{ccc}
1 & 0 & 0 \\
0 & c_{23} & s_{23}\\
0 & -s_{23} & c_{23}
\end{array}
\right)
\left(
\begin{array}{ccc}
c_{13} & 0 & s_{13}\mathrm{e}^{-i\delta} \\
0 & 1 & 0\\
-s_{13}\mathrm{e}^{i\delta} & 0 & c_{13}
\end{array}
\right)
\left(
\begin{array}{ccc}
c_{12} & s_{13} & 0 \\
-s_{13} & c_{23} & 0\\
0 & 0 & 1
\end{array}
\right),
\end{eqnarray}
where $s_{ij}=\sin\theta_{ij}$ and $c_{ij}=\cos\theta_{ij}$. Except for $\delta$, we generally adopt the last global fit results in Table \ref{tab:global_fit_parameters}, taken from \cite{Esteban:2016qun}, for the true values and the priors. For the consistency, we should assume a flavour model for both oscillation and NSI parameters. However, we do not expect that it makes a large difference since the flavour model should be allowed by global fit results. Further, as the current global result is not changed significantly after including NO$\nu$A data, of which results may have the impact of NSIs, our results do not lose predictability. Except for $\delta$, we implement priors; we assume Gaussian distribution, centred at the true value with the width taken as the $1\sigma$ bound from the last global fit results, shown in Table \ref{tab:global_fit_parameters}. For $\delta$, we do not implement a prior. 

\begin{table}[h!]
\centering
\begin{tabular}{ |c|c|c| } 
 \hline
 Parameter & Normal ordering & Inverted ordering \\
  [0.5ex] 
 \hline\hline
 $\theta_{12}$ [$^\circ$] & $33.56^{+0.77}_{-0.75}$ & $33.56^{+0.77}_{-0.75}$ \\ 
 $\theta_{13}$ [$^\circ$] & $8.46^{+0.15}_{-0.15}$ & $8.49^{+0.15}_{-0.15}$ \\
 $\theta_{23}$ [$^\circ$] & $41.6^{+1.5}_{-1.2}$ & $50.0^{+1.1}_{-1.4}$ \\
 $\Delta m_{21}^2$ [$\times10^{-5}$ eV$^2$]& $7.49^{+0.19}_{-0.17}$  & $7.49^{+0.19}_{-0.17}$ \\
 $\Delta m_{3l}^2$ [$\times10^{-3}$ eV$^2$]& $+2.524^{+0.039}_{-0.040}$ & $-2.514^{+0.038}_{-0.041}$ \\
 $\delta$ [$^\circ$]& $270$ & $270$ \\
 \hline
\end{tabular}

\caption{\label{tab:global_fit_parameters}The true values used in this work,
unless otherwise stated explicitly, with their uncertainties (the $1\sigma$
range of the priors we have used in our fit). These are based on NuFit 3.0
(2016)~\cite{Esteban:2016qun}. The definition of $\Delta m_{3l}^2$ are the same in NuFit 3.0,
for normal ordering $\Delta m^2_{3l}=\Delta m^2_{31}$, while for inverse one, $\Delta m^2_{3l}=\Delta m^2_{32}$.}

\end{table}
 
 \subsection{Parameter Setting for $A_4$ symmetry study\label{app:A4b_tab}}

In Section \ref{sec:test_A4}, we study the potential to exclude the hypothesis preserving the $A_4$ symmetry for DUNE. The setting for oscillation and NSI parameters in the simulation is summarised in Table \ref{A4b}.

\begin{table}[h!]
\centering
\begin{tabular}{|l|l|l|l|}
\hline
\backslashbox[22mm]{}{}             & Osc. Para.       & ${\alpha_{12},\ \alpha_{13}}$ & ${\alpha_{2n},\ \alpha_{3n}}$   \\ \hline
true values   & fix them at $b.f.$ & fix them at $0$                 & change one; fix the other at $0$  \\ \hline
tested values & all fixed or free    & allow them varying            & fix all at $0$                   \\ \hline
\end{tabular}
\caption{The summary of the setting for the true and tested values, used for studying $\Delta\chi^2_{A_4}$. The oscillation parameters (Osc. Para.) are fixed at the best fit ($b.f.$) of the global fit results in Tab.~\ref{tab:global_fit_parameters} for the true values. We study both scenarios with fixed and varying oscillation parameters with priors, for considering all possible correlations. The width of priors for oscillation parameters are used the size of $1\sigma$ uncertainty of global fit results in Tab.~\ref{tab:global_fit_parameters}. The flavour symmetry $A_4$ only allows $\{\alpha_{12},\ \alpha_{13}\}$, which are fixed at $0$ for true values, but allowed to freely vary for tested values. The parameters $\{\alpha_{2n}$, $\alpha_{3n}\}$ are not allowed by $A_4$. For their true values, we study each of them by changing its value from $-0.3$ to $0.3$, but fix the other at $0$. For the tested values, we fix all of them at $0$.}
\label{A4b}
\end{table}

\section{Textures of NSIs at the source and detector predicted by $A_4$ \label{app:NSI}}

In this appendix, we list the textures of NSIs at the source and detector in the framework of $A_4$ symmetry. These textures are directly dependent upon which representations the fermions are arranged in the flavour symmetry. 

NSIs at the source and detector are expressed as $3\times3$ complex matrices $\epsilon^\text{s}$ and $\epsilon^\text{d}$, respectively, contributing to superpositions of flavour states,
\begin{eqnarray} 
|\nu^{\text{s}}_\alpha\rangle = \frac{1}{n^{\text{s}}_\alpha}
\Big(|\nu_\alpha\rangle+\sum_{\beta}
\epsilon^{\text{s}}_{\alpha\beta}|\nu_\beta\rangle \Big)\,, \quad
\langle\nu^{\text{d}}_\beta| = \frac{1}{n^{\text{d}}_\beta}
\Big(\langle\nu_\beta|+\sum_{\alpha}
\epsilon^{\text{d}}_{\alpha\beta}\langle\nu_\alpha| \Big)\,, 
\label{eq:superposition}
\end{eqnarray}
where $n^{\text{s}}_\alpha=\sqrt{\sum_\beta
|\delta_{\alpha\beta}+\epsilon^{\text{s}}_{\alpha\beta}|^2}$, 
$n^{\text{d}}_\beta=\sqrt{\sum_\alpha
|\delta_{\alpha\beta}+\epsilon^{\text{d}}_{\alpha\beta}|^2}$
(for $\alpha\neq\beta\neq\gamma\neq\alpha$) are normalisation factors. 
Replacing $A$ with $-A$ and $\epsilon^\text{m,d,s}$ with $\epsilon^{\text{m,d,s}*}$, we obtain those for antineutrinos. The effective operators describing NSIs for neutrino production at the source and measured at the detector can be expressed as \begin{eqnarray}
\mathcal{L}_{\text{NSI}} = 2 \sqrt{2} G_F \sum_{p=7}^{12} c^{p}_{\alpha\beta\gamma\delta} \mathcal{O}_{\alpha\beta\gamma\delta}^{p} + \text{h.c.} \,.
\label{eq:sum_Op}
\end{eqnarray}
Given the higher-dimensional operators in Eq.~\eqref{eq:sum_O},
the relation between the NSI parameters at the source and the detector $\epsilon^{\text{s}}_{\alpha\beta}$, $\epsilon^{\text{d}}_{\alpha\beta}$ and the higher-dimensional operators is given by
\begin{eqnarray}
\epsilon^{\text{s}}_{\alpha\beta} = \sum_{p=7}^{12} n^{\text{s},p}c^{p}_{\alpha\beta11}   \,,\quad
\epsilon^{\text{d}}_{\alpha\beta} = \sum_{p=7}^{12} n^{\text{d},p}c^{p}_{\alpha\beta11} \,,
\end{eqnarray}
where $n^{\text{s},p}$ and $n^{\text{d},p}$ are order-one coefficients, related to the number densities of electron and neutron.

We only require the lepton doublets $L=(L_1,L_2,L_3)^T$ to be a triplet $\mt$ of $A_4$ ($L\mt$) for realise large mixing angles, but we do not specify representations of $A_4$ for the rest fermions. In other words, they could be any cases in the following: 
\begin{itemize}
\item
Three right-handed charged leptons $E_{1\RH}, E_{2\RH}, E_{3\RH}$ are arranged as different singlets of $A_4$ or form a triplet $\mt$. The former case is helpful for realising hierarchical charged lepton masses. Without lose of generality, we consider two cases $(E_\RH\ms)$ and $(E_\RH\mt)$ for right-handed charged leptons:
\begin{eqnarray}
&&(E_\RH\ms)\quad
E_{1\RH}\sim \ms, E_{2\RH} \sim \ms', E_{3\RH}\sim \ms''\,,\nonumber\\ 
&&(E_\RH\mt)\quad
E_\RH=(E_{1\RH},E_{2\RH},E_{3\RH})\sim \mt\,. 
\end{eqnarray}
\item
The left-handed quarks $Q_1$, $Q_2$, $Q_3$ may also be arranged as different singlets or form a triplet. We consider four cases: 
\begin{eqnarray}
&&(Q\ms)\;\,\quad Q_1\sim \ms\,,\nonumber\\ 
&&(Q\ms')\,\quad Q_1\sim \ms'\,,\nonumber\\
&&(Q\ms'')\quad Q_1\sim \ms''\,,\nonumber\\
&&(Q\mt)\;\;\quad Q=(Q_1,Q_2,Q_3)^T\sim \mt\,.
\end{eqnarray}
Since $Q_2$ and $Q_3$ do not contribute to NSIs in neutrino oscillations, we do not care about their representations.  
\item
Similarly, we will consider two cases for up-type and down-type right-handed quarks, respectively: 
\begin{eqnarray}
&&(U_\RH\ms)\,\,\quad U_{1\RH}\sim \ms\,, \hspace{4cm} (D_\RH\ms)\,\,\quad D_{1\RH}\sim \ms\,, \nonumber\\ 
&&(U_\RH\ms')\,\quad U_{1\RH}\sim \ms'\,, \hspace{4cm} \!\!(D_\RH\ms')\,\quad D_{1\RH}\sim \ms'\,, \nonumber\\
&&(U_\RH\ms'')\quad U_{1\RH}\sim \ms''\,, \hspace{4cm} \!\!\!(D_\RH\ms'')\quad D_{1\RH}\sim \ms''\,, \nonumber\\
&&(U_\RH\mt)\,\,\,\quad U_\RH=(U_{1\RH},U_{2\RH},U_{3\RH})^T\sim \mt
\,, \hspace{0.8cm} (D_\RH\mt)\,\,\,\quad D_\RH=(D_{1\RH},D_{2\RH},D_{3\RH})^T\sim \mt \,.
\end{eqnarray}
\end{itemize}
All the above possibilities are considered in this appendix. 

\subsection{$A_4$-invariant operators}

We scan all $A_4$-invariant operators $c^{7-12}_{\alpha\beta\gamma\delta}\mathcal{O}^{7-12}_{\alpha\beta\gamma\delta}$, which contribute to NSIs at the source and detector. Besides $\bT_{11}$, $\bT_{12}$, $\bT_{13}$ in Eq.~\eqref{eq:sumruleI}, we have found six more NSI textures: 
\begin{eqnarray}
&&\bT_{11}' = \left(\begin{array}{ccc}
 0 & 1 & 0 \\
 0 & 0 & 1 \\
 1 & 0 & 0 \\
\end{array}\right) \,,  \qquad
\bT_{12}' = 
\left(\begin{array}{ccc}
 0 & -1 & 0 \\
 0 & 0 & 2 \\
 -1 & 0 & 0 \\
\end{array}
\right) \,, \qquad
\bT_{13}' = 
\left(\begin{array}{ccc}
 0 & -1 & 0 \\
 0 & 0 & 0 \\
 1 & 0 & 0 \\
\end{array}
\right) \,, \nonumber\\
&&\bT_{11}'' = \left(\begin{array}{ccc}
 0 & 0 & 1 \\
 1 & 0 & 0 \\
 0 & 1 & 0 \\
\end{array}\right) \,,  \qquad
\bT_{12}'' = 
\left(\begin{array}{ccc}
 0 & 0 & -1 \\
 -1 & 0 & 0 \\
 0 & 2 & 0 \\
\end{array}
\right) \,, \qquad
\bT_{13}'' = 
\left(\begin{array}{ccc}
 0 & 0 & 1 \\
 -1 & 0 & 0 \\
 0 & 0 & 0 \\
\end{array}
\right) \,. 
\label{eq:sumruleIp}
\end{eqnarray}
The operators that may result in these correlations are listed in Table~\ref{tab:A4_NSI}. 


\begin{table}[h!]
\renewcommand\arraystretch{1}
\addtolength{\tabcolsep}{-2pt}
\begin{center}
\begin{tabular}{|l|l|L{5.2cm}|c|c|}
\hline\hline

& Representations & $A_4$-invariant operators  &NSI textures \\ \hline

\multirow{3}{*}{$\mathcal{O}^{7-9}$} & $(L\mt)$ &
$\LL_\ms \FF_\ms$ & 
$\bT_{11}$  \\\cline{2-4}

& \multirow{2}{4cm}{$(L\mt, F\mt)$} &
$\LL_{\mt_\Symm} \FF_{\mt_\Symm}$ &
$\bT_{12}$  \\\cline{3-4}

& & 
$\LL_{\mt_\AntiS} \FF_{\mt_\Symm}$ &
$\bT_{13}$  \\\hline

\multirow{11}{*}{$\mathcal{O}^{10,12}$} & \multirow{3}{4cm}{$(L\mt, E_\RH\mt, Q\mt, U_\RH\mt)$} 
& $\LE_{\ms} \QU_{\ms}$ & $\bT_{11}$ \\\cline{3-4} 
& & $\LE_{\mt_\Symm} \QU_{\mt_\Symm}$ & 
$\bT_{12}$  \\\cline{3-4} 

& & $\LE_{\mt_\AntiS} \QU_{\mt_\Symm}$ &$\bT_{13}$ \\\cline{2-4}

& \multirow{2}{5.4cm}{$(L\mt, E_\RH\mt, Q\mt, U_\RH\ms^{(\prime,\prime\prime)})$ or $(L\mt, E_\RH\mt, Q\ms^{(\prime\prime,\prime)},U_\RH\mt)$} &
 $\LE_{\mt_\Symm} \QU_\mt$ & 
$\bT_{12}^{(\prime,\prime\prime)}$ \\\cline{3-4}

& &
$\LE_{\mt_\AntiS} \QU_\mt$ & 
$\bT_{13}^{(\prime,\prime\prime)}$ \\\cline{2-4}

& \multirow{3}{7cm}{$(L\mt, E_\RH\mt,\,Q\ms,U_\RH\ms^{(\prime,\prime\prime)})$, $(L\mt, E_\RH\mt,\,Q\ms',U_\RH\ms^{\prime (\prime\prime,0)})$ or $(L\mt, E_\RH\mt,\,Q\ms'',U_\RH\ms^{\prime\prime (0,\prime)})$} &
\multirow{3}{*}{$\LE_{\ms^{(\prime\prime,\prime)}} \QU_{\ms^{(\prime,\prime\prime)}}$} & 
\multirow{3}{*}{$\bT_{11}^{(\prime,\prime\prime)}$} \\  
& & & \\
& & & \\\cline{2-4}

& \multirow{2}{*}{$(L\mt, E_\RH\ms, Q\mt, U_\RH\mt)$} &
\multirow{2}{*}{$\LE_\mt \QU_{\mt_\Symm}$} & \multirow{2}{*}{$D_1\bT_{11}$} \\
& & & \\\cline{2-4}

& \multirow{2}{7cm}{$(L\mt, E_\RH\ms, Q\ms^{(\prime\prime,\prime)}, U_\RH\mt)$ or $(L\mt, E_\RH\ms, Q\mt, U_\RH\ms^{(\prime,\prime\prime)})$} &
\multirow{2}{*}{$\LE_\mt \QU_\mt$} & \multirow{2}{*}{$D_2 \bT_{11}^{(\prime,\prime\prime)} $}  \\  
& & & \\

\hline

\multirow{2}{*}{$\mathcal{O}^{11}$} & \multicolumn{3}{c|}{\multirow{2}{*}{Results are obtained from those of $\mathcal{O}^{10,12}$ after the replacements $\overline{Q}\to\overline{D_\RH}$ and $U_\RH\to Q$. } }  \\
& \multicolumn{3}{c|}{} \\\hline\hline

\end{tabular}
\caption{\label{tab:A4_NSI}Operators preserving $A_4$ symmetry and the predicted NSI textures at the neutrino source and detector, where $F$ represents any fermion content in the SM and $\ms^0\equiv\ms$, $D_i$ are  arbitrary diagonal matrices. The notations of the representations are understood as follows. For instance, $(L\mt, E\mt, Q\ms^{(\prime,\prime\prime)},U\mt)$ means $L\sim\mt, e\sim\mt, Q\sim\ms^{(\prime,\prime\prime)},u\sim\mt$ and $D_\RH$ can take arbitrary representations of $A_4$. The textures $\bT_{1n}^{(\prime,\prime\prime)}$ are shown in Eq.~\eqref{eq:sumruleIp}. }
\end{center}
\end{table}


For $c^{7-9}_{\alpha\beta\gamma\delta}\mathcal{O}^{7-9}_{\alpha\beta\gamma\delta}$, the same discussions on $c^{2}_{\alpha\beta\gamma\delta}\mathcal{O}^{2}_{\alpha\beta\gamma\delta}$ apply to these operators. 
$c^{10-12}_{\alpha\beta\gamma\delta}\mathcal{O}^{10-12}_{\alpha\beta\gamma\delta}$ provides more textures for NSIs at the source and detector. Here we take $\mathcal{O}^{12}_{\alpha\beta\gamma\delta}$ as an example to obtain these textures in details. 
\begin{itemize}
\item If $L \sim E_\RH\sim Q\sim U_\RH \sim \mt$, the $A_4$-invariant combinations $\LE_{\mt_\Symm} \QU_{\mt_\Symm}$ and $\LE_{\mt_\AntiS} \QU_{\mt_\Symm}$ result in $\bT_{12}$ and $\bT_{13}$, respectively. 

\item If $L \sim E_\RH\sim Q\sim\mt$ and $U_{1\RH}\sim\ms'$, the $A_4$-invariant combinations $\LE_{\mt_\Symm} \QU_\mt$ and $\LE_{\mt_\AntiS} \QU_\mt$ result in $\bT_{12}'$ and $\bT_{13}'$, respectively. Replacing $U_\RH\sim\ms'$ by $U_\RH\sim\ms''$ leads to another two textures $\bT_{12}''$ and $\bT_{13}''$, respectively. 
These relations are also valid for $L \sim E_\RH\sim U_\RH\sim\mt$, $Q\sim\ms''$ and $\ms'$, respectively. 

\item If $L \sim  E_\RH \sim\mt$ and $Q_1\sim U_{1\RH} \sim \ms, \ms',\ms''$, the $A_4$-invariant combinations $\LE_\ms \QU_\ms$ result in $\bT_{11}$. If $Q_1$ and $U_{1\RH}$ belong to different singlets of $A_4$, we obtain $\bT_{11}'$ and $\bT_{11}''$ for $\overline{Q_1}U_{1\RH} \sim \ms'$ and $\ms''$, respectively. 

\item If $L \sim  Q\sim U_\RH\sim\mt$, $E_{1\RH}\sim\ms, E_{2\RH}\sim\ms', E_{3\RH}\sim\ms''$, we obtain the $A_4$-invariant combinations $\sum_i y_i(\overline{L}E_{i\RH})_{\mt} \QU_{\mt}$ and $\sum_i y_i'(\overline{L}E_{i\RH})_{\mt} \QU_{\mt_\AntiS}$, which we denote as $\LE_{\mt} \QU_{\mt}$ and $\LE_{\mt} \QU_{\mt_\AntiS}$, respectively. Here, $y_i$ and $y_i'$ are arbitrary parameters. We find for the first term
\begin{eqnarray}
c_{\alpha\beta11} = 0 ~\text{ for } \alpha\neq\beta \,.
\end{eqnarray}
Then the NSI matrix $\epsilon^{\text{s,d}}$ can be re-expressed as $D_1\bT_{11}$, where $D_1$ is an arbitrary diagonal matrix. The second operator does not contribute to NSIs.

\item If $L \sim  U_\RH \sim\mt$, $E_{1\RH}\sim\ms, E_{2\RH}\sim\ms', E_{3\RH}\sim\ms''$ and $Q \sim \ms$, the $A_4$-invariant combinations $\LE_{\mt} \QU_{\mt_\Symm}$ only result in an arbitrary diagonal matrix, just like the former item, and we express the NSI matrix $\epsilon^{\text{s,d}}$ as $D_2 \bT_{11}$, where $D_2$ is an arbitrary matrix.  
Once we change the representation of $Q$ to be $\ms^{\prime\prime (\prime)}$, the order of the three components of the triplet $\QU_{\mt_\Symm}$ will be changed, and we arrive at $D_2 \bT_{11}^{\prime (\prime\prime)}$. 

\end{itemize}
Since $\mathcal{O}^{10}_{\alpha\beta\gamma\delta}$ is only different from $\mathcal{O}^{12}_{\alpha\beta\gamma\delta}$ by the Lorentz indices, it gives the same types of correlations as the latter. $\mathcal{O}^{11}_{\alpha\beta\gamma\delta}$ has a different particle arrangement from $\mathcal{O}^{12}_{\alpha\beta\gamma\delta}$. Performing the replacements $\overline{Q}\to \overline{D_\RH}$ and $Q\to U_\RH$, all the discussions on $\mathcal{O}^{12}_{\alpha\beta\gamma\delta}$ apply to $\mathcal{O}^{11}_{\alpha\beta\gamma\delta}$. 

The textures in Eq.~\eqref{eq:sumruleIp} only appear at the neutrino source and detector and the NSI matrices $\epsilon^\text{s,d}$ may be combinations of some of $I_i$, $I_i'$ and $I_i''$, depending the choices of representations of $A_4$ to which $E_\RH$, $Q$, $U_\RH$ and $D_\RH$ belong. For instance, if $E_{1\RH}\sim \ms$, $E_{2\RH}\sim \ms'$, $E_{3\RH}\sim \ms''$, $Q\sim \mt$, $U_{1\RH}\sim \ms$, $D_{1\RH} \sim \ms$, we get a combination of NSI textures at the source and the detector as that in matter,
\begin{eqnarray}
\epsilon^\text{s,d} = \bT_{11}\alpha^\text{s,d}_{11} +\bT_{12} \alpha^\text{s,d}_{12} +\bT_{13} \alpha^\text{s,d}_{13} \,.
\end{eqnarray}
where $\alpha^\text{s,d}_{1n}$ are complex parameters. 
Changing the representation of $U_{1\RH}$ to $\ms'$, we arrive at
\begin{eqnarray}
\epsilon^\text{s,d} = \bT_{11}\alpha^\text{s,d}_{11} +\bT_{12} \alpha^\text{s,d}_{12} +\bT_{13} \alpha^\text{s,d}_{13} + \bT_{11}'\alpha^{\prime \text{s,d}}_{11} +\bT_{12} \alpha^{\prime \text{s,d}}_{12} +\bT_{13} \alpha^{\prime \text{s,d}}_{13}\,, 
\end{eqnarray}
where $\alpha^{(\prime) \text{s,d}}_{1n}$ are complex parameters. 

\subsection{$Z_2$-invariant operators}


\begin{table}[h!]
\renewcommand\arraystretch{1}
\addtolength{\tabcolsep}{-2pt}
\begin{center}
\begin{tabular}{|l|l|L{5.2cm}|C{2.5cm}|}
\hline\hline

& Representations & $Z_2$-invariant operators  &NSI textures \\ \hline

\multirow{6}{*}{$\chi\mathcal{O}^{7-9}$} & \multirow{2}{*}{$(L \mt)$} &
$\chi\LL_{\mt_\Symm} \FF_\ms$ & 
$\bT_{12}+\bT_{22}$  \\\cline{3-4}

& &
$\chi\LL_{\mt_\AntiS} \FF_\ms$ & 
$\bT_{13}+\bT_{23}$  \\\cline{2-4}

& \multirow{4}{4cm}{$(L\mt, F\mt)$} &
$\chi \big( \LL_{\mt_\Symm} \FF_{\mt_\Symm} \big)_{\mt_\Symm}$ &
$2\bT_{12}-\bT_{22}$  \\\cline{3-4}

& &
$\chi \big( \LL_{\mt_\AntiS} \FF_{\mt_\Symm} \big)_{\mt_\Symm}$ &
$2\bT_{13}-\bT_{23}$  \\\cline{3-4}

& &
$\chi \big( \LL_{\mt_\Symm} \FF_{\mt_\Symm} \big)_{\mt_\AntiS}$ &
$\bT_{32}$  \\\cline{3-4}

& & 
$\chi \big( \LL_{\mt_\AntiS} \FF_{\mt_\Symm} \big)_{\mt_\AntiS}$ &
$\bT_{33}$  \\\hline

\multirow{17}{*}{$\chi\mathcal{O}^{10,12}$} & \multirow{4}{*}{$(L\mt, E_\RH\mt, Q\mt, U_\RH\mt)$} &
$\chi \big( \LE_{\mt_\Symm} \QU_{\mt_\Symm} \big)_{\mt_\Symm}$ &
$2\bT_{12}-\bT_{22}$  \\\cline{3-4}

& &
$\chi \big( \LE_{\mt_\AntiS} \QU_{\mt_\Symm} \big)_{\mt_\Symm}$ &
$2\bT_{13}-\bT_{23}$  \\\cline{3-4}

& &
$\chi \big( \LE_{\mt_\Symm} \QU_{\mt_\Symm} \big)_{\mt_\AntiS}$ &
$\bT_{32}$  \\\cline{3-4}

& & 
$\chi \big( \LE_{\mt_\AntiS} \QU_{\mt_\Symm} \big)_{\mt_\AntiS}$ &
$\bT_{33}$  \\\cline{2-4}

& \multirow{4}{4cm}{$(L\mt, E_\RH\mt, Q\mt, U_\RH\ms^{(\prime,\prime\prime)})$ \\ or \\ $(L\mt, E_\RH\mt, Q\ms^{(\prime,\prime\prime)}, U_\RH\mt)$} &
 $\chi \big( \LE_{\mt_\Symm} \QU_\mt \big)_{\mt_\Symm}$ & 
$2\bT_{12}-\bT_{22}$ \\\cline{3-4}

& &
$\chi \big( \LE_{\mt_\AntiS} \QU_\mt \big)_{\mt_\Symm}$ & 
$2\bT_{13}-\bT_{23}$ \\\cline{3-4}

& &
$\chi \big( \LE_{\mt_\Symm} \QU_\mt \big)_{\mt_\AntiS}$ & 
$\bT_{32}$ \\\cline{3-4}

& &
$\chi \big( \LE_{\mt_\AntiS} \QU_\mt \big)_{\mt_\AntiS}$ & 
$\bT_{33}$ \\\cline{2-4}

& \multirow{4}{5.5cm}{$(L\mt, E_\RH\mt,\,Q\ms,U_\RH\ms^{(\prime,\prime\prime)})$, $(L\mt, E_\RH\mt,\,Q\ms',U_\RH\ms^{\prime (\prime\prime,0)})$ or $(L\mt, E_\RH\mt,\,Q\ms'',U_\RH\ms^{\prime\prime (0,\prime)})$} &
\multirow{2}{*}{$\chi \big( \LE_{\mt_\Symm} \QU_{\ms,\ms',\ms''} \big)_\mt$} & 
\multirow{2}{*}{$\bT_{12}+\bT_{22}$} \\
& & & \\\cline{3-4}  
& & 
\multirow{2}{*}{$\chi \big( \LE_{\mt_\Symm} \QU_{\ms,\ms',\ms''} \big)_\mt$} & 
\multirow{2}{*}{$\bT_{13}+\bT_{23}$} \\ 
& & & \\\cline{2-4}

& \multirow{3}{4cm}{$(L\mt, E_\RH\ms, Q\mt, U_\RH\mt)$} &
$\chi \LE_\mt \QU_{\ms}$ & 
$D_3 \bT_1' \bT_{11}$ \\\cline{3-4}

& &
$\chi \big(\LE_\mt \QU_{\mt_\Symm}\big)_{\mt_\Symm}$ & 
$D_4 \bT_1' \bT_{12}$ \\\cline{3-4}

& &
$\chi \big(\LE_\mt \QU_{\mt_\Symm}\big)_{\mt_\AntiS}$ & 
$D_5 \bT_1' \bT_{13}$ \\\cline{2-4}

& \multirow{2}{4.4cm}{$(L\mt, E_\RH\ms, Q\ms^{(\prime,\prime\prime)}, U_\RH\mt)$ or $(L\mt, E_\RH\ms, Q\mt, U_\RH\ms^{(\prime,\prime\prime)})$} &
$\chi \big(\LE_\mt \QU_\mt \big)_{\mt_\Symm}$ & $D_6 \bT_1' \bT_{12}$ \\\cline{3-4}  
& &
$\chi \big(\LE_\mt \QU_\mt \big)_{\mt_\AntiS}$ & $D_7 \bT_1' \bT_{13}$ \\\cline{2-4}

\hline

\multirow{2}{*}{$\chi\mathcal{O}^{11}$} & \multicolumn{3}{c|}{\multirow{2}{*}{Results are obtained from those of $\chi\mathcal{O}^{10,12}$ after the replacements $\overline{Q}\to\overline{D_\RH}$ and $U_\RH\to Q$. } }  \\
& \multicolumn{3}{c|}{} \\\hline\hline

\end{tabular}
\caption{\label{tab:Z2_NSI}Operators preserving the residual symmetry $Z_2$, $Z_2\subset A_4$, and the resulted NSI textures at the neutrino source and detector, where $F$ represents any fermion content in the SM. The NSI parameter correlations $\bT_{2n}$ and $\bT_{3n}$ are shown in Eq.~\eqref{eq:sumruleII}. $D_i$ are arbitrary diagonal matrices. }
\end{center}
\end{table}


Once the operators $\mathcal{O}^{7-12}$ couple to the flavon VEV, $\chi =(1,1,1)^T v_\chi$, new NSI textures at the source and detector are predicted, as summarized in Table~\ref{tab:Z2_NSI}. $\chi_{\alpha'} \mathcal{O}^{7-9}_{\alpha\beta\gamma\delta}$ give rise to the same textures as in Eq.~\eqref{eq:sumruleII}. For $\chi_{\alpha'} \mathcal{O}^{10-12}_{\alpha\beta\gamma\delta}$, we follow the same procedure as that in the last section,  
taking  $\chi_{\alpha'} \mathcal{O}^{12}_{\alpha\beta\gamma\delta}$ as an example: 
\begin{itemize}
\item If $L \sim E_\RH\sim Q\sim U_\RH \sim \mt$, the $Z_2$-invariant operators $\chi \big( \LE_{\mt_\Symm} \QU_{\mt_\Symm} \big)_{\mt_\Symm}$, $\chi \big( \LE_{\mt_\AntiS} \QU_{\mt_\Symm} \big)_{\mt_\Symm}$, $\chi \big( \LE_{\mt_\Symm} \QU_{\mt_\Symm} \big)_{\mt_\AntiS}$ and $\chi \big( \LE_{\mt_\AntiS} \QU_{\mt_\Symm} \big)_{\mt_\AntiS}$
 result in the textures $3\bT_{12}-\bT_{22}$, $3\bT_{13}+\bT_{23}$, $\bT_{32}$ and $\bT_{33}$, respectively. Changing the representations to $L \sim E_\RH\sim Q\sim\mt$ and $U_{1\RH}\sim\ms^{(\prime,\prime\prime)}$, or $L \sim E_\RH\sim U_\RH \sim\mt$ and $Q_{1}\sim\ms^{(\prime,\prime\prime)}$, we arrive at the same textures. 

\item If $L \sim  E_\RH \sim\mt$ and $Q_1,U_{1\RH} \sim \ms, \ms',\ms''$, the $Z_2$-invariant combinations $\chi \big(\LE_{\mt_\Symm} \QU_{\ms,\ms',\ms''} \big)_{\mt}$, $\chi \big(\LE_{\mt_\AntiS} \QU_{\ms,\ms',\ms''} \big)_{\mt}$ result in $\bT_{21}$ and $\bT_{22}$, respectively. 

\item If $L \sim Q \sim  U_\RH \sim\mt$, $E_{1\RH}\sim\ms, E_{2\RH}\sim\ms', E_{3\RH}\sim\ms''$, the operator $\sum_i y_i'' \chi (\overline{L}E_{i\RH})_{\mt} \QU_{\ms}$ requires 
\begin{eqnarray}
c_{ee11} = c_{e\mu11} = c_{e\tau11} \,,~
c_{\mu e11} = c_{\mu \mu11} = c_{\mu\tau11} \,,~
c_{\tau e11} = c_{\tau \mu11} = c_{\tau\tau11} \,,
\end{eqnarray}
where there is no correlation between $c_{\alpha \beta 11}$ and $c_{\alpha' \beta' 11}$ once $\alpha\neq\alpha'$. 
It gives rise to the NSI texture 
\begin{eqnarray}
\left(
\begin{array}{ccc}
 y_1'' & y_1'' & y_1'' \\
 y_2'' & y_2'' & y_2'' \\
 y_3'' & y_3'' & y_3'' \\
\end{array}
\right)=\left(
\begin{array}{ccc}
 y_1'' & 0 & 0 \\
 0 & y_2'' & 0 \\
 0 & 0 & y_3'' \\
\end{array}
\right) \bT_1' \bT_{11}\,,
\end{eqnarray}
where 
\begin{eqnarray}
\bT_1'=\left(
\begin{array}{ccc}
 1 & 1 & 1 \\
 1 & 1 & 1 \\
 1 & 1 & 1 \\
\end{array}
\right) \,.
\end{eqnarray}
$\chi \big( \LE_{\mt} \QU_{\mt_\Symm} \big)_{\mt_\Symm}$ and $\chi \big( \LE_{\mt} \QU_{\mt_\Symm} \big)_{\mt_\AntiS}$ lead to
\begin{eqnarray}
&c_{ee11} = -2c_{e\mu11} = -2c_{e\tau11} \,,~
c_{\mu e11} = -2c_{\mu \mu11} = -2c_{\mu\tau11} \,,~
c_{\tau e11} = -2c_{\tau \mu11} = -2c_{\tau\tau11} \,;\nonumber\\
&c_{e\mu11} = -c_{e\tau11} \,,\qquad
c_{\mu \mu11} = -c_{\mu\tau11} \,,\qquad
c_{\tau \mu11} = -c_{\tau\tau11} \,,
\end{eqnarray}
respectively, also no correlation between $c_{\alpha \beta 11}$ and $c_{\alpha' \beta' 11}$ for $\alpha\neq\alpha'$ in each case. From these two operators, we obtain the NSI textures
\begin{eqnarray}
D_4 \bT_1'\bT_{12}\,,~  D_5 \bT_1'\bT_{13} \,,
\end{eqnarray}
respectively, where $D_i$ are independently arbitrary diagonal matrices. Replacing the representation of $Q$ to be any singlet $\ms$, $\ms'$ or $\ms''$, we obtain the $Z_2$-invariant operators $\chi \big( \LE_{\mt} \QU_{\mt} \big)_{\mt_\Symm}$ and $\chi \big( \LE_{\mt} \QU_{\mt} \big)_{\mt_\AntiS}$, which give the similar textures $D_6 \bT_1' \bT_{12}$ and $D_7 \bT_1' \bT_{13}$, respectively, with $D_6$ and $D_7$ being arbitrary diagonal matrices. 

\end{itemize}

\section{Mathematical properties of $\bT_i$ \label{sec:math}}
The textures $\bT_i$ satisfy the following interesting mathematical properties. They are helpful for our discussion in Section \ref{sec:UV}. 
\begin{itemize}
\item $\bT_i$ (for $i=1,2,3,4$) form the following ``closed'' algebras, 
\begin{eqnarray}
&\bT_i^2 = \bT_1\,, \quad \bT_1 \bT_i = \bT_i \,,\quad
\bT_2 \bT_3 = -i \bT_4\,, \quad
\bT_2 \bT_4 = i \bT_3\,,\quad 
\bT_3 \bT_4 = -i \bT_2\,. 
\label{eq:algebra}
\end{eqnarray}

\item Given two $3\times3$ coupling matrices or mass matrices $M_1=\alpha_0 \bI + \sum_{i=1}^4 \alpha_i \bT_i$ and $M_2=\beta_0 \bI + \sum_{i=1}^4 \beta_i \bT_i$,
their product $M_1 M_2$ is a linear combination of $\bI$ and $\bT_i$,
\begin{eqnarray} 
M_1 M_2 = \alpha_0\beta_0 \bI &+& (\alpha_0\beta_1+\alpha_1\beta_0+\alpha_1\beta_1+\alpha_2\beta_2+\alpha_3\beta_3+\alpha_4\beta_4) \bT_1 \nonumber\\ 
&+& ( \alpha_0 \beta_2 + \alpha_2 \beta_0 + \alpha_1 \beta_2 + \alpha_2 \beta_1 + i\alpha_4 \beta_3 -i \alpha_3 \beta_4) \bT_2  \nonumber\\ 
&+& ( \alpha_0 \beta_3 + \alpha_3 \beta_0 + \alpha_1 \beta_3 + \alpha_3 \beta_1 + i\alpha_2 \beta_4 - i\alpha_4 \beta_2) \bT_3  \nonumber\\ 
&+& ( \alpha_0 \beta_4 + \alpha_4 \beta_0 + \alpha_1 \beta_4 + \alpha_4 \beta_1 + i\alpha_3 \beta_2 - i\alpha_2 \beta_3) \bT_4 \,.
\label{eq:AB}
\end{eqnarray}
\item If $M_1$ is reversible, the inverse matrix $M_1^{-1}$
\begin{eqnarray} 
M_1^{-1} &=& \frac{\alpha_0}{\det A}\left[\frac{\det A}{\alpha_0^2} \, \bI + \left(\alpha_0 + \alpha_1  - \frac{\det A}{\alpha_0^2} \right) \bT_1 - \alpha_2 \bT_2 - \alpha_3 \bT_3 - \alpha_4 \bT_4 \right] \,,
\label{eq:invA}
\end{eqnarray}
where $\det M_1= \alpha_0 (\alpha_0^2 + 2 \alpha_0 \alpha_1 + \alpha_1^2 - \alpha_2^2 - \alpha_3^2 - \alpha_4^2)$, is also a linear combination of $\bI$ and $\bT_i$. 
\end{itemize}
By setting some of $\alpha_i$ or $\beta_i$ to zero, the following corollaries are obtained: 
\begin{itemize}

\item $\bI$ and $\bT_1$ form a closed algebra, if $M_1$, $M_2$ are linear combinations of $\bI$ and $\bT_1$, their product and inverse matrices (if reversible) are also linear combinations of $\bI$ and $\bT_1$. 

\item $\bI$, $\bT_1$ and $\bT_2$ form a closed algebra, if $M_1$, $M_2$ are linear combinations of $\bI$, $\bT_1$ and $\bT_2$, their product and inverse matrices (if reversible) are also linear combinations of $\bI$, $\bT_1$ and $\bT_2$. 

\item $\bI$, $\bT_1$ and $\bT_3$ form a closed algebra, if $M_1$, $M_2$ are linear combinations of $\bI$, $\bT_1$ and $\bT_2$, their product and inverse matrices (if reversible) are also linear combinations of $\bI$, $\bT_1$ and $\bT_3$. 
\end{itemize}

\section{Oscillation probabilities with matter-effect NSIs} \label{sec:NSIprob} 

To understand the impact of $\alpha^\text{m}_{mn}$ (in the following, we simply use $\alpha_{mn}$) on neutrino oscillation probabilities, we are based on the knowledge of the probabilities with non-zero $\epsilon^\text{m}_{\alpha\beta}$ (in the following, we simply use $\epsilon_{\alpha\beta}$). Therefore we firstly study the probability including the NSI matter effects in terms of $\epsilon_{\alpha\beta}$, and then by using 
the relations between two parameter sets Table\ \ref{tab:prob_coeff_texture}, we can extend our understanding on how flavour symmetry model realises at oscillation probability through matter-effect NSI. 

Assuming $\sqrt{\frac{\Delta m^2_{21}}{\Delta m^2_{31}}}\sim\sqrt{|\epsilon_{\alpha\beta}|}\sim s_{13}$  as the 1st order perturbation terms $\xi$, we expand the disappearance oscillation probability $P(\nu_\mu\rightarrow \nu_\mu)$ and appearance oscillation probability $P(\nu_\mu\rightarrow \nu_e)$. These equations are given with the leading-ordering coefficient for each $\epsilon_{\alpha\beta}$ to understand how each elements affect to the probability at the leading order\footnote{Our result is consistent with those in Ref.~\cite{Kikuchi:2008vq}.}, 
\begin{eqnarray}\label{Prob_disapp_NSI} 
P({\nu_\mu\rightarrow\nu_\mu})&=&P_{0}({\nu_\mu\rightarrow\nu_\mu})
+\delta P_{\text{NSI}}({\nu_\mu\rightarrow\nu_\mu}) \nonumber\\
&\approx & P_{0}({\nu_\mu\rightarrow\nu_\mu}) \nonumber\\
&&-A\epsilon_{\mu\tau}\cos\phi_{\mu\tau}\left(\sin^32\theta_{23}\frac{L}{2E}
\sin 2 \Delta_{31}L+4\sin2\theta_{23}\cos^22\theta_{23} \frac{1}{\Delta 
m^2_{31}}\sin^2\Delta_{31}L\right) \nonumber\\
&&-A\tilde{\epsilon}_{\tau\tau}c^2_{23}s^2_{23}(c^2_{23}-s^2_{23})\left(\frac{L}{8E}
\sin 2 \Delta_{31}L-\frac{1}{\Delta m^2_{31}}\sin^2 \Delta_{31}L\right) \nonumber\\
&&+\mathcal{C}^{1}_{\mu\rightarrow e; e\mu}  |\epsilon_{e\mu}|
+\mathcal{C}^{1}_{\mu\rightarrow e; e\tau}  |\epsilon_{e\tau}|
+\mathcal{C}^{2}_{\mu\rightarrow e; ee}  \tilde{\epsilon}_{ee},
\end{eqnarray}
\begin{eqnarray}\label{Prob_app_NSI} 
P({\nu_\mu\rightarrow\nu_e})&=&P_{0}({\nu_\mu\rightarrow\nu_e})
+\delta P_{\text{NSI}}({\nu_\mu\rightarrow\nu_e}) \nonumber\\
&\approx& P_{0}({\nu_\mu\rightarrow\nu_e}) \nonumber\\
&& +8s_{13}|\epsilon_{e\mu}| s_{23}\frac{ \Delta m^2_{31}}{\Delta m^2_{31}-A}\sin\Delta_{31}^AL \nonumber \\ 
&& \times
\left(
s^2_{23}\frac{A}{\Delta m^2_{31}-A}\cos\left(\delta+\phi_{e\mu}\right)
\sin\Delta_{31}^AL
+
c^2_{23}\sin\frac{AL}{4E}\cos\left(\delta+\phi_{e\mu}-\Delta_{31}L \right)
\right) \nonumber\\
&&+8s_{13}|\epsilon_{e\tau}|c_{23} s^2_{23} \frac{\Delta m_{31}^2}{\Delta m^2_{31}
-A}\sin\Delta_{31}^AL \nonumber\\ 
&&\times \left(
\frac{A}{\Delta m^2_{31}-A}\cos\left(\delta+\phi_{e\tau}\right)\sin\Delta_{31}^AL
-\sin\frac{AL}{4E}\cos\left(\delta+\phi_{e\tau}-\Delta_{31}L\right)
\right) \nonumber\\
&& +\mathcal{C}^{2}_{\mu\rightarrow e; \mu\tau} |\epsilon_{\mu\tau}|
+\mathcal{C}^{2}_{\mu\rightarrow e; ee}  \tilde{\epsilon}_{ee}
+\mathcal{C}^{2}_{\mu\rightarrow e; \tau\tau} \tilde{\epsilon}_{\tau\tau},
\end{eqnarray}
where $P_{0}(\nu_\alpha\rightarrow \nu_\beta)$ is the transition probability for $\nu_\alpha\rightarrow\nu_\beta$ without NSI matter effects, $\Delta_{31} \equiv \frac{\Delta m^2_{31}}{4E}$, $\Delta_{31}^A\equiv\frac{\Delta m^2_{31}-A}{4E}$. Here, for the coefficient $\mathcal{C}^{\text{order}}_{\text{channel};\ \text{element}}$, the upper index gives the order of this coefficient, and the lower one gives the channel and the element.

\begin{table}[h]
\newcommand{\tabincell}[2]{\begin{tabular}{@{}#1@{}}#2\end{tabular}}
  \centering
  \begin{tabular}{|c|c|c|}\hline
channel & $\nu_\mu\rightarrow\nu_\mu$ 
& $\nu_\mu\rightarrow\nu_e$  \\\hline \hline
$\tilde{\epsilon}_{ee}$& $\mathcal{C}^{2}_{\mu\rightarrow \mu; ee}$ &
$\mathcal{C}^{2}_{\mu\rightarrow e; ee}$ \\\hline
$\tilde{\epsilon}_{\tau\tau}$& $\mathcal{C}^{0}_{\mu\rightarrow \mu; \tau\tau}$  & 
$\mathcal{C}^{2}_{\mu\rightarrow e; \tau\tau}$ \\\hline
$\epsilon_{e\mu}$& $\mathcal{C}^{1}_{\mu\rightarrow \mu; e\mu}$ & 
$\mathcal{C}^{1}_{\mu\rightarrow e; e\mu}$\\\hline
$\epsilon_{e\tau}$& $\mathcal{C}^{1}_{\mu\rightarrow \mu; e\tau}$  & 
$\mathcal{C}^{1}_{\mu\rightarrow e; e\tau}$ \\\hline
$\epsilon_{\mu\tau}$& $\mathcal{C}^{0}_{\mu\rightarrow \mu; \mu\tau}$ & 
$\mathcal{C}^{2}_{\mu\rightarrow e; \mu\tau}$\\\hline \hline
$\alpha_{12}$&  $\mathcal{C}^{2}_{\mu\rightarrow \mu; ee}$ & 
$\mathcal{C}^{2}_{\mu\rightarrow e; ee}$\\\hline
$\alpha_{13}$&  $-\sqrt{2}\mathcal{C}^0_{\mu\rightarrow\mu,\tau\tau}$ & 
$\frac{1}{2}\mathcal{C}^{2}_{\mu\rightarrow e; ee}
-\sqrt{2}\mathcal{C}^{2}_{\mu\rightarrow e; \tau\tau}$
\\\hline
$\alpha_{21}$&  $\frac{1}{\sqrt{6}}\mathcal{C}^{0}_{\mu\rightarrow \mu; \mu\tau}$ & 
$\frac{1}{\sqrt{6}}\mathcal{RC}^{1}_{\mu\rightarrow e; e\mu}
+\frac{1}{\sqrt{6}}\mathcal{RC}^{1}_{\mu\rightarrow e; e\tau}$
\\\hline
$\alpha_{22}$& $\frac{1}{\sqrt{3}}\mathcal{C}^{0}_{\mu\rightarrow \mu; \mu\tau}$& 
$\frac{1}{\sqrt{12}}\mathcal{RC}^{1}_{\mu\rightarrow e; e\mu}
+\frac{1}{\sqrt{12}}\mathcal{RC}^{1}_{\mu\rightarrow e; e\tau}$
\\\hline
$\alpha_{23}$& $-\frac{1}{2}\mathcal{RC}^1_{\mu\rightarrow\mu,e\mu}
+\frac{1}{2}\mathcal{RC}^1_{\mu\rightarrow\mu,e\tau}$& 
$-\frac{1}{2}\mathcal{RC}^{1}_{\mu\rightarrow e; e\mu}
+\frac{1}{2}\mathcal{RC}^{1}_{\mu\rightarrow e; e\tau}
$
\\\hline
$\alpha_{31}$& $-\frac{1}{\sqrt{6}}\mathcal{IC}^1_{\mu\rightarrow\mu,e\mu}
+\frac{1}{\sqrt{6}}\mathcal{IC}^1_{\mu\rightarrow\mu,e\tau}$& 
$-\frac{1}{\sqrt{6}}\mathcal{IC}^{1}_{\mu\rightarrow e; e\mu}
+\frac{1}{\sqrt{6}}\mathcal{IC}^{1}_{\mu\rightarrow e; e\tau}
$
\\\hline
$\alpha_{32}$& $\frac{1}{\sqrt{12}}\mathcal{IC}^1_{\mu\rightarrow\mu,e\mu}
-\frac{1}{\sqrt{12}}\mathcal{IC}^1_{\mu\rightarrow\mu,e\tau}$&  
$\frac{1}{\sqrt{12}}\mathcal{IC}^{1}_{\mu\rightarrow e; e\mu}
-\frac{1}{\sqrt{12}}\mathcal{IC}^{1}_{\mu\rightarrow e; e\tau}
$
\\\hline
$\alpha_{33}$& $\frac{1}{2}\mathcal{IC}^1_{\mu\rightarrow\mu,e\mu}
+\frac{1}{2}\mathcal{IC}^1_{\mu\rightarrow\mu,e\tau}$& 
$\frac{1}{2}\mathcal{IC}^{1}_{\mu\rightarrow e; e\mu}
+\frac{1}{2}\mathcal{IC}^{1}_{\mu\rightarrow e; e\tau}
$
\\\hline
\end{tabular}
  \caption{\label{tab:coef_comb}. The leading coefficient of each 
  $\epsilon_{\alpha\beta}$ and $\alpha_{ij}$, for $\nu_\mu\rightarrow\nu_\mu$ and 
  $\nu_\mu\rightarrow\nu_e$. $\mathcal{RC}^{x}_{\alpha\rightarrow\beta; \gamma\delta}$ ($\mathcal{IC}^{x}_{\alpha\rightarrow\beta; \gamma\delta}$ is the coefficient for real (image) part of $\gamma\delta$ in $\alpha\rightarrow\beta$, which is of the order $x$.}
\end{table}

In Eq.~\eqref{Prob_disapp_NSI}, coefficients of $\epsilon_{\mu\tau}$ and $\tilde{\epsilon}_{\tau\tau}$ appear at leading order, i.e., at the order $\mathcal{C}^{0}_{\mu \rightarrow \mu;\ \text{element}}$. 
However, the coefficient of $\tilde{\epsilon}_{\tau\tau}$ is proportional to the factor $(c_{23}^2-s_{23}^2)$, which is suppressed since $\theta_{23}\sim 45^\circ$.
Coefficients of $\tilde{\epsilon}_{ee}$, $\epsilon_{e\mu}$, $\epsilon_{e\tau}$, which are of the 2nd, 1st and 1st order, respectively, have 
less influence on $P(\nu_\mu\rightarrow\nu_\mu)$.
Therefore, the impact of NSIs on the disappearance channel is dominated by $\epsilon_{\mu\tau}$. 
On the other hand, from Eq.~\eqref{Prob_app_NSI}, it is obvious that the largest contributions to the transition probability are from $\epsilon_{e\mu}$ and $\epsilon_{e\tau}$, with coefficients of the 1st order.
In Table~\ref{tab:coef_comb}, we present coefficients for $\alpha_{mn}$, based on Eqs.~\eqref{Prob_disapp_NSI} and \eqref{Prob_app_NSI} and Table \ref{tab:prob_coeff_texture}.

\end{appendix}

\end{document}